\RequirePackage[british]{babel}
\documentclass[reqno,a4paper,10pt]{amsart}
\usepackage{etex}
\usepackage{fixltx2e}
\usepackage[utf8x]{inputenc}
\usepackage[T1]{fontenc}
\usepackage{tgpagella}
\usepackage[small]{eulervm}
\usepackage{amsmath,amssymb,amstext,amsthm,amscd,mathrsfs,eucal}
\usepackage[hmargin=2.5cm,vmargin=2.5cm]{geometry}
\usepackage{graphicx,color}
\usepackage{multicol}
\usepackage {array,colortbl,xtab,framed,subfigure}
\usepackage[all]{xy}
\usepackage[noadjust]{cite}
\usepackage{tikz}
\usepackage{hyperref}
\hypersetup{%
  pdftitle   = {Symmetric M-theory backgrounds},
  pdfkeywords = {supergravity, isometries, homogeneity, M-theory, symmetric},
  pdfauthor  = {José Figueroa-O'Farrill},
  pdfcreator = {\LaTeX\ with package \flqq hyperref\frqq}
}
\newcommand{\half}{\tfrac{1}{2}}

\newcommand{\fg}{\mathfrak{g}}

\newcommand{\fk}{\mathfrak{k}}

\newcommand{\fp}{\mathfrak{p}}

\newcommand{\fS}{\mathfrak{S}}

\newcommand{\fsl}{\mathfrak{sl}}
\newcommand{\fso}{\mathfrak{so}}
\newcommand{\fsp}{\mathfrak{sp}}

\newcommand{\fsu}{\mathfrak{su}}
\newcommand{\fu}{\mathfrak{u}}
\newcommand{\fusp}{\mathfrak{usp}}

\newcommand{\SO}{\mathrm{SO}}

\ifx\Sp\UnDeFiNeD
\newcommand{\Sp}{\mathrm{Sp}}
\else
\renewcommand{\Sp}{\mathrm{Sp}}
\fi
\renewcommand{\Im}{\mathrm{Im}}

\newcommand{\SL}{\mathrm{SL}}
\newcommand{\SU}{\mathrm{SU}}
\newcommand{\U}{\mathrm{U}}

\newcommand{\RR}{\mathbb{R}}
\newcommand{\CC}{\mathbb{C}}
\newcommand{\HH}{\mathbb{H}}

\newcommand{\ZZ}{\mathbb{Z}}

\newcommand{\eM}{\mathcal{M}}

\DeclareMathOperator{\ad}{ad}
\DeclareMathOperator{\CW}{CW}

\DeclareMathOperator{\AdS}{AdS}
\DeclareMathOperator{\dS}{dS}

\DeclareMathOperator{\dvol}{dvol}

\DeclareMathOperator{\Mat}{Mat}

\DeclareMathOperator{\tr}{tr}
\DeclareMathOperator{\Ric}{Ric}

\DeclareMathOperator{\SLAG}{SLAG}
\DeclareMathOperator{\ASSOC}{ASSOC}
\newcommand{\be}{\boldsymbol{e}}

\newcommand{\CP}{\mathbb{C}\text{P}}
\newcommand{\CH}{\mathbb{C}\text{H}}
\newcommand{\HP}{\mathbb{H}\text{P}}
\definecolor{gray}{rgb}{0.5,0.5,0.5}
\newcommand{\MUNCH}[1]{\relax}

\allowdisplaybreaks[1]
\begin{document}
\title{Symmetric M-theory backgrounds}
\author{José Figueroa-O'Farrill}
\address{Maxwell and Tait Institutes, School of Mathematics, The University of Edinburgh, King's Buildings, Edinburgh EH9 3JZ, Scotland, United Kingdom}
\thanks{EMPG-11-28}
\begin{abstract}
  We classify symmetric backgrounds of eleven-dimensional supergravity up to local isometry.  In other words, we classify triples $(M,g,F)$, where $(M,g)$ is an eleven-dimensional lorentzian locally symmetric space and $F$ is an invariant $4$-form, satisfying the equations of motion of eleven-dimensional supergravity.  The possible $(M,g)$ are given either by (not necessarily nondegenerate) Cahen--Wallach spaces or by products $\AdS_d \times M^{11-d}$ for $2\leq d \leq 7$ and $M^{11-d}$ a not necessarily irreducible riemannian symmetric space.  In most cases we determine the corresponding $F$-moduli spaces.
\end{abstract}
\maketitle
\tableofcontents

\section{Introduction and motivation}
\label{sec:intr-motiv}

It is now a theorem that 10- and 11-dimensional supergravity backgrounds preserving more than one half of the supersymmetry --- hereafter referred to as $\half+$ backgrounds --- are homogeneous \cite{FMPHom,EHJGMHom,HomogThm}.  In fact, those those symmetries of the background which are a consequence of the supersymmetry already act transitively.  The more precise statement is that the tangent space to the spacetime at any point is spanned by Killing vectors which are constructed by squaring the Killing spinors of the background.  This result implies that classifying homogeneous supergravity backgrounds automatically classifies $\half+$ backgrounds --- this being one of the main goals in the study of supergravity backgrounds.

Let us say an eleven-dimensional supergravity background $(M,g,F)$ is \textbf{homogeneous} if $(M,g)$ is a homogeneous eleven-dimensional lorentzian manifold, $F$ is an invariant closed $4$-form, and both $g$ and $F$ satisfy the field equations of eleven-dimensional supergravity with zero gravitino, to be reviewed in Section \ref{sec:symm-supergr-backgr}.  If in addition $(M,g)$ is a symmetric space, then we say that $(M,g,F)$ is a \textbf{symmetric} background.  For a homogeneous background, the field equations, which in general are nonlinear PDEs, become algebraic; whereas for a symmetric background these equations are further simplified due to the fact that an invariant form on a symmetric space is automatically closed and coclosed.

This paper represents a first step in the classification of homogeneous supergravity backgrounds of eleven-dimensional supergravity.  In it, we will classify those homogeneous supergravity backgrounds which are symmetric.  A second paper \cite{FOHIIB} will report on the similar classification for type IIB supergravity, whereas a third paper \cite{FOUM2} will report on a classification of homogeneous backgrounds which are candidate dual geometries to M2 branes.  One of the lessons learnt from this body of work is that it is possible to make real progress on the classification problem of homogeneous supergravity backgrounds, thanks in part to the ongoing progress in our understanding of homogeneous lorentzian geometry.  This is perhaps not so evident in this paper, but underlies the results in \cite{FOUM2}.

This paper is organised as follows.  In Section \ref{sec:local-geometries} we discuss the basic geometric ingredients: namely the lorentzian symmetric spaces.  Since we do not assume that the space be irreducible (or even indecomposable), we must take into account products of lorentzian symmetric spaces with riemannian symmetric spaces, which themselves need not be irreducible.  Therefore in Section~\ref{sec:irred-riem-symm} we review Cartan's classification of irreducible riemannian symmetric spaces, whereas in Section \ref{sec:indec-lorentz-symm} we review the classification of indecomposable lorentzian symmetric spaces due to Cahen and Wallach.  In Section \ref{sec:stat-symm-backgr} we put the previous results together and count the number of (families of) eleven-dimensional lorentzian symmetric spaces.  We find that there are 1978 such families which, in principle, should be checked to see whether they can support the necessary fluxes to become a supergravity background.  Luckily, after a little work, one can rule out many of those geometries.  This is done in Section \ref{sec:symm-supergr-backgr}, which starts in Section \ref{sec:supergr-field-equat} with a review of the supergravity field equations.  In Section \ref{sec:cases-where-f=0} we deal with the case of backgrounds with zero flux and show that they are all given by (possibly decomposable) Cahen--Wallach pp-waves.  In Section \ref{sec:case-d=1} we look at those cases where the geometry is static with a metric of the form $-dt^2 + h$, where $h$ is a riemannian (locally) symmetric metric.  In Section \ref{sec:no-de-sitter} we rule out backgrounds where the lorentzian factor is locally isometric to de Sitter space.  Finally, in Section \ref{sec:cases-where-m_0} we consider those backgrounds where the lorentzian factor is a Cahen--Wallach pp-wave and conclude that all such backgrounds are given by (possibly decomposable) Cahen--Wallach spacetimes.  This leaves the problem of classifying those backgrounds whose geometry is of the form $\AdS_{2\leq d\leq 7} \times M^{11-d}$.  There are 568 possible geometries, many of which are ruled out and the rest are examined in detail in Section \ref{sec:AdS-backgrounds}, where we list the possible backgrounds and in most cases determine the $F$-moduli space of allowed fluxes.  The paper ends with two appendices.  In Appendix \ref{sec:geom-some-symm} we collect detailed information on the geometry of certain riemannian symmetric spaces which we needed in the calculations of Section~\ref{sec:AdS-backgrounds} and in Appendix \ref{sec:geometries-which-do} we list, for completeness, those anti~de~Sitter geometries which cannot support symmetric supergravity backgrounds due to the lack of suitable fluxes.

The symmetric backgrounds are to be found throughout the paper in boxed equations, such as equation \eqref{eq:CWbgs} for the Cahen--Wallach backgrounds.  The many anti~de~Sitter backgrounds in Section \ref{sec:AdS-backgrounds} are also similarly adorned.  There are many figures in the paper summarising the moduli space of fluxes for some of the different anti~de~Sitter geometries.  We should emphasise that we only work with local metrics, so that topology will not play a significant rôle.  In particular, when we write $S^1$ or $T^n$ we do not necessarily mean a circle or a flat torus, but simply a flat space of that dimension.  We could have equally well written $\RR$ or $\RR^n$.

\subsection*{A note to the reader}

Given the vast literature on supergravity solutions, it is very likely that some of the symmetric backgrounds constructed in this paper coincide with backgrounds which have appeared in earlier work, perhaps as special cases of a more general family of backgrounds.  This is most certainly the case with the Freund--Rubin $\AdS_4$ backgrounds in Section \ref{sec:ads_4-backgrounds} and the Cahen--Wallach backgrounds in Section \ref{sec:cases-where-m_0}, but I expect more examples to surface.  I would appreciate being notified of such cases so that I may both properly attribute the discovery of the background and place this work in a broader context.

\section{The local geometries}
\label{sec:local-geometries}

In this section we list the basic ingredients of the classification: namely, lorentzian symmetric spaces.  A lorentzian (locally) symmetric space $(M,g)$ is locally isometric to a product
\begin{equation}
  M = M_0 \times M_1 \times \cdots \times M_k
\end{equation}
where $M_0$ is an indecomposable lorentzian symmetric space and the $M_i$ for $i>0$ are irreducible riemannian symmetric spaces.  We now proceed to list these basic ingredients.

\subsection{Irreducible riemannian symmetric spaces}
\label{sec:irred-riem-symm}

Irreducible riemannian symmetric spaces were classified by Élie Cartan (see, e.g., \cite{Helgason}).  Every symmetric space is determined locally by a pair of real Lie algebras $(\fg, \fk)$ with $\fk \subset \fg$.  Letting $\fp$ denote an $\fk$-invariant complement of $\fk$ in $\fg$, so that $\fg = \fk \oplus \fp$, we have that
\begin{equation}
  [\fk,\fk] \subset \fk \qquad [\fk,\fp] \subset \fp \qquad
  [\fp,\fp]\subset \fk~.
\end{equation}
The holonomy algebra is $\fk$ and the holonomy representation is the action of $\fk$ on $\fp$ induced from the adjoint representation of $\fg$.  Therefore the space of parallel forms in the symmetric space corresponding to $(\fg,\fk)$ is isomorphic to the $\fk$-invariant subspace of $\Lambda^*\fp$.

Table~\ref{tab:rss} lists all the irreducible riemannian symmetric spaces of dimension less than or equal to $10$, together with the ranks of the parallel forms.  Each row in the diagram corresponds to two symmetric spaces: one compact and one noncompact.  The names in the last column correspond to the compact spaces and use the following notation $G^+_\RR(p,n)$ denotes the grassmannian of oriented real $p$-planes in $\RR^n$, $G_\CC(p,n)$ is the grassmannian of complex $p$-planes in $\CC^n$, $\ASSOC$ is the grassmannian of associative $3$-planes in $\RR^7$ and $\SLAG_n$ in the grassmannian of special lagrangian planes in $\CC^n$.

\begin{table}[h!]
  \centering
  \renewcommand{\arraystretch}{1.3}
  \caption{Irreducible $d$-dimensional riemannian symmetric spaces with $d\leq 10$.  The names are those of the compact forms.}\label{tab:rss}
  \begin{tabular}{>{$}r<{$}|>{$}c<{$}|>{$}c<{$}|>{$}c<{$}|>{$}l<{$}|>{$}l<{$}}
    \multicolumn{1}{c|}{dim} & \fg~\text{(compact)} & \fg~\text{(noncompact)} & \fk & \multicolumn{1}{c|}{$\fk$-inv. forms} & \multicolumn{1}{c}{Name}\\
    \hline
    2 & \fu(2) & \fu(1,1) & \fu(1) \oplus \fu(1) & 0,2 & S^2\\[3pt]
    3 & \fsu(2) \oplus \fsu(2) & \fsl(2,\CC) & \fsu(2) & 0,3 & S^3\\[3pt]
    4 & \fu(3) & \fu(2,1) & \fu(2) \oplus \fu(1) & 0,2,4 & \CP^2\\
    4 & \fsp(2) & \fsp(1,1) & \fsp(1) \oplus \fsp(1) & 0,4 & S^4\\[3pt]
    5 & \fsu(3) & \fsl(3,\RR) & \fso(3) & 0,5 & \SLAG_3\\ 
    5 & \fsu(4) & \fsl(2,\HH) & \fsp(2) & 0,5 & S^5\\[3pt]
    6 & \fu(4) & \fu(3,1) & \fu(3) \oplus \fu(1) & 0,2,4,6 & \CP^3\\
    6 & \fso(7)  & \fso(6,1)  & \fso(6) & 0,6 & S^6\\
    6 & \fsp(2) & \fsp(2,\RR) & \fu(2) & 0,2,4,6 & G^+_\RR(2,5)\\[3pt]
    7 & \fso(8)  & \fso(7,1)  & \fso(7) & 0,7 & S^7\\[3pt]
    8 & \fu(4) & \fu(2,2) & \fu(2) \oplus \fu(2) & 0,2,4^2,6,8 & G_\CC(2,4)\\
    8 & \fu(5) & \fu(4,1) & \fu(4) \oplus \fu(1) & 0,2,4,6,8 & \CP^4 \\
    8 & \fso(9)  & \fso(8,1)  & \fso(8) & 0,8 & S^8\\
    8 & \fsp(3) & \fsp(2,1) & \fsp(2) \oplus \fsp(1) & 0,4,8 & \HP^2\\
    8 & \fg_{2(-14)} & \fg_{2(2)} & \fsp(1) \oplus \fsp(1) & 0,4,8 & \ASSOC \\
    8 & \fsu(3) \oplus \fsu(3) & \fsl(3,\CC) & \fsu(3) & 0,3,5,8 & \SU(3)\\[3pt]
    9 & \fsu(4) & \fsl(4,\RR) & \fso(4) & 0,4,5,9 & \SLAG_4\\
    9 & \fso(10) & \fso(9,1) & \fso(9) & 0,9 & S^9\\[3pt]
    10 & \fu(6) & \fu(5,1) & \fu(5) \oplus \fu(1) & 0,2,4,6,8,10 & \CP^5 \\
    10 & \fso(11) & \fso(10,1) & \fso(10) & 0,10 & S^{10}\\
    10 & \fso(7)  & \fso(5,2)  & \fso(5) \oplus \fso(2) & 0,2,4,6,8,10 & G^+_\RR(2,7)\\
    10 & \fsp(2) \oplus \fsp(2) & \fsp(2,\CC) & \fsp(2) & 0,3,7,10 & \Sp(2)
  \end{tabular}
\end{table}

There are several cases where two spaces have the same invariant forms.  One could ask whether they are diffeomorphic, but have different metrics (perhaps related by ``squashing'').  There are no diffeomorphic pairs, though.

It is instructive to understand why $G:=G_\RR^+(2,7) \not\cong \CP^5$.  I owe the following beautiful argument to Elmer Rees.  Let $V:=V_2^+(\RR^7)$ denote the Stiefel manifold of oriented orthonormal $2$-frames on $\RR^7$.  Clearly $V$ fibres over $G$ with fibres $\SO(2)$.  This is the principal circle bundle associated to the tautological bundle of $G$, whose first Chern class is the generator of $H^2(G)$.  In contrast the principal circle bundle associated to the line bundle with first Chern class the generator of $H^2(\CP^5)$ is the $11$-sphere $S^{11}$.  Thus we will have shown that $G \not\cong \CP^5$ if we show that $V \not\cong S^{11}$.  To see this we notice that $V$ fibres over $S^6$ with fibre $S^5$.  The map $V \to S^6$ is given explicitly by sending the ordered $2$-frame $(\be_1, \be_2)$ to $\be_1 \in S^6 \subset \RR^7$.  The fibre consists of all unit norm vectors on the $\RR^6 \subset \RR^7$ perpendicular to $\be_1$.  In other words, the fibration $V \to S^6$ is nothing but the sphere tangent bundle of $S^6$.  We now use the Gysin sequence associated to this fibration, which contains a short exact sequence of the form
\begin{equation}
  \begin{CD}
    0 @>>> H^5(V) @>\pi_*>> H^0(S^6) @>e\wedge>> H^6(S^6) @>\pi^*>>
    H^6(V) @>>> 0
  \end{CD} \end{equation}
where $e$ is the Euler class of the fibration.  Because the Euler characteristic of $S^6$ is equal to $2$, the middle map is actually
\begin{equation}
  \begin{CD}
    \ZZ @>2\times>> \ZZ
  \end{CD}
\end{equation}
whence $H^5(V) = 0$ and $H^6(V) \cong \ZZ/2\ZZ$, and hence $V \not\cong S^{11}$ since $H^6(S^{11}) = 0$.  A similar argument shows that $G_\RR^+(2,5) \not\cong \CP^3$.  Similarly, one can show that $\ASSOC \not\cong \HP^2$.  In fact, they are not even homotopy equivalent \cite{MO61521}.  Finally, $\SLAG_3 \not\cong S^5$; although it is a homology sphere.

\subsection{Indecomposable lorentzian symmetric spaces}
\label{sec:indec-lorentz-symm}

Indecomposable lorentzian symmetric spaces have also been classified \cite{CahenWallach,CahenParker}.  Apart from the space forms (i.e., de~Sitter and anti de~Sitter spaces), there is a third family of indecomposable lorentzian symmetric spaces with solvable transvection group.  The corresponding symmetric pair of Lie algebras is defined as follows.

Let $V$ be a $(d{-}2)$-dimensional real vector space with euclidean inner product $\left<-,-\right>$.  Let $V^*$ be the dual space and let $A\in S^2V^*$ be a symmetric bilinear form on $V$.  The inner product induces musical isomorphisms
\begin{equation}
  \flat : V \to V^* \quad\text{and}\quad \sharp : V^* \to V~.
\end{equation}
Similarly, $A$ defines a linear map $V\to V$, also denoted $A$ and defined by $\left<A(v_1),v_2\right> = A(v_1,v_2)$.  Let $Z$ be a one-dimensional real vector space with basis $e_+$ and let $Z^*$ be its dual, with canonical dual basis $e_-$.  Let $\fg_A = V \oplus V^* \oplus Z \oplus Z^*$ with the following nonzero Lie brackets:
\begin{equation}
  [e_-, v] = v^\flat \qquad [e_-,\alpha] = A(\alpha^\sharp) \qquad [\alpha,v] = A(v, \alpha^\sharp)\, e_+~,
\end{equation}
where $v\in V$, $\alpha\in V^*$, $e_+\in Z$ and $e_- \in Z^*$ its dual.  Notice that, in particular, $Z$ is central and $V$ and $V^*$ are abelian.  These brackets define a Lie algebra by virtue of $A$ being symmetric.  It is evident, moreover, that $\fg_A''$ is central, whence $\fg_A$ is solvable.

Let $\fk_A = V^*$ and $\fp_A = V \oplus Z \oplus Z^*$.  This is a symmetric split of $\fg_A$ and the corresponding symmetric space inherits the metric from the $\fk_A$-invariant symmetric bilinear form $B$ on $\fp_A$ defined by
\begin{equation}
  B(v_1,v_2) = \left<v_1,v_2\right> \quad\text{and}\quad B(e_+,e_-) = 1~,
\end{equation}
for $v_1,v_2\in V$.  In particular, $Z$ and $Z^*$ are isotropic subspaces.  This means that $B$ has signature $(d-1,1)$.

Let $G_A$ denote the simply-connected Lie group with Lie algebra $\fg_A$ and let $K_A$ denote the Lie subgroup corresponding to $\fk_A$.  Then the bilinear form $B$ on $\fp_A$ gives rise to a lorentzian metric on $M_A = G_A/K_A$.  It is easy to find an explicit coordinate expression for this metric.  Let $e_i$ be an orthonormal basis for $V$ (with corresponding dual basis $e_i^*$ for $V^*$), and introduce local coordinates $x^+, x^-, x^i$ associated to the basis $\{e_+, e_-, e_i\}$ for $\fp_A$.  This allows us to embed $M_A$ in $G_A$ via $\sigma: M_A \to G_A$, where
\begin{equation}
  \sigma(x^+, x^-, x^i) = \exp(x^+ e_+) \exp(x^- e_-) \exp\left(\sum_i x^i e_i\right)~.
\end{equation}
Pulling back the left-invariant Maurer--Cartan form from $G_A$ to $M_A$ via $\sigma$, we obtain
\begin{equation}
  \sigma^{-1} d\sigma = \omega + \theta~,
\end{equation}
where the $\fk_A$-connection $\omega$ and the $\fp_A$-valued soldering form $\theta$ are given by
\begin{equation}
  \omega = dx^- \sum_i x^i e^*_i
\end{equation}
and
\begin{equation}
  \theta = dx^- e_- + \sum_i dx^i e_i + \left(dx^+ + \half \sum_{i,j} A_{ij}x^i x^j dx^-\right) e_+~.
\end{equation}
The $G_A$-invariant metric on $M_A$ is given, in these coordinates, by 
\begin{equation}
  \label{eq:CWmetric}
  ds^2 = B(\theta, \theta) = 2 dx^+ dx^- + \sum_{i,j} A_{ij} x^i x^j (dx^-)^2 +  \sum_i dx^i dx^i
\end{equation}
It is not hard to show that the resulting lorentzian symmetric space is indecomposable if and only if $A$ is nondegenerate.

It is also clear that if $A' = c\, O A O^t$, where $c$ is a positive scalar and $O:V \to V$ is an orthogonal transformation, then $M_A$ and $M_{A'}$ are isometric.  Indeed, the diffeomorphism is given explicitly by rotating the $x^i$ by $O$ and then rescaling $x^\pm \mapsto \sqrt{c^{\pm1}} x^\pm$.

This allows us to diagonalise $A$, choose an ordering for the eigenvalues and rescale them by a positive constant, and in this way parametrise the family of metrics.  Doing so one sees that for $d=\dim M_A \geq 3$, the indecomposable such metrics live in a $(d-3)$-dimensional family, parametrised by $\eM_d = (S^{d-3} - \Delta)/\fS_{d-2}$, where
\begin{equation}
  \Delta = \left\{(\lambda_1,\lambda_2,\ldots,\lambda_{d-2}) \in S^{d-3}\subset \RR^{d-2} \mid \lambda_1 \lambda_2 \cdots \lambda_{d-2} = 0 \right\}
\end{equation}
is the subset of $S^{d-3}$ consisting of points where at least one of the coordinates is $0$, and $\fS_{d-2}$ is the permutation group acting in the obvious way in $\RR^{d-2}$.  Every point in $\eM_d$ is thus uniquely determined by an ordered $(d-2)$-tuple of nonzero real numbers $\{\lambda_i\}$ normalised to having unit norm.  The corresponding metric can be read off from \eqref{eq:CWmetric}:
\begin{equation}
  ds^2 = 2 dx^+ dx^- + \sum_{i=1}^{d-2} \lambda_i\, (x^i)^2 (dx^-)^2 + \sum_{i=1}^{d-2} (dx^i)^2~.
\end{equation}
Let us refer to these spaces as \emph{Cahen--Wallach} spaces, and denote them by $\CW_d(\lambda)$, where $\lambda \in \eM_d$ or, more invariantly, by $\CW_d(A)$, with the understanding that $\CW_d(A) = \CW_d(A')$ if and only if $A'= c\, O A O^t$ as before.

The Riemann curvature and Ricci tensors of the Cahen--Wallach space $\CW_d(A)=\CW_d(\lambda)$ have the following nonzero components
\begin{equation}
  R_{-i-j}=-A_{ij} = -\lambda_i \delta_{ij} \qquad\text{and}\qquad R_{--}= -\tr A = -\sum_{i=1}^{d-2} \lambda_i~,
\end{equation}
whence the scalar curvature vanishes.

The holonomy of the Cahen--Wallach space $\CW_d(A)$ is isomorphic to $\fk_A \cong \RR^{d-2}$ and the representation is induced by the adjoint action of $\fk_A$ on $\fp_A$.  It is not hard to show that the $\fk_A$-invariant subspace of $\Lambda \fp_A$ is spanned by the constants together with polyvectors of the form $e_+ \wedge \phi$, where $\phi \in \Lambda V$.  This means that the corresponding parallel forms on $\CW_d(A)$ are the constants together with $dx^- \wedge \varphi$ where $\varphi$ is any constant-coefficient form
\begin{equation}
  \varphi = \sum_{1\leq i_1<i_2<\cdots<i_p \leq d-2} c_{i_1i_2\cdots i_p} dx^{i_1} \wedge dx^{i_2} \wedge \cdots \wedge dx^{i_p}~.
\end{equation}

Table~\ref{tab:lss} lists the indecomposable lorentzian symmetric spaces together with the ranks of the parallel forms.  This information is encoded in the Poincaré polynomial:
\begin{equation}
  P(t) = \sum_{i=0}^d b_i t^i~,\qquad\text{where $b_i = \dim_\RR \left(\Lambda^i \fp\right)^{\fk}$.}
\end{equation}
In the Table, the notations $\fg(\lambda)$ and $\fk(\lambda)$ refer to the Lie algebras $\fg_A$ and $\fk_A$ defined above, where $\lambda \in \eM_d$ characterises the symmetric bilinear form $A$ up to the equivalence relation defined above.

\begin{table}[h!]
  \centering
  \renewcommand{\arraystretch}{1.3}
  \caption{Indecomposable $d$-dimensional lorentzian symmetric spaces.}
  \label{tab:lss}
  \begin{tabular}{>{$}c<{$}|>{$}c<{$}|>{$}c<{$}|>{$}l<{$}}
    \text{type} & \fg & \fk & \multicolumn{1}{c}{$\fk$-invariant forms}\\
    \hline
    \dS_d & \fso(d,1) & \fso(d{-}1,1) & 1+t^d\\
    \AdS_d & \fso(d{-}1,2) & \fso(d{-}1,1) & 1+t^d\\
    \CW_d(\lambda) & \fg(\lambda) & \fk(\lambda) & 1+t(1+t)^{d-2} + t^d
  \end{tabular}
\end{table}

\subsection{Statistics of symmetric backgrounds}
\label{sec:stat-symm-backgr}

Let us count the number of (families of) eleven-dimensional lorentzian symmetric spaces.  Every indecomposable symmetric space (except for the Cahen--Wallach spaces) has a parameter corresponding to rescaling the metric.  As discussed above, indecomposable $d$-dimensional Cahen--Wallach spaces come in ($d-3$)-parameter families.  We will ignore these parameters in the counting, whence we will count families of geometries and not the geometries themselves.  Each such geometry is of the form $L_d \times R_{11-d}$, where $L$ is a $d$-dimensional indecomposable lorentzian symmetric space and $R$ is an ($11-d$)-dimensional riemannian symmetric space which is made out of the ingredients in Table~\ref{tab:rss}.  Let $i_d$ denote the number of irreducible $d$-dimensional riemannian symmetric spaces up to local isometry.  Clearly $i_1 = 1$ since $\RR$ and $S^1$ are locally isometric.  The other values of $i_d$ can be read from Table~\ref{tab:rss}:

\begin{table}[h!]
  \centering
  \caption{Number of irreducible riemannian symmetric spaces up to local isometry}
  \label{tab:nirss}
  \renewcommand{\arraystretch}{1.3}
  \begin{tabular}{c|cccccccccc}
   $d$ & 1 & 2 & 3 & 4 & 5 & 6 & 7 & 8 & 9 & 10\\
    \hline
    $i_d$ & 1 & 2 & 2 & 4 & 4 & 6 & 2 & 12 & 4 & 8
 \end{tabular}
\end{table}

We now let $r_d$ denote the number of $d$-dimensional riemannian symmetric spaces up to local isometry.  Clearly,
\begin{equation}
  \prod_{d=1}^\infty \frac{1}{1 - i_d t^d} = \sum_{d=1}^\infty r_d t^d~.
\end{equation}
Since we are interested only in $d\leq 10$, we can simply compute the first 10 terms in the left-hand side
\begin{equation}
  \prod_{d=1}^{10} \frac{1}{1 - i_d t^d} = 1 + t + 3 t^2 + 5 t^3 + 13 t^4 + 21 t^5 + 47 t^6 + 73 t^7 + 161 t^8 + 253 t^9 + 497 t^{10} +
  O\left(t^{11}\right)~,
\end{equation}
from where we can read off the values of $r_d$.  These are tabulated in Table~\ref{tab:nrss}.

\begin{table}[h!]
  \centering
  \renewcommand{\arraystretch}{1.3}
  \caption{Number of riemannian symmetric spaces up to local isometry}
  \label{tab:nrss}
  \begin{tabular}{c|cccccccccc}
    $d$ & 1 & 2 & 3 & 4 & 5 & 6 & 7 & 8 & 9 & 10\\
    \hline
    $r_d$ & 1 & 3 & 5 & 13 & 21 & 47 & 73 & 161 & 253 & 497
  \end{tabular}
\end{table}

Finally we let $\ell_d$ denote the number of indecomposable lorentzian symmetric spaces up to local isometry.  The total number of possible geometries is then
\begin{equation}
  N = \sum_{d=1}^{11} \ell_d r_{11-d}~.
\end{equation}
We notice that $\ell_1 = 1$, $\ell_2 = 2$ and $\ell_{d>2} = 3$, whence
\begin{equation}
  \begin{aligned}
    N &= 3 ( 1 + r_1 + r_2 + \cdots + r_8 ) + 2 r_9 + r_{10}\\
    &= 3 ( 1 + 1 + 3 + 5 + 13 + 21 + 47 + 73 + 161 ) + 2 \times 253 + 497\\
    &= \boldsymbol{1978}~.
  \end{aligned}
\end{equation}
Luckily we will not have to check them all.

\section{Symmetric supergravity backgrounds}
\label{sec:symm-supergr-backgr}

\subsection{The supergravity field equations}
\label{sec:supergr-field-equat}

Following the conventions of \cite{DS2brane}, the bosonic part of the action of $d{=}11$ supergravity is (setting Newton's constant to $1$)
\begin{equation}
  \label{eq:lag}
  \int_M \left( \half R \dvol - \tfrac14 F\wedge\star F + \tfrac1{12} F \wedge F \wedge A \right)~,
\end{equation}
where $F=dA$ locally, $R$ is the scalar curvature of $g$ and $\dvol$ is the (signed) volume element
\begin{equation}
  \dvol := \sqrt{|g|}\, dx^0 \wedge dx^1 \wedge \cdots \wedge
  dx^{10}~.
\end{equation}
The Euler-Lagrange equations following from \eqref{eq:lag} are
\begin{equation}
  \label{eq:EL}
  \begin{aligned}
    d \star F &= \half F\wedge F\\
    \Ric(X,Y) &= \half \left< \iota_X F, \iota_Y F \right> - \tfrac16 g(X,Y) |F|^2~,
  \end{aligned}
\end{equation}
for all vector fields $X,Y$ on $M$.  In this equation we have introduced the inner product $\left<-,-\right>$ on differential forms, defined by
\begin{equation}
   \left<\theta,\omega\right> \, \dvol = \theta \wedge \star \omega~,
\end{equation}
and the associated norm
\begin{equation}
  |\theta|^2 = \left<\theta, \theta\right>~,
\end{equation}
which in a lorentzian manifold is \emph{not} positive-definite.

It is well known that the field equations \eqref{eq:EL} are invariant under the homothetic action of $\RR^+$: $(g,F) \mapsto (e^{2t}g, e^{3t}F)$, where $t\in \RR$.  Indeed, under $g \mapsto e^{2t} g$, the Levi-Civita connection, consisting of terms of the form $g^{-1}dg$, does not change.  This means that the $(3,1)$ Riemann curvature tensor is similarly invariant, and so is any contraction such as the Ricci tensor.  Under $F \mapsto e^{3t}F$, the tensor in the right-hand side of the Einstein equation is similarly invariant, since the $e^{6t}$ coming from the two $F$s cancels the $e^{-6t}$ coming from the three $g^{-1}$s.  On the other hand, the Bianchi identity $dF=0$ is clearly invariant under homotheties and the Maxwell-like equation is as well.  Indeed, using that the Hodge $\star$ acting on $p$-forms in a $D$-dimensional manifold, scales like $e^{(D-2p)t}$ under $g \mapsto e^{2t}g$, we see that $\star$ acting on $4$-forms in $11$-dimensions scales like $e^{3t}$, just like $F$, whence both sides of the Maxwell-like equation scale in the same way: namely, $e^{6t}$.

This means that the moduli space of eleven-dimensional supergravity backgrounds admits the action of $\RR^+$ and hence it is a cone.  Some backgrounds, of course, might actually be invariant under this action.  This is the case with the Minkowski vacuum or more generally any vacuum of the form $\CW_d(A)\times \RR^{11-d}$, as we will see just before the start of Section \ref{sec:AdS-backgrounds}.

The homothetic action of $\RR^+$ can be extended to an action of $\RR^\times$ by letting $-1 \in\RR^\times$ act by simultaneously reversing the orientation of the spacetime and changing the sign of $F$.  This operation is often called ``skew-whiffing'' in the early Kaluza--Klein literature (see, e.g., \cite{DNP}).  This accounts for an additional parity symmetry in the resulting moduli spaces.

We now specialise to symmetric backgrounds where both $g$ (and hence the curvature tensors) and $F$ are invariant under the isometry group.  Every homogeneous (pseudo)riemannian manifold has a canonical connection and symmetric spaces are precisely those for which the canonical connection coincides with the Levi-Civita connection.  This means that invariant tensors are actually parallel with respect to a torsion-free connection and, in particular, this means that $dF = d\star F = 0$, reducing the field equation for $F$ to the algebraic equation $F \wedge F = 0$.

For a symmetric space $M= G/K$ with $\fg = \fk \oplus \fp$, the Ricci tensor at $o \in M$ is given by a very simple expression derived, for example, in \cite{Besse}.  If $X,Y \in \fp$, then at $o$
\begin{equation}
  \Ric(X,Y) = - \tr (\ad_X \circ \ad_Y)\bigr|_\fp~.
\end{equation}
In practice, though, we will not need to compute the Ricci tensor in this way.  The fact that for an irreducible symmetric space with symmetric pair $(\fg,\fk)$, the linear isotropy representation of $\fk$ on its complement $\fp$ is irreducible, implies that any two invariant symmetric bilinear forms are proportional.  Hence the Ricci tensor, being an invariant symmetric bilinear form, must be proportional to the metric.  The proportionality constant is then determined in terms of $F$ from the supergravity Einstein equation.

We now proceed to study the field equations.  We will see that as simple consequences of them we shall be able to discard all backgrounds except for those of the form $\CW_{d\geq 3}(A) \times \RR^{11-d}$ and $\AdS_{2\leq d\leq 7}\times M^{11-d}$.

\subsection{Cases where $F=0$}
\label{sec:cases-where-f=0}

The equations of motion for $F=0$ say that $g$ is Ricci-flat.  It is not hard to see from our list that the only Ricci-flat lorentzian symmetric spaces are of the form $\CW_d(A) \times \RR^{11-d}$, where $\tr A = 0$.  Therefore any other geometry which forces $F=0$ cannot appear.  These are listed in Appendix \ref{sec:geom-with-suit}.

\subsection{The case of $d=1$}
\label{sec:case-d=1}

In this case, the metric takes the form
\begin{equation}
  g = -dt^2 + \overline g~.
\end{equation}
The flux can be of one of three types:
\begin{enumerate}
\item $F = \overline F$,
\item $F = dt \wedge \overline F$, and
\item $F = dt \wedge \overline G + \overline F$,
\end{enumerate}
where here and in the sequel a bar denotes the absence of a $dt$ component.  Since $t$ is a flat direction, the $tt$ component of the Ricci curvature vanishes, whence
\begin{equation}
  R_{tt} = \half |F_t|^2 + \tfrac16 |F|^2 = 0~,
\end{equation}
where $F_t$ is the $3$-form defined by $F_t(X,Y,Z) = F(\partial_t, X,Y,Z)$.  Applying this equation to each of the above three cases in turn, we find
\begin{enumerate}
\item $0 = \tfrac16 |\overline F|^2 \implies |\overline F|=0 \implies \overline F=0 \implies F=0$,
\item $0 = \half |\overline F|^2 - \tfrac16 |\overline F|^2 \implies |\overline F| = 0 \implies \overline F = 0 \implies F =0$, and
\item $0 = \half |\overline G|^2 + \tfrac16 |F|^2$, but $|F|^2 = |\overline F|^2 - |\overline G|^2$, whence $\tfrac13 |\overline G|^2 + \tfrac16 |\overline F|^2 = 0 \implies \overline G = \overline F = 0 \implies F =0$.
\end{enumerate}
Therefore the only solution is such that $F = 0$, which means that $\overline g$ is Ricci-flat and hence flat.  The only solution is therefore the Minkowski vacuum of eleven-dimensional supergravity.  

\subsection{No de Sitter backgrounds}
\label{sec:no-de-sitter}

We will now show that no de Sitter background can occur.  Let $\mu,\nu,...$ be lorentzian indices and $a,b,...$ be riemannian indices.  If $F = \overline F$, meaning that it has no lorentzian components, the Einstein equations are
\begin{equation}
  R_{\mu\nu} = - \tfrac16 |F|^2 g_{\mu\nu}~,
\end{equation}
which implies that the lorentzian factor is $\AdS_d$.  This gets rid of any case where $F = \overline F$ is forced and yet the lorentzian factor is not $\AdS_d$.  This includes any background of the form $\dS_d \times M^{11-d}$ for $d \geq 5$.  We can do better, though, and in fact discard all the de Sitter backgrounds.

Indeed, it is easy to dispose of the backgrounds of the form $\dS_{d\leq 4} \times M^{11-d}$ as well.  For such backgrounds the most general $4$-form $F$ is of the form $F = \nu_d \wedge \omega + \overline F$, where $\nu_d$ is the volume form of $\dS_d$, $\omega \in \Omega^{4-d}(M)$ and $\overline F \in \Omega^4(M)$.  It follows that $|F|^2 = |\overline F|^2 - |\omega|^2$, whence the
$\dS_d$-components of the Einstein equation are
\begin{equation}
  R_{\mu\nu} = \half \left<F_\mu,F_\nu\right> - \tfrac16 g_{\mu\nu} (|\overline F|^2 - |\omega|^2)~.
\end{equation}
However $\left<F_\mu,F_\nu\right> = - |\omega|^2 g_{\mu\nu}$, whence
\begin{equation}
    R_{\mu\nu} = - \tfrac16 g_{\mu\nu} (2|\omega|^2 + |\overline F|^2)~,
\end{equation}
which is absurd, since $\dS_d$ has positive scalar curvature.

\subsection{Cases where $M_0 = \CW_d(A)$}
\label{sec:cases-where-m_0}

The near Ricci-flatness of Cahen--Wallach spaces and the non-Ricci flatness of riemannian symmetric spaces can be very useful in eliminating many other candidates from the list.  Let us assume a geometry of the type $\CW_d(A) \times M^{11-d}$.

\subsubsection{The case $d>4$}
\label{sec:case-dgt4}

First we shall assume that $d>4$.  In that case, letting $i,j,\dots$ denote transverse indices in $\CW_d(A)$, the $4$-form $F$ can only have the following nonzero components: $F_{-ijk}$, $F_{-ija}$, $F_{-iab}$, $F_{-abc}$ and $F_{abcd}$.  The riemannian components of the Einstein equations in this case are
\begin{equation}
  \label{eq:einsteinRiem}
  R_{ab} = \tfrac12 \left<F_a,F_b\right> - \tfrac16 |F|^2 g_{ab}~.
\end{equation}
Now, if $F_{abcd} = 0$ then $|F|=0$ and hence the Einstein equations become
\begin{equation}
  R_{ab} = \half \left<F_a,F_b\right> = 0~,
\end{equation}
whence $M$ is flat and the solution is locally isometric to $\CW_d(A) \times \RR^{11-d}$.  On the other hand, if $F_{abcd} \neq 0$, then $|F|\neq 0$ and the Einstein equations say that
\begin{equation}
  R_{ij} = \tfrac16 |F|^2 g_{ij} \neq 0~,
\end{equation}
which is absurd, unless $d=2$ (so that there are no transverse dimensions).  But $\CW_2$ is flat, so we are actually back to the $d=1$ case discussed in Section~\ref{sec:case-d=1}, resulting in the Minkowski vacuum.  This deals with all $\CW_d \times M^{11-d}$ from the list, except for $d\leq 4$.

\subsubsection{The case $d=4$}
\label{sec:case-d=4}

Here we can have one extra component for the $4$-form, namely $f = F_{+-12}$.  The Einstein equation together with the vanishing of the $R_{+-}$ component of the Ricci tensor become
\begin{equation}
  \label{eq:leftf_+-f_-right}
  \left<F_+,F_-\right> = \tfrac13 g_{+-} |F|^2~.
\end{equation}
This equation is still true relative to a Witt frame, in which case both parts of this equation become
\begin{equation}
  \left<F_+,F_-\right> = -f^2  \qquad\text{and}\qquad  |F|^2 = |\overline F|^2 - f^2~,
\end{equation}
where $\overline F$ has components $F_{abcd}$.  Into equation \eqref{eq:leftf_+-f_-right}, we now see that
\begin{equation}
  \tfrac13 |\overline F|^2 + \tfrac23 f^2 = 0 \implies f = \overline F = 0~.
\end{equation}
Thus we are reduced to the case $d>4$ treated above, with the same conclusion.

\subsubsection{The case $d=3$}
\label{sec:case-d=3}

The $4$-form can have the following components: $F_{+-1a}$, $F_{-1ab}$, $F_{-abc}$ and $F_{abcd}$.  The component $F_{+-1a}$ is nonzero if and only there is an $S^1$ factor in $M$.  But this is then basically a degenerate case of $\CW_4$.  We did not use nondegeneracy when discussing the $d=4$ case, whence the same result obtains.

In summary, we have that the only possible backgrounds with a Cahen--Wallach space are $\CW_d(A) \times \RR^{11-d}$ and $d\geq 3$.  As we have seen above, the $4$-form is $F = dx^- \wedge \varphi$.  It follows that $|F|=0$ and $F \wedge F = 0$, so the only field equation which is not trivially satisfied is the ${-}{-}$ component of the Einstein equation:
\begin{equation}
  R_{--} = \half \left<F_-,F_-\right> = \half |\varphi|^2~.
\end{equation}
This simply imposes the condition $\tr A = - \half |\varphi|^2$ on the matrix $A$ defining the Cahen--Wallach space.

In summary, we have the following family of backgrounds, with $\varphi$ a transverse constant-coefficient $3$-form (relative to the flat coordinates):
\begin{equation}\label{eq:CWbgs}
  \boxed{
    \CW_{d>2}(A) \times \RR^{11-d} \qquad F = dx^- \wedge \varphi \qquad  \tr A = -\half |\varphi|^2
  }
\end{equation}
Such backgrounds are invariant under the homothetic action of $\RR^+$.  Indeed, if we let $(g,F)\mapsto(e^{2t}g,e^{3t}F)$, then we may undo this via the following diffeomorphism: $x^- \mapsto x^-$, $x^i \mapsto e^t x^i$ and $x^+ \mapsto e^{2t} x^+$; that is,
\begin{equation}
  e^{2t} g(x^+,x^-,x^i) = g(e^{2t}x^+,x^-,e^t x^i) \qquad\text{and}\qquad
  e^{3t} F(x^+,x^-,x^i) = F(e^{2t}x^+,x^-,e^t x^i)~.
\end{equation}

Symmetric Cahen--Wallach backgrounds have been considered in earlier work \cite{FOPflux,Limits,ChrisJerome,Michelson26}.  In fact, since the property of being symmetric is preserved under geometric limits \cite{Geroch,Limits}, the plane-wave limits of the AdS backgrounds to be discussed in the next section are symmetric Cahen--Wallach backgrounds.  This was, in fact, the original interest in Cahen--Wallach backgrounds.  Although we shall not do this in this paper, we should remark that the possible plane-wave limits of symmetric spaces are easy to determine.  In fact, symmetric spaces are particular examples of geodesic orbit spaces, which are manifolds whose geodesics coincide with the orbits of one-parameter subgroups of isometries.  In particular, the geodesics are homogeneous, for which there is a well-developed theory of plane-wave limits \cite{MR2240408,FMPHomPL} including an explicit Lie-theoretic formula relating the choice of geodesic to the geometry of the plane wave which is very easy to implement.

\section{Anti de Sitter backgrounds}
\label{sec:AdS-backgrounds}

It remains to study those backgrounds of the form $\AdS_{d\geq 2} \times M^{11-d}$, such as the well-known Freund--Rubin backgrounds $\AdS_4 \times S^7$ and $\AdS_7 \times S^4$, and we do this now.  The first observation is that we must restrict to $d\leq 7$, since if $d>8$ there are no invariant $4$-forms: the only invariant forms in $\AdS_d$ are the constant functions and constant multiples of the volume form and there are not enough riemannian dimensions.  Hence we are left with backgrounds of the form $\AdS_{2\leq d\leq 7} \times M^{11-d}$.  There are 568 possible such geometries, but thankfully many of them can be ruled out because they do not admit invariant $4$-forms compatible with the field equations.  Such geometries are described for completeness in Appendix \ref{sec:geometries-which-do}.  In the rest of this section we list the possible backgrounds and as far as possible we determine their $F$-moduli.

\subsection*{A word about notation}

We label the possible geometries using the compact forms of the riemannian symmetric spaces.  The Einstein equations will determine whether it is the compact form or the noncompact dual which arises for a particular $F$.  Notice however that since we are only concerned with a classification up to local isometry, the topology will play no rôle.  In particular, when we write $T^n$ we only mean an $n$-dimensional flat riemannian manifold and not necessarily the $n$-torus; although we might indeed refer to it by name as a torus.

Unless otherwise noted, we use the following notations:
\begin{itemize}
\item $\nu$ denotes the volume form in $\AdS_d$;
\item $\omega$ usually denotes the Kähler form of a hermitian symmetric space, such as $\CP^n$ or $G_\RR^+(2,7)$;
\item $\sigma$ usually denotes the volume form of a sphere;
\item the flat directions on $T^n$ are usually denoted $\vartheta^a$;
\item $AdS$ indices are usually denoted $\mu,\nu,\dots$, whereas riemannian indices are $a,b,\dots$;
\item coframes are usually denoted $\theta^a$ and the shorthand $\theta^{ab\cdots c}$ is used for $\theta^a \wedge \theta^b \wedge \cdots \wedge \theta^c$.
\end{itemize}

\subsection{A useful heuristic}
\label{sec:useful-heuristic}

There is a useful heuristic which allows us to discard some candidate backgrounds and at the same time provides a useful check of our computations.  This is the fact that the Einstein equation implies an overall balance between the scalar curvatures of the $\AdS$ and riemannian factors.  Although, the precise nature of the balance depends on the explicit form of $F$, it is not hard to show that in a background of the type $\AdS_p \times K^{11-p}$, where $K$ is not necessarily irreducible, the scalar curvature of $K$ has to be positive.  Let us show this. We shall adhere to the notation explained above.

First let us consider the case of $p>4$.  Since the only invariant forms on $\AdS_p$ are the constants and the volume $p$-form, if $p>4$, $F$ must have no legs along $\AdS_p$.  From the Einstein equation we have that $R_{\mu\nu} = -\tfrac16 g_{\mu\nu}|F|^2$ where $|F|^2>0$, whence the Ricci scalar of the $\AdS$ factor is $-\tfrac16 p |F|^2$, which is indeed negative.  On the other hand $R_{ab} = \tfrac{p+1}{6} g_{ab}|F|^2$ which means that $K$ has positive scalar curvature.

Let $p=4$, so that $F = f \nu + \overline F$, with $\overline F$ a 4-form on $K$.  It follows that $|F|^2 = -f^2 + |\overline F|^2$.  From the Einstein equations $g^{\mu\nu}R_{\mu\nu} = -\tfrac43 (f^2 + |\overline F|^2)$, whereas $g^{ab}R_{ab} = \tfrac56 |\overline F|^2 + \tfrac76 f^2$, which is again positive.

If $p=3$, $F = \nu \wedge \alpha + \overline F$, where $\alpha$ is a 1-form on $K$.  The Einstein equations now imply that
$g^{\mu\nu}R_{\mu\nu} = - |\alpha|^2 - \tfrac16 |\overline F|^2 < 0$, whereas $g^{ab}R_{ab} = \tfrac56 |\alpha|^2 + \tfrac23 |\overline F|^2$, which is positive.

Finally, if $p=2$, $F = \nu \wedge \beta + \overline F$, where $\beta$ is a 2-form on $K$.  The Einstein equations now imply that
$g^{\mu\nu}R_{\mu\nu} = - \tfrac23 |\beta|^2 - \tfrac13 |\overline F|^2 < 0$, whereas $g^{ab}R_{ab} = \half\left( |\beta|^2 + |\overline F|^2\right)$, which is again positive.

\subsection{$\AdS_7$ backgrounds}
\label{sec:ads_7-backgrounds}

The possible backgrounds of the form $\AdS_7 \times M^4$ are given below where the first corresponds to the well-known Freund--Rubin background.

\begin{multicols}{3}
  \begin{enumerate}
  \item $\AdS_7 \times S^4$
  \item $\AdS_7 \times \CP^2$
  \item $\AdS_7 \times S^2 \times S^2$
 \end{enumerate}
\end{multicols}

They solve the field equations provided that the metric on the $\AdS_7$ obeys $R_{\mu\nu} = -\tfrac16 f^2 g_{\mu\nu}$ whereas that of the four-dimensional riemannian manifold obeys $R_{ab} = \tfrac13 f^2 g_{ab}$.  This means that we must take the compact forms.  The only subtlety is that $\CP^2$ does not admit a spin structure.  So strictly speaking we cannot perhaps include it as a supergravity background; although it does satisfy the equations of motion.  The radii of curvature of the two $2$-spheres in $\AdS_7 \times S^2 \times S^2$ are equal.

\begin{equation}\label{eq:ads7backgrounds}
  \boxed{\AdS_7 \times S^4 \qquad \AdS_7 \times \CP^2 \qquad \AdS_7 \times S^2 \times S^2 \qquad F = f \omega_4}
\end{equation}

\subsection{$\AdS_6$ backgrounds}
\label{sec:ads_6-backgrounds}

There are no $\AdS_6 \times M^5$ backgrounds.  All possible such geometries are ruled out either because they have no nonzero $F$ or if they do because they have a flat direction and this is incompatible with the Einstein equation as described in Appendix \ref{sec:geometries-with-no}.

\subsection{$\AdS_5$ backgrounds}
\label{sec:ads_5-backgrounds}

The $\AdS_5 \times M^6$ backgrounds are listed below and described in the sequel.

\begin{multicols}{3}
  \begin{enumerate}
  \item $\AdS_5 \times \CP^3$
  \item $\AdS_5 \times G_\RR^+(2,5)$
  \item $\AdS_5 \times \CP^2 \times S^2$
  \item $\AdS_5 \times S^4 \times S^2$
  \item $\AdS_5 \times S^2 \times S^2 \times S^2$
  \item $\AdS_5 \times S^2 \times S^2 \times T^2$
  \end{enumerate}
\end{multicols}

\subsubsection{$\AdS_5 \times \CP^3$ and $\AdS_5 \times G_\RR^+(2,5)$ backgrounds}
\label{sec:ads_5-times-cp3}

The geometry $\AdS_5 \times \CP^3$ supports a background.  Indeed, $F = \half f \omega\wedge \omega$, where $\omega$ is the Kähler form on $\CP^3$.  As shown in Appendix~\ref{sec:cp3}, we may write $\omega = \theta^{12} + \theta^{34} + \theta^{56}$, relative to an orthonormal coframe.  Then $F = f (\theta^{1234} + \theta^{1256} + \theta^{3456})$, from where we see that $|F|^2 = 3 f^2$.  Similarly, $F_1 = f (\theta^{234} + \theta^{256})$, etc... from where $\left<F_i,F_j\right> = 2 f^2 g_{ij}$.  The Einstein equations become
\begin{equation}
  R_{\mu\nu} = -\half f^2 g_{\mu\nu} \qquad\text{and}\qquad R_{ab} = \half f^2 g_{ab}~.
\end{equation}
The invariant metric on $\CP^3$ is Einstein due to the irreducibility of the linear isotropy representation, hence these equations are satisfied simply by taking the right scalar multiple of the metric.  Exactly the same argument shows, using the results in Appendix \ref{sec:g_rr+2-5}, that $\AdS_5 \times G_\RR^+(2,5)$ also supports a background.
\begin{equation}
  \label{eq:ads5cp3}
  \boxed{\AdS_5 \times \CP^3 \qquad \AdS_5 \times G_\RR^+(2,5) \qquad F = \half f \omega^2~.}
\end{equation}

\subsubsection{$\AdS_5 \times S^4 \times S^2$ and $\AdS_5 \times \CP^2 \times S^2$ backgrounds}
\label{sec:ads_5-times-s4}

The geometries $\AdS_5 \times S^4 \times S^2$ and $\AdS_5 \times \CP^2 \times S^2$ themselves do not admit solutions of the Einstein equations, but do if we consider the hyperbolic plane $H^2$ instead of $S^2$.  In either of the two cases, we must have $F = f \omega_4$, where $\omega_4$ is the volume form of $S^4$ or $\CP^2$, whence $|F|^2 = f^2$.  Therefore letting $\mu,\nu,\dots$, $i,j,\dots$ and $a,b,\dots$ denote indices in $\AdS_5$, $H^2$ and $S^4$ (or $\CP^2$), respectively, we find the following Einstein equations:
\begin{equation}
  R_{\mu\nu} = - \tfrac16 f^2 g_{\mu\nu} \qquad R_{ij} = - \tfrac16 f^2 g_{ij} \qquad\text{and}\qquad
  R_{ab} = \tfrac13 f^2 g_{ab}~,
\end{equation}
where we have used that $\left<F_a,F_b\right> = f^2 g_{ab}$.  The usual caveat about $\CP^2$ not admitting a spin structure applies.

\begin{equation}\label{eq:ads5h2s4}
  \boxed{\AdS_5 \times H^2 \times S^4  \qquad  \AdS_5 \times H^2 \times \CP^2 \qquad F = f \omega_4~.}
\end{equation}

\subsubsection{$\AdS_5 \times S^2 \times S^2 \times S^2$ and $\AdS_5 \times S^2 \times S^2 \times T^2$ backgrounds}
\label{sec:ads_5-times-s2}

The geometry $\AdS_5 \times S^2 \times S^2 \times S^2$ admits a three-parameter family of backgrounds; although for some values of those parameters one of the $S^2$ should be substituted by $H^2$ or $T^2$.  Let us write $F = f_1 \sigma_1 \wedge \sigma_2 + f_2 \sigma_2 \wedge \sigma_3 + f_3 \sigma_1 \wedge \sigma_3$, where $\sigma_i$ are the volume 2-forms of the three ``spheres''.  Then $|F|^2 = f_1^2 + f_2^2 + f_3^2$ and the Einstein equations for the AdS factor say that $R_{\mu\nu} = -\tfrac16 (f_1^2 + f_2^2 + f_3^2) g_{\mu\nu}$, which just sets the scale.  In particular, not all $f_i$ can be zero.  On the other hand, the Einstein equations for the three two-dimensional factors are of the form $\Ric = \tfrac13 \lambda g$, where
\begin{equation}
  \lambda_1 = f_1^2 + f_3^2 - \half f_2^2 \qquad
  \lambda_2 = f_1^2 + f_2^2 - \half f_3^2 \qquad\text{and}\qquad
  \lambda_3 = f_2^2 + f_3^2 - \half f_1^2~.
\end{equation}
The sum of any two of the $\lambda$s is positive, hence at most one of them must be non-positive.  This means that we have three possible geometries:
\begin{equation}
  \label{eq:ads5s2s2s2}
  \boxed{\AdS_5 \times S^2 \times S^2 \times S^2 \qquad \lambda_i > 0~\forall i}
\end{equation}

\begin{equation}
  \label{eq:ads5s2s2t2}
  \boxed{\AdS_5 \times S^2 \times S^2 \times T^2 \qquad \lambda_i = 0~\exists! i}
\end{equation}

\begin{equation}
  \label{eq:ads5s2s2h2}
  \boxed{\AdS_5 \times S^2 \times S^2 \times H^2 \qquad \lambda_i < 0~\exists! i}
\end{equation}

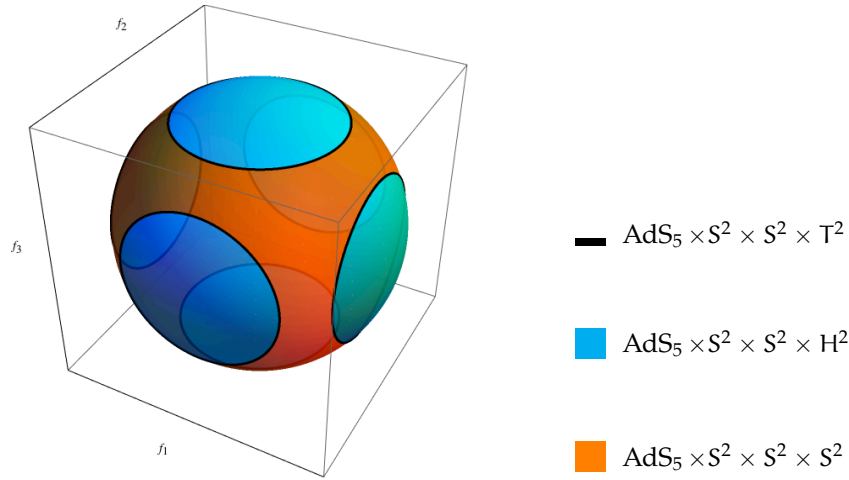
\begin{figure}[h!]
  \centering
  \includegraphics[width=6cm]{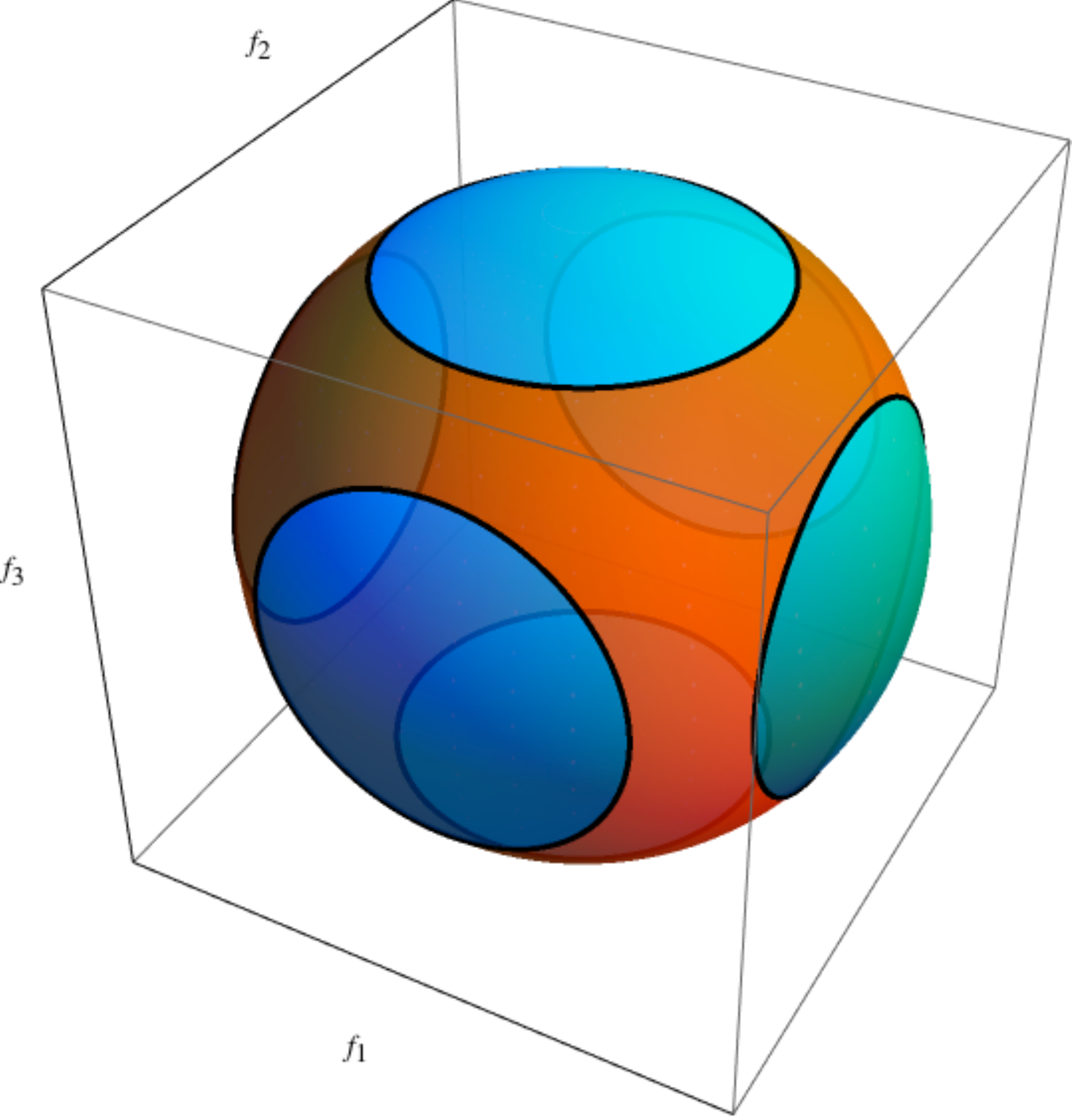}\qquad\qquad
  \begin{tikzpicture} [scale=1]
    \def\rectanglepath{-- ++(4mm,0mm) -- ++(0mm,4mm) -- ++(-4mm,0mm) -- cycle}
    \def\thicklinepath{-- ++(4mm,0mm) -- ++(0mm,1mm) -- ++(-4mm,0mm) -- cycle}
    \fill [orange] (0,0) \rectanglepath;
    \draw (0,2mm) node [black,right=5mm] {$\AdS_5 \times S^2 \times S^2 \times S^2$};
    \fill [cyan] (0,1.5) \rectanglepath;
    \draw (0,1.7) node [black,right=5mm] {$\AdS_5 \times S^2 \times S^2 \times H^2$};    
    \fill [black] (0,3) \thicklinepath;
    \draw (0,3.15) node [black,right=5mm] {$\AdS_5 \times S^2 \times S^2 \times T^2$};
  \end{tikzpicture}
  \caption{$F$-moduli for $\AdS_5 \times S^2 \times S^2 \times S^2$ geometries}
  \label{fig:FmoduliAdS5S2S2S2}
\end{figure}

The regions in $\RR^3$ defined by the above (in)equalities are easy to describe: for the case of $S^2 \times S^2 \times S^2$ it is the outside of the double cones defined by $f_1^2 = 2(f_2^2 + f_3^2)$ and cyclic permutations, whereas for the case of $S^2 \times S^2 \times H^2$ it is the interior of any of these double cones.  The boundary of these regions is the union of the three double cones themselves.  All such regions are conical (meaning that they admit an action of $\RR^+$) due to the homothety invariance of the equations of motion of 11-dimensional supergravity: $g\mapsto e^{2t} g$ and $F \mapsto e^{3t} F$.  This means that they are defined by their intersection with the unit sphere, its \emph{link}.  Figure~\ref{fig:FmoduliAdS5S2S2S2} illustrates the $F$-moduli for this geometry.

\subsection*{A comment on this and similar figures}

Figures like \ref{fig:FmoduliAdS5S2S2S2} depict a \emph{compactification} of the F-moduli space.  The boundary geometry, $\AdS_5 \times S^2 \times S^2 \times T^2$ in this example, is reached from either $\AdS_5 \times S^2 \times S^2 \times S^2$  or $\AdS_5 \times S^2 \times S^2 \times H^2$ by letting the radius of curvature of the $S^2$ or the $H^2$ tend to infinity.  It is therefore infinitely far away and one can therefore \emph{not} cross the boundary, just approach it from either side.

\subsection{$\AdS_4$ backgrounds}
\label{sec:ads_4-backgrounds}

In any such background, $F = f \nu + \overline F$, where $\nu$ is the volume form on $\AdS_4$.  The condition $F\wedge F =0$ becomes $f \overline F = 0$, whence either $f=0$ or else $\overline F = 0$.  The latter case is the standard Freund--Rubin ansatz.  Here $|F|^2 = -f^2$ and the Einstein equations become
\begin{equation}
  R_{\mu\nu} = - \tfrac13 f^2 g_{\mu\nu} \qquad\text{and}\qquad  R_{ab} = \tfrac16 f^2 g_{ab}~.
\end{equation}
This last equation means that only compact riemannian symmetric spaces can appear and moreover no flat directions are allowed.  The case where $f=0$ is where $F = \overline F$.  The Einstein equation for the $\AdS$ factor simply sets the radius of curvature: $R_{\mu\nu} = -\frac16 |F|^2 g_{\mu\nu}$.  The only conditions come from the Einstein equation for the riemannian factor:
\begin{equation}
  R_{ab} = \half \left<F_a,F_b\right> - \tfrac16 |F|^2 g_{ab}~,
\end{equation}
whence tracing $R = \tfrac43 |F|^2$.  This means that there has to be at least one compact non-flat factor.   Notice also that any geometry with flat directions but where $F$ has no legs along the flat directions cannot occur.  The reason is that the Einstein equations along the flat directions say that $R_{ij} = -\tfrac16 |F|^2 g_{ij}$, whence since $R_{ij}=0$ force $F=0$.  These considerations result in the following six classes of backgrounds.  Those backgrounds with $F = f \nu$ have been known since the early 1980s in the context of homogeneous Freund--Rubin backgrounds \cite{CastellaniRomansWarner}.

\subsubsection{$\AdS_4 \times S^7$, $\AdS_4 \times S^5 \times S^2$ and $\AdS_4 \times \SLAG_3 \times S^2$ backgrounds}
\label{sec:ads_4-times-s7}

The first is of course the well-known Freund--Rubin vacuum.  In all cases, $F = f\nu$, and 
\begin{equation}
  R_{\mu\nu} = - \tfrac13 f^2 g_{\mu\nu} \qquad\text{and}\qquad  R_{ab} = \tfrac16 f^2 g_{ab}~.
\end{equation}
In particular, the radii of curvature of all the compact riemannian symmetric spaces are equal.

\begin{equation}
  \label{eq:ads4FR}
  \boxed{\AdS_4 \times S^7\qquad \AdS_4 \times S^5 \times S^2 \qquad \AdS_4 \times \SLAG_3 \times S^2 \qquad F = f \nu}
\end{equation}

\subsubsection{$\AdS_4 \times S^4 \times S^3$,  $\AdS_4 \times \CP^2 \times S^3$ and $\AdS_4 \times S^2\times S^2 \times S^3$ backgrounds}
\label{sec:ads_4-times-s4}

Here we have $F = f_0 \nu + f_1 \sigma$, where $\sigma$ is the volume form on $S^4$ or $\CP^2$ or $S^2 \times S^2$.  The condition $F \wedge F = 0$ says that $f_0 f_1 =0$.  If $f_1=0$ we have the standard Freund--Rubin background, whereas if $f_0=0$, then since $F$ has no legs along the $S^3$, those components of the Einstein equation say that it has negative scalar curvature and hence we must take $H^3$.  This gives rise to two distinct classes of backgrounds.

\begin{equation}
  \label{eq:ads4FR2}
  \boxed{\AdS_4 \times S^4 \times S^3\qquad \AdS_4 \times \CP^2 \times S^3 \qquad \AdS_4 \times S^2\times S^2 \times S^3 \qquad F = f\nu}
\end{equation}

\begin{equation}
  \label{eq:ads4nonFR}
  \boxed{\AdS_4 \times S^4 \times H^3\qquad \AdS_4 \times \CP^2 \times H^3 \qquad \AdS_4 \times S^2\times S^2 \times H^3 \qquad F = f\sigma}
\end{equation}

Notice that the boundary cases, where the curvature of $S^3$ vanishes, cannot occur because they force $F=0$.

\subsection{$\AdS_3$ backgrounds}
\label{sec:ads_3-backgrounds}

These are the $\AdS_3 \times M^8$ backgrounds:

\begin{multicols}{3}
  \begin{enumerate}
  \item $\AdS_3 \times \CP^3 \times S^2$
  \item $\AdS_3 \times G_\RR^+(2,5) \times S^2$
  \item $\AdS_3 \times S^4 \times S^4$
  \item $\AdS_3 \times S^4 \times \CP^2$
  \item $\AdS_3 \times \CP^2 \times \CP^2$
  \item $\AdS_3 \times S^4 \times S^2 \times S^2$
  \item $\AdS_3 \times \CP^2 \times S^2 \times S^2$
  \item $\AdS_3 \times \CP^2 \times S^2 \times T^2$
  \item $\AdS_3 \times S^3 \times S^3 \times T^2$
  \item $\AdS_3 \times S^3 \times T^5$
  \item $\AdS_3 \times S^2 \times S^2 \times S^2 \times S^2$
  \item $\AdS_3 \times S^2 \times S^2 \times S^2 \times T^2$
  \item $\AdS_3 \times S^2 \times S^2 \times T^4$
  \item $\AdS_3 \times S^2 \times T^6$
  \end{enumerate}
\end{multicols}

\subsubsection{$\AdS_3 \times \CP^3 \times S^2$ and $\AdS_3 \times G_\RR^+(2,5) \times S^2$ backgrounds}
\label{sec:ads_3-times-cp3}

Let us consider the geometries $\AdS_3 \times \CP^3 \times S^2$ and $\AdS_3 \times G_\RR^+(2,5) \times S^2$. Let $M$ stand for either $\CP^3$ or $G_\RR^+(2,5)$.  In this case, $F = f_1 \half \omega^2 + f_2 \omega \wedge \sigma$, where $\omega$ is the Kähler form on $M$ and $\sigma$ is the area form on $S^2$.  Then $F\wedge F = f_1 f_2 \omega^3 \wedge \sigma$, which vanishes if and only if $f_1f_2=0$.  Also $|F|^2 = 3 (f_1^2 + f_2^2)$, whereas along the sphere $\left<F_a,F_b\right> = 3 f_2^2 g_{ab}$ and along $M$, $\left<F_i,F_j\right> = (2f_1^2 + f_2^2) g_{ij}$.  This implies the following Einstein equations:
\begin{equation}
  R_{\mu\nu} = -\half (f_1^2 + f_2^2) g_{\mu\nu} \qquad
  R_{ij} = \half f_1^2 g_{ij} \qquad
  R_{ab} = (f_2^2 - \half f_1^2) g_{ab}~.
\end{equation}
Since $M$ is not Ricci-flat, we must have $f_1\neq 0$, whence $f_2=0$ and hence we must have $H^2$ instead of the sphere.  In summary,
\begin{equation}
  \boxed{\AdS_3 \times \CP^3 \times H^2 \qquad \AdS_3 \times G_\RR^+(2,5) \times H^2 \qquad F = f \half \omega^2}
\end{equation}
with Einstein equations $\Ric = \half \lambda g$ for each factor, where $\lambda_M = - \lambda_{\AdS_3} = - \lambda_{H^2} = f^2$

\subsubsection{$\AdS_3 \times M^4 \times N^4$ backgrounds}
\label{sec:ads_3-times-m4}

Let us consider backgrounds with the geometry $\AdS_3 \times M^4 \times N^4$, where $M,N$ are 4-dimensional irreducible, or $M=S^4$ and $N=S^2\times S^2$.  (The case $\AdS_3 \times \CP^2 \times S^2 \times S^2$ is treated below.)  In these cases, $F= f_1 \nu_M + f_2 \nu_N$ is a linear combination of the volume forms.  Since $F\wedge F = 2f_1f_2 \nu_M \wedge\nu_N$, we see that $f_1f_2=0$.  Notice now that $|F|^2 = f_1^2 + f_2^2$, whence the $\AdS$ components of the Einstein equations are $R_{\mu\nu} = -\tfrac16 (f_1^2 + f_2^2) g_{\mu\nu}$, whereas the components along $M$ and $N$ are
\begin{equation}
  \Ric^{(M)} = \tfrac16 (2 f_1^2 - f_2^2) g^{(M)} \qquad\text{and}\qquad   \Ric^{(N)} = \tfrac16 (2 f_2^2 - f_1^2) g^{(N)}~.
\end{equation}
This means that $M$ and $N$ can be compact or noncompact depending on whether $2f_1^2 - f_2^2$ and $2f_2^2 - f_1^2$ are positive or negative.  (They cannot be zero, because we have assumed that neither $M$ nor $N$ are flat.)  Since $f_1f_2=0$, then precisely one of them must be zero.  If $f_1=0$, then $M$ is noncompact and $N$ is compact, whereas if $f_2=0$ the it is the other way around.  In other words, $F$ is proportional to the volume form of the compact factor.  Letting $\CH^2$ --- the complex hyperbolic plane --- denote the noncompact dual to $\CP^2$, we have the following backgrounds:

\begin{gather*}
  \boxed{\AdS_3 \times S^4 \times H^4 \quad \AdS_3 \times S^4 \times \CH^2 \quad \AdS_3 \times \CP^2 \times H^4 \quad \AdS_3 \times \CP^2 \times \CH^2\qquad F= f \nu_{\text{compact}}}\\
\boxed{\AdS_3 \times S^4 \times H^2 \times H^2 \quad \AdS_3 \times S^2 \times S^2 \times H^4 \qquad F= f \nu_{\text{compact}}}
\end{gather*}

\subsubsection{$\AdS_3 \times S^3 \times S^3 \times T^2$ and $\AdS_3 \times S^3 \times T^5$ backgrounds}
\label{sec:ads_3-times-s3}

\begin{figure}[h!]
  \centering
  \includegraphics[width=6cm]{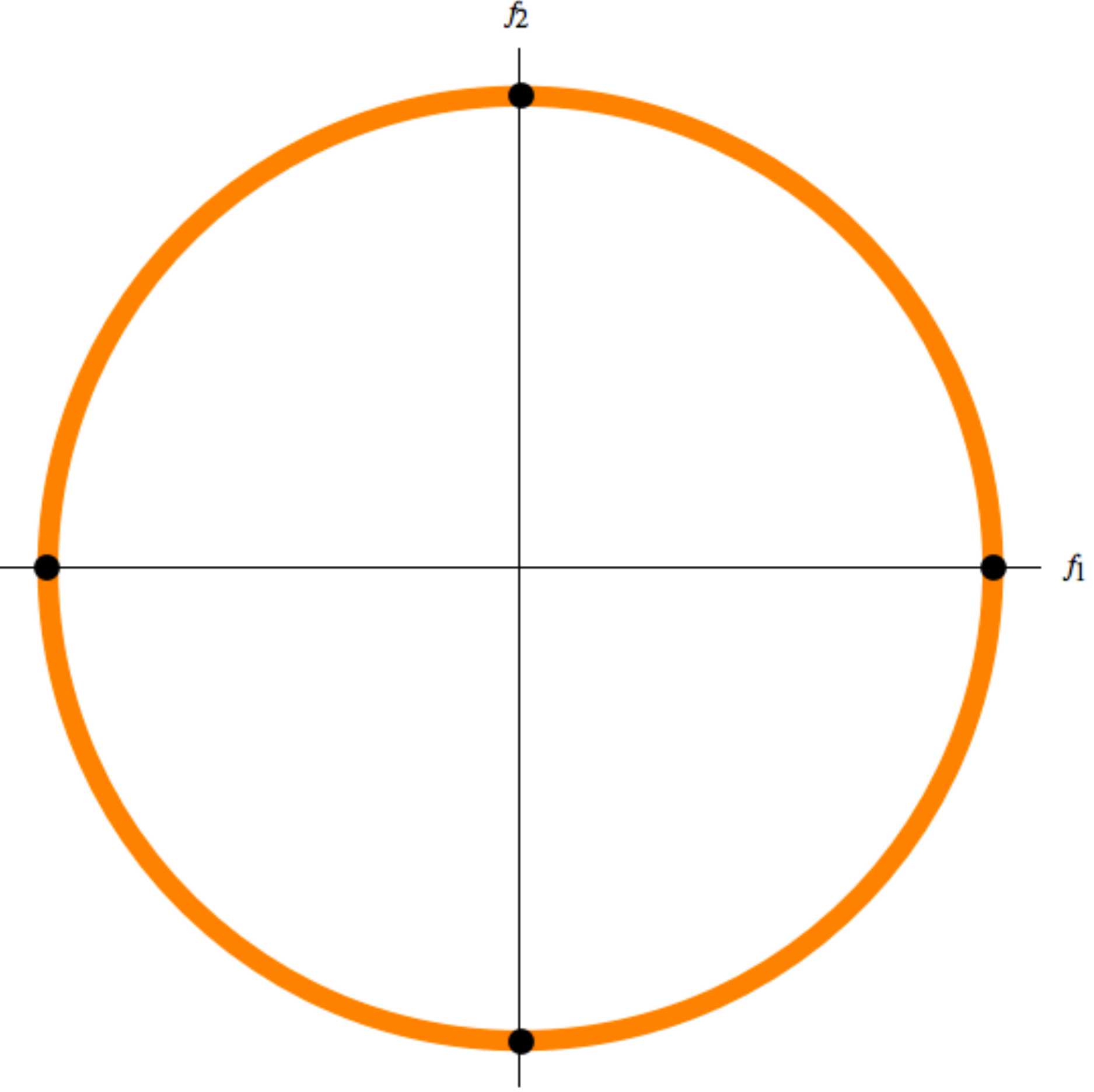}\qquad\qquad
  \begin{tikzpicture} [scale=1]
    \def\thicklinepath{-- ++(4mm,0mm) -- ++(0mm,1mm) -- ++(-4mm,0mm) -- cycle}
    \fill [orange] (0,0) \thicklinepath;
    \draw (0,0.15) node [black,right=5mm] {$\AdS_3 \times S^3 \times S^3 \times T^2$};
    \fill [black] (0.2,1) circle (1mm);
    \draw (0,1.1) node [black,right=5mm] {$\AdS_3 \times S^3 \times T^5$};
  \end{tikzpicture}
  \caption{$F$-moduli for $\AdS_3 \times S^3 \times S^3 \times T^2$ geometries}
  \label{fig:FmoduliAdS3S3S3T2}
\end{figure}
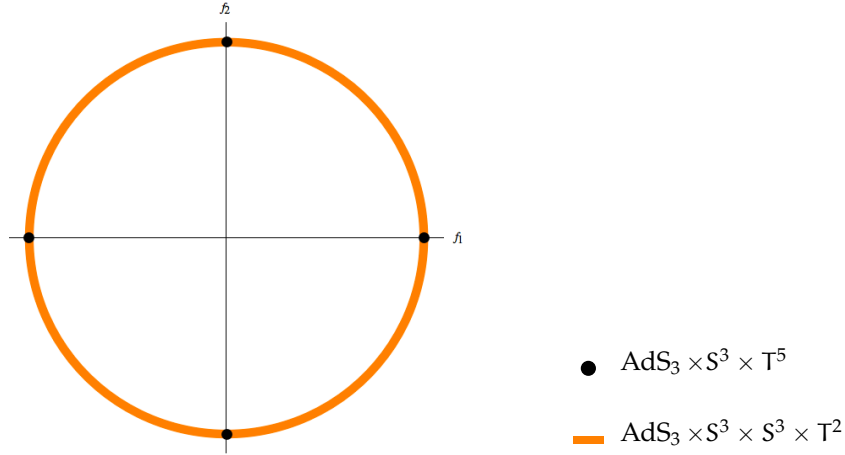

The $F$-moduli space for these backgrounds is illustrated in Figure \ref{fig:FmoduliAdS3S3S3T2}.

Let us first consider the $\AdS_3 \times S^3 \times T^5$ geometry.  Here the most general $F$ is given by $F = \nu_3 \wedge \alpha + \sigma_3 \wedge \beta + \gamma$, where $\alpha,\beta \in \Omega^1(T^5)$ and $\gamma \in \Omega^4(T^5)$.  The equations $F \wedge F =0$ impose the relations $\alpha\wedge\beta = 0$, $\alpha\wedge \gamma =0$ and $\beta\wedge \gamma = 0$.  We have two cases to study depending on whether at least one of $\alpha,\beta$ is nonzero or both are zero.

If both $\alpha$ and $\beta$ are zero, then we can take $\gamma = f d\vartheta^{1234}$ without loss of generality, but then the Einstein equation for $\vartheta^5$ says that $f^2=0$ and hence $F=0$.  Therefore, we must have at least one of $\alpha,\beta$ nonzero.  In that case, $\alpha$ and $\beta$ are proportional and $\star\gamma$ is perpendicular to both.  In other words, we can bring $F$ to the form
\begin{equation}
  F = d\vartheta^1 \wedge (f_0 \nu  + f_1 \sigma_3 + f_2 d\vartheta^{234})~.
\end{equation}
The $T^5$ Einstein equations become $-f_0^2 + f_1^2 + f_2^2 = f_2^2 = 0$, whence $f_2=0$ and $f_1^2 = f_0^2$.  This means that $|F|^2=0$, and hence the $S^3$ Einstein equation says that $\Ric = \half f_1^2 g$, whereas that of $\AdS_3$ is $\Ric = - \half f_1^2 g$.  In summary, we have
\begin{equation}
  \boxed{\AdS_3 \times S^3 \times T^5 \qquad F = f (\nu \pm \sigma_3) \wedge d\vartheta^5~.}
\end{equation}

For the geometry $\AdS_3 \times S^3 \times S^3 \times T^2$, the most general $F$ is
\begin{equation}
  F= \nu_3 \wedge \alpha + \sigma_3 \wedge \beta + \sigma'_3 \wedge \gamma~,
\end{equation}
where $\alpha,\beta,\gamma$ are invariant $1$-forms on $T^2$ not all of which are zero.  The equation $F \wedge F = 0$ implies that $\alpha \wedge \beta = 0$, $\alpha \wedge \gamma = 0$ and $\beta \wedge \gamma = 0$.  This means that $\alpha,\beta,\gamma$ are proportional and hence $F = d\vartheta^1 \wedge (f_0 \nu_3 + f_1 \sigma_3 + f_2 \sigma'_3)$, whence $|F|^2 = -f_0^2 + f_1^2 + f_2^2$.  Since $F$ has no $d\vartheta^2$, the $\theta^2$ Einstein equation says that $|F|^2=0$, whence $f_0^2 = f_1^2 + f_2^2$.  The Einstein equations for $\AdS_3$ and the two 3-spheres are of the form $\Ric = \half \lambda g$, where
\begin{equation}
  \lambda_{\AdS_3} = - (f_1^2+f_2^2) \qquad \lambda_{S^3} = f_1^2 \qquad \lambda_{{S'}^3} = f_2^2~.
\end{equation}
Notice that if $f_2=0$ or $f_1=0$, we recover precisely the $\AdS_3 \times S^3 \times T^5$ background discussed above.  In summary, we have

\begin{equation}
  \boxed{\AdS_3 \times S^3 \times S^3 \times T^2 \qquad F = d\vartheta^1 \wedge (f_0 \nu_3 + f_1 \sigma_3 + f_2 \sigma'_3) \qquad f_0^2 = f_1^2 + f_2^2}
\end{equation}

\subsubsection{$\AdS_3 \times \CP^2 \times S^2 \times S^2$, $\AdS_3 \times \CP^2 \times S^2 \times T^2$, $\AdS_3 \times \CP^2 \times T^4$, $\AdS_3 \times S^2 \times S^2 \times T^4$ and $\AdS_3 \times S^2 \times T^6$  backgrounds}
\label{sec:ads_3-times-cp2}

\begin{figure}[h!]
  \centering
 \includegraphics[width=6cm]{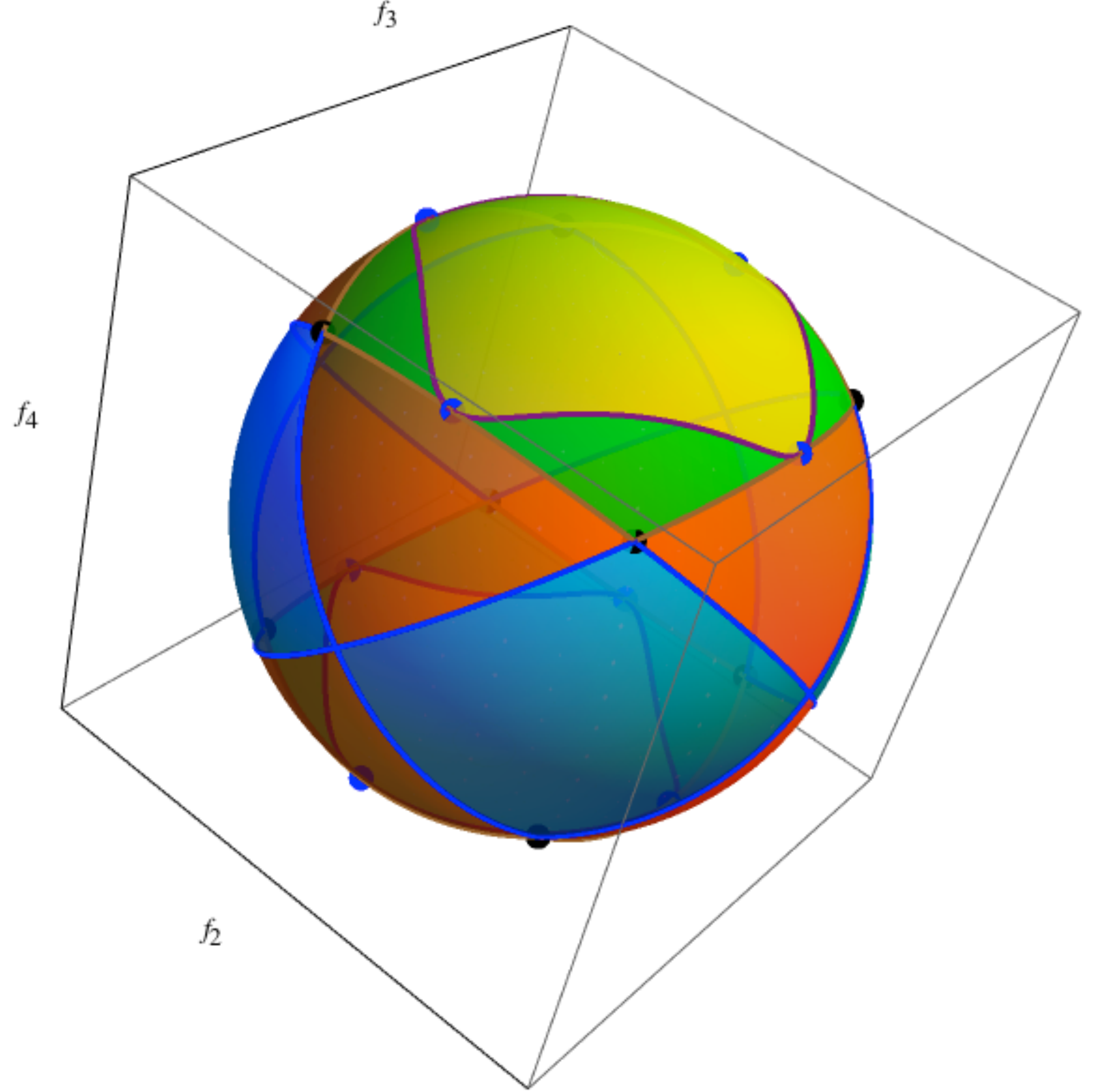}\qquad\qquad
  \begin{tikzpicture} [scale=0.75]
    \def\rectanglepath{-- ++(4mm,0mm) -- ++(0mm,4mm) -- ++(-4mm,0mm) -- cycle}
    \def\thicklinepath{-- ++(4mm,0mm) -- ++(0mm,1mm) -- ++(-4mm,0mm) -- cycle}
    \fill [green] (0,0) \rectanglepath;
    \draw (0,0.2) node [black,right=5mm] {$\AdS_3 \times \CP^2 \times S^2 \times S^2$};
    \fill [orange] (0,1) \rectanglepath;
    \draw (0,1.2) node [black,right=5mm] {$\AdS_3 \times \CP^2 \times S^2 \times H^2$};
    \fill [yellow] (0,2) \rectanglepath;
    \draw (0,2.2) node [black,right=5mm] {$\AdS_3 \times \CH^2 \times S^2 \times S^2$};
    \fill [cyan] (0,3) \rectanglepath;
    \draw (0,3.2) node [black,right=5mm] {$\AdS_3 \times \CP^2 \times H^2 \times H^2$};
    \fill [blue] (0,4) \thicklinepath;
    \draw (0,4.15) node [black,right=5mm] {$\AdS_3 \times \CP^2 \times H^2 \times T^2$};
    \fill [brown] (0,5) \thicklinepath;
    \draw (0,5.15) node [black,right=5mm] {$\AdS_3 \times \CP^2 \times S^2 \times T^2$};
    \fill [purple] (0,6) \thicklinepath;
    \draw (0,6.15) node [black,right=5mm] {$\AdS_3 \times T^4 \times S^2 \times S^2$};
    \fill [black] (0.2,7) circle (1mm);
    \draw (0,7.1) node [black,right=5mm] {$\AdS_3 \times \CP^2 \times T^4$};
    \fill [blue] (0.2,8) circle (1mm);
    \draw (0,8.1) node [black,right=5mm] {$\AdS_3 \times S^2 \times T^6$};
\end{tikzpicture}
\caption{$F$-moduli for $\AdS_3 \times \CP^2 \times S^2 \times S^2$ backgrounds}
\label{fig:FmoduliAdS3CP2S2S2}
\end{figure}

All these backgrounds can be obtained from $\AdS_3 \times \CP^2 \times S^2 \times S^2$ by allowing the curvature of some of the factors to vanish.  The resulting $F$-moduli space is illustrated in Figure \ref{fig:FmoduliAdS3CP2S2S2}.

Let us start with $\AdS_3 \times \CP^2 \times S^2 \times S^2$.  In this geometry, the general $F$ is given by
\begin{equation}
 F=f_1 \half \omega^2 + f_2 \omega \wedge \sigma_1 + f_3 \omega \wedge \sigma_2 + f_4 \sigma_1 \wedge \sigma_2~,
\end{equation}
where $\omega$ is the Kähler form on $\CP^2$ (or its noncompact dual) and $\sigma_{1,2}$ are the area forms on the spheres (or hyperbolic planes).  The equation $F\wedge F =0$ imposes the relation $f_1f_4 + 2 f_2f_3=0$.  The Einstein equations for each of the factors take the form $\Ric = \tfrac16 \lambda g$, where $\lambda_{\AdS_3} = -|F|^2 = - (f_1^2 + 2 f_2^2 + 2 f_3^2 + f_4^2)$ and
\begin{equation}
  \begin{aligned}[m]
    \lambda_{\CP^2} &= 2 f_1^2 + f_2^2 + f_3^2 - f_4^2\\
    \lambda_{S^2} &= - f_1^2 + 4 f_2^2 - 2 f_3^2 + 2 f_4^2 \\
    \lambda_{{S'}^2} &= - f_1^2 - 2 f_2^2 + 4 f_3^2 + 2 f_4^2
  \end{aligned}
\end{equation}
Depending on the value of the $\lambda$s we obtain different backgrounds.  First of all, we notice that the sum of all the riemannian $\lambda$s is non-negative, hence not all three can be negative.  Furthermore, if $\lambda_{\CP^2} \leq 0$, then $\lambda_{S^2}\geq 0$ and $\lambda_{{S'}^2} \geq 0$ and similarly if all inequalities are strict.  Assuming for the moment that no $\lambda$ vanishes, we have the following possible backgrounds.

\begin{equation}
  \boxed{\AdS_3 \times \CP^2 \times S^2 \times S^2 \qquad f_1^2 - f_4^2 + \tfrac13 |F|^2 > 0,\quad \tfrac23 |F|^2 - f_1^2 - 2 f_3^2>0,\quad \tfrac23 |F|^2 - f_1^2 - 2 f_2^2 >0}
\end{equation}

\begin{equation}
  \boxed{\AdS_3 \times \CP^2 \times S^2 \times H^2 \qquad f_1^2 - f_4^2 + \tfrac13 |F|^2 > 0,\quad
    \begin{cases}
      \tfrac23 |F|^2 - f_1^2 - 2 f_3^2<0\quad\text{and}\quad \tfrac23 |F|^2 - f_1^2 - 2 f_2^2 >0\\
      \text{or}\\
      \tfrac23 |F|^2 - f_1^2 - 2 f_3^2>0\quad\text{and}\quad \tfrac23 |F|^2 - f_1^2 - 2 f_2^2 <0
    \end{cases}
}
\end{equation}

\begin{equation}
  \boxed{\AdS_3 \times \CP^2 \times H^2 \times H^2 \qquad f_1^2 - f_4^2 + \tfrac13 |F|^2 > 0,\quad \tfrac23 |F|^2 - f_1^2 - 2 f_3^2<0,\quad \tfrac23 |F|^2 - f_1^2 - 2 f_2^2 <0}
\end{equation}

\begin{equation}
  \boxed{\AdS_3 \times \CH^2 \times S^2 \times S^2 \qquad f_1^2 - f_4^2 + \tfrac13 |F|^2 < 0,\quad \tfrac23 |F|^2 - f_1^2 - 2 f_3^2>0,\quad \tfrac23 |F|^2 - f_1^2 - 2 f_2^2 >0}
\end{equation}

Allowing either $\lambda_{S^2}$ or $\lambda_{{S'}^2}$ to vanish, we obtain the geometry $\AdS_3 \times \CP^2 \times S^2 \times T^2$.  However, the existence of the flat direction means that our Ansatz for $F$ is not the most general one and hence it is conceivable that this geometry might allow for a branch of backgrounds different from the degeneration of the previous backgrounds.  We will see, though, that this does not happen.
Indeed, the most general $F$ here is given by
\begin{equation}
  F= f_0 \nu_3 \wedge d\vartheta^1 + f_1  \half \omega^2 + f_2  \omega \wedge \sigma + f_3  \omega \wedge d\vartheta^{12} + f_4  \sigma \wedge d\vartheta^{12}~.
\end{equation}
The equation $F \wedge F = 0$ leads to $f_0f_2 =0$, $f_0f_1=0$ and $2f_3f_2 + f_4f_1=0$, whereas $|F|^2 = -f_0^2 + 2 f_3^2 + f_4^2 + 2 f_2^2 + f_1^2$.  The $T^2$ components of the Einstein equations say that $|F_1|^2 = |F_2|^2 = \tfrac13 |F|^2$, which translates into $f_0=0$ and $4f_3^2 + 2 f_4^2 = 2f_2^2 + f_1^2$.  Since $f_0=0$, we are back in the previous Ansatz for $F$ and hence all the $\AdS_3 \times \CP^2 \times S^2 \times T^2$ backgrounds are obtained by letting the curvature of either one of the two $S^2$s go to zero.  There is no $\AdS_3 \times \CH^2 \times S^2 \times T^2$ background, since as we saw before if $\lambda_{\CP^2}<0$ then $\lambda_{S^2}>0$ and $\lambda_{{S'}^2}>0$.  In summary, we obtain the following backgrounds
\begin{equation}
  \boxed{\AdS_3 \times \CP^2 \times S^2 \times T^2 \qquad f_1^2 - f_4^2 + \tfrac13 |F|^2 > 0,\quad
    \begin{cases}
      \tfrac23 |F|^2 - f_1^2 - 2 f_3^2=0\quad\text{and}\quad \tfrac23 |F|^2 - f_1^2 - 2 f_2^2 >0\\
      \text{or}\\
      \tfrac23 |F|^2 - f_1^2 - 2 f_3^2>0\quad\text{and}\quad \tfrac23 |F|^2 - f_1^2 - 2 f_2^2 =0
    \end{cases}}
\end{equation}
\begin{equation}
  \boxed{\AdS_3 \times \CP^2 \times H^2 \times T^2 \qquad f_1^2 - f_4^2 + \tfrac13 |F|^2 > 0,\quad
    \begin{cases}
      \tfrac23 |F|^2 - f_1^2 - 2 f_3^2<0\quad\text{and}\quad \tfrac23 |F|^2 - f_1^2 - 2 f_2^2 =0\\
      \text{or}\\
      \tfrac23 |F|^2 - f_1^2 - 2 f_3^2=0\quad\text{and}\quad \tfrac23 |F|^2 - f_1^2 - 2 f_2^2 <0
    \end{cases}
}
\end{equation}

Setting $\lambda_{S^2}=0=\lambda_{{S'}^2}$ we obtain backgrounds with geometry $\AdS_3 \times \CP^2 \times T^4$, but as before the flat directions allow for a more general $F$ and we need to investigate the existence of other branches of backgrounds sharing the same geometry but having this more general $F$.  Again we will see that there are none.  Indeed, in this geometry, the most general $F$ can be written as
\begin{equation}
  F = \nu_3 \wedge \alpha + f_1 \half \omega^2 + f_4 d\vartheta^{1234} + \omega \wedge \beta 
\end{equation}
for some $\alpha \in \Omega^1(T^4)$ and $\beta \in \Omega^2(T^4)$.  The equation $F \wedge F = 0$ implies the relations $\alpha \wedge \beta = 0$, $f_1 \alpha = 0$ and $\beta^2 + f_1f_4 d\vartheta^{1234} = 0$.  We must distinguish two cases, depending on whether or not $\alpha = 0$.
  \begin{enumerate}
  \item If $\alpha \neq 0$, then $f_1 =0$ and $\beta = \alpha \wedge \zeta$, for some $\zeta \in \Omega^1(T^4)$.  This allows us to write $\alpha = f_0 d\vartheta^1$ and $\beta = f_2 d\vartheta^{12}$ without loss of generality.  The $T^4$ equations of motion become the relations
    \begin{equation}
      -f_0^2 + 2 f_2^2 + f_4^2 = 2 f_2^2 + f_4^2 = f_4^2 = 0~,
    \end{equation}
    which easily implies that $f_0=f_4=f_2=0$ and hence $F=0$.
  \item On the other hand, if $\alpha = 0$, then we can write $\beta = f_2 d\vartheta^{12} + f_3 d\vartheta^{34}$ without loss of generality, resulting in the following $F$:
    \begin{equation}
      F = f_1 \half \omega^2 + f_2 \omega \wedge d\vartheta^{12} + f_3 \omega \wedge d\vartheta^{34} + f_4 d\vartheta^{1234}~.
    \end{equation}
    But this is the original Ansatz and hence the backgrounds are obtained from the previous ones by setting $\lambda_{S^2}=0$ and $\lambda_{{S'}^2}=0$.
  \end{enumerate}
 
  These equations are given by
    \begin{equation}
- f_1^2 + 4 f_2^2 - 2 f_3^2 + 2 f_4^2  = 0 \qquad\text{and}\qquad - f_1^2 - 2 f_2^2 + 4 f_3^2 + 2 f_4^2 = 0~.
   \end{equation}
   Subtracting one equation from the other, yields that $f_2^2 = f_3^2$ and then $f_1^2 = 2(f_2^2 + f_4^2)$.  Squaring the relation $f_1f_4 = - 2 f_2 f_3$ and using the above equations, we see that $f_2^4 = f_4^2 (f_2^2 + f_4^2)$, whose only (real) solutions obey $f_2^2 = f_4^2$.  In other words, $f_2^2 = f_3^2 = f_4^2$ and $f_1^2 = 4 f_2^2$.  In summary, we have
   \begin{equation}
      \boxed{\AdS_3 \times \CP^2 \times T^4 \qquad F = f ( 2 \varepsilon_1 \half \omega^2 + \varepsilon_2 \omega \wedge d\vartheta^{12} - \varepsilon_1\varepsilon_2 \omega \wedge d\vartheta^{34} + d\vartheta^{1234})~,}
    \end{equation}
with $\varepsilon_i$ signs.  This accounts for the black dots in Figure~\ref{fig:FmoduliAdS3CP2S2S2}.

Setting $\lambda_{\CP^2}=0$ we obtain backgrounds with underlying geometry $\AdS_3 \times S^2 \times S^2 \times T^4$, but again there might be extra branches of backgrounds withe same underlying geometry due to the possibility of more general $F$s than the ones considered above.  The most general $F$ here is given by
\begin{equation}
  F = \nu_3 \wedge \alpha + \sigma_1 \wedge \beta + \sigma_2 \wedge \gamma + f_1 d\vartheta^{1234} + f_4 \sigma_1 \wedge \sigma_2~,
\end{equation}
where $\alpha\in\Omega^1(T^4)$ and $\beta,\gamma \in \Omega^2(T^4)$.  The equation $F\wedge F = 0$ says that $f_4\alpha=0$, $\alpha\wedge\beta=0$, $\alpha\wedge\gamma=0$ and $\beta\wedge\gamma + f_1f_4 d\vartheta^{1234}=0$.  We must distinguish two cases, depending on whether or not $\alpha = 0$.
  \begin{enumerate}
  \item If $\alpha\neq 0$, we can bring it to the form $\alpha = f_0 d\vartheta^1$ without loss of generality.  Then the equations $F \wedge F = 0$ imply that $f_4 = 0$, and that $\alpha \wedge \beta = 0$, $\alpha \wedge \gamma =0$ and $\beta \wedge \gamma = 0$.  The most general solution can be chosen to be $\beta = f_2 d\vartheta^{12}$ and $\gamma = f_5 d\vartheta^{12} + f_6 d\vartheta^{13}$.  The $T^4$ Einstein equations impose the following additional relations
    \begin{equation}
      -f_0^2 + f_1^2 + f_2^2 + f_5^2 + f_6^2 = f_1^2 + f_2^2 + f_5^2 = f_1^2 + f_6^2 = f_1^2 = f_5 f_6 = 0~,
    \end{equation}
    which clearly imply that $F=0$.

  \item If $\alpha = 0$, then the most general solution is $\beta = f_2 d\vartheta^{12} + f_3 d\vartheta^{34}$ and $\gamma = f_5 d\vartheta^{12} + f_6 d\vartheta^{13} + f_7 d\vartheta^{14} + f_8 d\vartheta^{23} + f_9 d\vartheta^{34}$, without loss of generality, so that
    \begin{equation}
      F = f_4 \sigma_1 \wedge \sigma_2 + f_1 d\vartheta^{1234} + \sigma_2 \wedge (f_2 d\vartheta^{12} + f_3 d\vartheta^{34}) + \sigma_1 \wedge (f_5 d\vartheta^{12} + f_6 d\vartheta^{13} + f_7 d\vartheta^{14} + f_8 d\vartheta^{23} + f_9 d\vartheta^{34})~.
    \end{equation}
    The equation $F \wedge F = 0$ becomes the relation $f_1 f_4 + f_2 f_9 + f_3 f_5 = 0$.  The $T^4$ Einstein equations, namely $3 \left<F_a,F_b\right> = |F|^2 g_{ab}$, give a number of further quadratic relations.  We compute
    \begin{equation}
      \begin{aligned}[m]
        F_1 &= f_1 d\vartheta^{234} + f_2 \sigma_2 \wedge d\vartheta^2 + f_5 \sigma_1 \wedge d\vartheta^2 + f_6 \sigma_1 \wedge d\vartheta^3 + f_7 \sigma_1 \wedge d\vartheta^4\\
        F_2 &= - f_1 d\vartheta^{134} - f_2 \sigma_2 \wedge d\vartheta^1 - f_5 \sigma_1 \wedge d\vartheta^1 + f_8 \sigma_1 \wedge d\vartheta^3\\
        F_3 &= f_1 d\vartheta^{124} + f_3 \sigma_2 \wedge d\vartheta^4 - f_6 \sigma_1 \wedge d\vartheta^1 - f_8 \sigma_1 \wedge d\vartheta^2 + f_9 \sigma_1 \wedge d\vartheta^4\\
        F_4 &= -f_1 d\vartheta^{123} - f_3 \sigma_2 \wedge d\vartheta^3 - f_7 \sigma_1 \wedge d\vartheta^1 - f_9 \sigma_1 \wedge d\vartheta^3
      \end{aligned}
    \end{equation}
    whence the Einstein equations become
    \begin{multline}
      \label{eq:f_12-+-f_22}
      f_1^2 + f_2^2 + f_5^2 + f_6^2 + f_7^2 = f_1^2 + f_2^2 + f_5^2 + f_8^2 = f_1^2 + f_3^2 + f_6^2 + f_8^2 + f_9^2 = f_1^2 + f_3^2 + f_7^2 + f_9^2\\
      = \tfrac13( f_1^2 + f_2^2 + f_3^2 + f_4^2 + f_5^2 + f_6^2 + f_7^2 + f_8^2 + f_9^2 )
   \end{multline}
   and in addition
   \begin{equation}
     f_6 f_8 = -f_5 f_8 + f_7 f_9 = f_6 f_9 = f_5 f_6 = f_5 f_7 - f_8 f_9 = f_6 f_7 = 0~.
   \end{equation}
   The first equality in equation \eqref{eq:f_12-+-f_22} says that $f_8^2 = f_6^2 + f_7^2$ whereas the third says that $f_7^2 = f_6^2 + f_8^2$, whence $f_6=0$ and $f_8^2 = f_7^2$.  The second equality then says $f_2^2 + f_5^2 = f_3^2 + f_9^2$ and the last says $f_4^2 = 2f_1^2 + f_2^2 + f_5^2 + f_7^2$.  The Einstein equations for the spheres are of the form $\Ric = \half \lambda g$, where
   \begin{equation}
     \lambda_1  = f_1^2 + f_5^2 + 2 f_7^2 + f_9^2 \qquad\text{and}\qquad \lambda_2  = f_1^2 + f_2^2 + f_3^2~,
   \end{equation}
both of which are non-negative.  Either can be zero, leading to a $\AdS_3 \times S^2 \times T^6$ background, to be discussed in more generality below.
\end{enumerate}

In summary, we have
\begin{equation}
 \boxed{\AdS_3 \times S^2 \times S^2 \times T^4  \qquad
  \begin{cases}
    f_1^2 + f_5^2 + 2 f_7^2 + f_9^2 > 0 \\
    f_1^2 + f_2^2 + f_3^2 > 0
  \end{cases}}
\end{equation}

There do appear to be extra branches of backgrounds for this geometry.  In fact, the original branch is the one where $f_7=f_8=0$ and $f_2^2=f_3^2$ and $f_5^2 = f_9^2$.  We will see in Section \ref{sec:ads_3-times-s2} that some of them (those with $f_7=f_8=0$) arise out of degenerations of $\AdS_2 \times S^2 \times S^2 \times S^2 \times S^2$ backgrounds where two of the $S^2$ become flat.

Finally, the geometry $\AdS_3 \times S^2 \times T^6$ supports backgrounds which are in the boundary between backgrounds we have considered previously.  It has possible $F = \nu_3 \wedge \alpha + \sigma_2 \wedge \beta + \gamma$, where $\alpha \in \Omega^1(T^6)$, $\beta \in \Omega^2(T^6)$, $\gamma \in \Omega^4(T^6)$ and $\sigma_2$ is the volume form on $S^2$.  Then the equation $F\wedge F = 0$ implies
\begin{equation}
  \nu_3 \wedge \sigma_2 \wedge \alpha \wedge \beta + \nu_3 \wedge \alpha \wedge \gamma + \sigma_2 \wedge \beta \wedge \gamma = 0~.
\end{equation}
Since all three terms have different types, they must each vanish separately.  Since neither $\nu_3$ nor $\sigma_2$ vanish, we must have that $\alpha \wedge\beta = 0$, $\alpha \wedge \gamma = 0$ and $\beta \wedge \gamma = 0$.  We have two cases to consider, depending on whether or not $\alpha = 0$.
\begin{enumerate}
\item  If $\alpha \neq 0$, the first two equations imply that $\beta = \alpha \wedge \delta$ and $\gamma = \alpha \wedge \eta$.  This means that $F$ can be brought to the form
\begin{equation}
  F = f_0 \nu_3 \wedge d\vartheta^1 + f_1 \sigma_2 \wedge d\vartheta^{12} + f_2 d\vartheta^{1456} + f_3 d\vartheta^{1256} + f_4 d\vartheta^{1236} + f_5 d\vartheta^{1234}~,
\end{equation}
which satisfies $|F|^2 = -f_0^2 + \sum_{i=1}^5 f_i^2$.  Let us now look at the $T^6$ components of the Einstein equations: $|F_i|^2 = \tfrac13 |F|^2$ for all $i=1,\dots,6$ and $f_2 f_3 = f_3 f_4 = f_4 f_5 = 0$.  Since $|F_1|^2 = |F|^2$, this says that $|F_i|^2 = 0$ for all $i$ and hence also that $|F|^2=0$, whence $f_0^2 = \sum_{i=1}^5 f_i^2$.  Now $|F_2|^2 = f_1^2 + f_3^2 + f_4^2 + f_5^2$, whence $f_1=f_3=f_4=f_5 =0$, but now $|F_4|^2 = f_2^2$, whence $f_2=0$ as well and thus $f_0=0$ too.

\item Now let $\alpha = 0$.  We claim that $\gamma = 0$.  To see this, consider the Einstein equations in the flat directions:
  \begin{equation}
    \left<\beta_a,\beta_b\right> + \left<\gamma_a,\gamma_b\right> = \tfrac13 |F|^2 g_{ab} = \tfrac13 (|\beta|^2 + |\gamma|^2) g_{ab}~.
  \end{equation}
Tracing with $g^{ab}$ we find
\begin{equation}
  2 |\beta|^2 + 4 |\gamma|^2 = 2 (|\beta|^2 + |\gamma|^2) \implies 2|\gamma|^2 =0 \implies \gamma =0~.
\end{equation}
We may now bring $\beta$ to the form $\beta = f_1 d\vartheta^{12} + f_2 d\vartheta^{34} + f_3 d\vartheta^{56}$, whence $|\beta|^2 = f_1^2 + f_2^2 + f_3^2$.  The Einstein field equations along the flat directions say that $f_1^2 = f_2^2 = f_3^2$, whereas along the $S^2$ direction, $\Ric = f_1^2 g$.
\end{enumerate}

\begin{equation}
  \label{eq:ads3s2t6}
  \boxed{\AdS_3 \times S^2 \times T^6 \qquad F = f \sigma_2 \wedge (d\vartheta^{12} \pm d\vartheta^{34} \pm d\vartheta^{56})}
\end{equation}

This is precisely the background we obtain from $\AdS_3 \times \CP^2 \times S^2 \times S^2$ after letting the curvatures of the $\CP^2$ and one of the $S^2$ vanish.

\subsubsection{$\AdS_3 \times S^2 \times S^2 \times S^2 \times S^2$ and $\AdS_3 \times S^2 \times S^2 \times S^2 \times T^2$ backgrounds}
\label{sec:ads_3-times-s2}

The geometry $\AdS_3 \times S^2 \times S^2 \times S^2 \times S^2$ admits several families of backgrounds.  Letting $\sigma_i$, $i=1,\dots,4$ denote the area forms of the four $2$-spheres, the most general $F$ is given by $F = \sum_{1\leq i < j \leq 4} f_{ij} \sigma_i \wedge \sigma_j$, with $|F|^2 = \sum_{i,j} f_{ij}^2$.  The equation $F \wedge F =0$ imposes the following relation $f_{12}f_{34} + f_{13} f_{24} + f_{14}f_{23} = 0$.  The Einstein equations for each of the factors are of the form $\Ric = \tfrac16 \lambda g$, where the $\lambda$s are given as follows: for $\AdS_3$, $\lambda = - |F|^2$, and for each of the $S^2$ factors $\lambda$ is given, respectively, by the following expressions:
  \begin{align*}
    \lambda_1 &= 2 f_{12}^2 + 2 f_{13}^2 + 2 f_{14}^2 - f_{23}^2 - f_{24}^2 - f_{34}^2\\
    \lambda_2 &= 2 f_{12}^2 + 2 f_{23}^2 + 2 f_{24}^2 - f_{13}^2 - f_{14}^2 - f_{34}^2\\
    \lambda_3 &= 2 f_{13}^2 + 2 f_{23}^2 + 2 f_{34}^2 - f_{12}^2 - f_{14}^2 - f_{24}^2\\
    \lambda_4 &= 2 f_{14}^2 + 2 f_{24}^2 + 2 f_{34}^2 - f_{12}^2 - f_{13}^2 - f_{23}^2
 \end{align*}
One can see by inspection that the sum of any three of the $\lambda$s is non-negative and if the sum of any three is zero --- particularly if each of the three is zero --- then the remaining $\lambda > 0$ provided that $F\neq 0$.  In summary, we have three possible backgrounds with nonzero curvatures:

\begin{equation}
  \boxed{\AdS_3 \times S^2 \times S^2 \times S^2 \times S^2 \qquad \text{all $\lambda>0$}}
\end{equation}

\begin{equation}
  \boxed{\AdS_3 \times S^2 \times S^2 \times S^2 \times H^2 \qquad \text{one $\lambda<0$ and the remaining $\lambda>0$}}
\end{equation}

\begin{equation}
  \boxed{\AdS_3 \times S^2 \times S^2 \times H^2 \times H^2 \qquad \text{two $\lambda<0$ and two $\lambda>0$}}
\end{equation}

Letting the curvature of one of the $S^2$ (or $H^2$) vanish we obtain a geometry of the type $\AdS_3 \times S^2 \times S^2 \times S^2 \times T^2$.  This geometry allows for a more general $F$, since we can have an extra term $\nu_3 \wedge \alpha$ for some $\alpha \in \Omega^1(T^2)$.  However that term is quickly discarded.  Indeed, the most general $F$ can be brought to the form
  \begin{equation}
    F = f_0 \nu_3 \wedge d\vartheta^1 + \sum_{1\leq i< j \leq 3} f_{ij} \sigma_i\wedge \sigma_j + \sum_{i=1}^3 f_i \sigma_i \wedge d\vartheta^{12}~.
  \end{equation}
The equation $F \wedge F = 0$ becomes the relations $f_0f_{ij} = 0$, $f_{12}f_3 + f_{13}f_2 + f_{23}f_1 = 0$.  We have $|F|^2 = -f_0^2 + \sum_{i<j} f_{ij}^2 + \sum_{i}f_i^2$.  The $T^2$ components of the Einstein equation are $|F_1|^2 = |F_2|^2 = \tfrac13 |F|^2$ and this imposes, in particular, $f_0=0$, whence we are back to the previous ansatz and hence all the $\AdS_3 \times S^2 \times S^2 \times S^2 \times T^2$ backgrounds can be obtained by letting the curvature of one of the spheres vanish.

\begin{equation}
  \boxed{\AdS_3 \times S^2 \times S^2 \times S^2 \times T^2 \qquad \text{one $\lambda=0$ and the remaining $\lambda>0$}}
\end{equation}

\begin{equation}
  \boxed{\AdS_3 \times S^2 \times S^2 \times H^2 \times T^2 \qquad \text{one $\lambda=0$, one $\lambda<0$ and two $\lambda>0$}}
\end{equation}

Letting two of the curvatures vanish we obtain backgrounds with geometry $\AdS_3 \times S^2 \times S^2 \times T^4$, which were discussed in Section \ref{sec:ads_3-times-cp2} in more generality.  In the notation of that section, the backgrounds appearing here are those with $f_7 = f_8=0$.  Similarly, letting three of the curvatures vanish we obtain backgrounds with geometry $\AdS_3 \times S^2 \times T^6$, which were also already discussed in in Section \ref{sec:ads_3-times-cp2}.

\subsection{$\AdS_2$ backgrounds}
\label{sec:ads_2-backgrounds}

Finally, these are the $\AdS_2 \times M^9$ backgrounds.

\begin{multicols}{3}
  \begin{enumerate}
  \item $\AdS_2 \times \SLAG_4$
  \item $\AdS_2 \times G_\CC(2,4) \times S^1$
  \item $\AdS_2 \times S^7 \times S^2$
  \item $\AdS_2 \times \CP^3 \times S^3$
  \item $\AdS_2 \times G_\RR^+(2,5) \times S^3$
  \item $\AdS_2 \times \CP^3 \times T^3$
  \item $\AdS_2 \times G_\RR^+(2,5) \times T^3$
  \item $\AdS_2 \times \SLAG_3 \times S^4$
  \item $\AdS_2 \times S^5 \times S^4$
  \item $\AdS_2 \times \SLAG_3 \times \CP^2$
  \item $\AdS_2 \times S^5 \times \CP^2$
  \item $\AdS_2 \times \SLAG_3 \times S^2 \times S^2$
  \item $\AdS_2 \times S^5 \times S^2 \times S^2$
  \item $\AdS_2 \times \SLAG_3 \times S^2 \times T^2$
  \item $\AdS_2 \times S^5 \times S^2 \times T^2$
  \item $\AdS_2 \times \SLAG_3 \times T^4$
  \item $\AdS_2 \times S^5 \times T^4$
  \item $\AdS_2 \times S^4 \times S^3 \times S^2$
  \item $\AdS_2 \times \CP^2 \times S^3 \times S^2$
  \item $\AdS_2 \times \CP^2 \times S^3 \times T^2$
  \item $\AdS_2 \times \CP^2 \times S^2 \times T^3$
  \item $\AdS_2 \times \CP^2 \times T^5$
  \item $\AdS_2 \times S^3 \times S^3 \times T^3$
  \item $\AdS_2 \times S^3 \times S^2 \times S^2 \times S^2$
  \item $\AdS_2 \times S^3 \times S^2 \times S^2 \times T^2$
  \item $\AdS_2 \times S^3 \times S^2 \times T^4$
  \item $\AdS_2 \times S^3 \times T^6$
  \item $\AdS_2 \times S^2 \times S^2 \times S^2 \times S^2 \times S^1$
  \item $\AdS_2 \times S^2 \times S^2 \times S^2 \times T^3$
  \item $\AdS_2 \times S^2 \times S^2 \times T^5$
  \item $\AdS_2 \times S^2 \times T^7$
  \end{enumerate}
\end{multicols}

\subsubsection{$\AdS_2 \times \SLAG_4$ backgrounds}
\label{sec:ads_2-times-slag_4}

In this geometry, the most general $F$ is proportional to the invariant $4$-form $\Omega$ on $\SLAG_4$ defined in equation \eqref{eq:slag4form}: $F = f \Omega$.  Therefore $\left<F_a,F_b\right> = 8f^2\delta_{ab}$ and $|F|^2 = 18f^2$.  The equation $F\wedge F = 0$ is satisfied because $\SLAG_4$ has no nonzero invariant $8$-forms.  The Einstein equation for $\AdS_2$ is $\Ric = -3f^2 g$, whereas that of $\SLAG_4$ is $\Ric= f^2 g$.

\begin{equation}
  \boxed{\AdS_2 \times \SLAG_4 \qquad F = f \Omega~.}
\end{equation}

\subsubsection{$\AdS_2 \times S^7 \times S^2$ backgrounds}
\label{sec:ads_2-times-s7}

In the geometry $\AdS_2 \times S^7 \times S^2$ the most general $4$-form is given by $F= f \nu \wedge \sigma$, where $\nu$ and $\sigma$ are the area forms on $\AdS_2$ and $S^2$, respectively.  The equation $F\wedge F =0$ is trivially satisfied and $|F|^2 = -f^2$.  The Einstein equation for $\AdS_2$ is $\Ric = -\tfrac13 f^2 g$ and similarly for $S^2$, whence we must take $H^2$ instead.  The one for $S^7$ is $\Ric = \tfrac16 f^2 g$.

\begin{equation}
  \boxed{\AdS_2 \times H^2 \times S^7 \qquad F = f \nu \wedge \sigma~.}
\end{equation}

\subsubsection{$\AdS_2 \times \SLAG_3 \times S^4$ and $\AdS_2 \times S^5 \times S^4$ backgrounds}
\label{sec:ads_2-times-slag_3}

In these geometries, $F = f \sigma_4$, where $\sigma_4$ is the volume form on $S^4$, whence $F\wedge F =0$ and $|F|^2=f^2$.  The Einstein equation for $\AdS_2$ says $\Ric = -\tfrac16 f^2 g$, whereas for $S^4$, $\Ric = \tfrac13 f^2 g$.  The Einstein equation for the $5$-dimensional space is the same as for $\AdS_2$, whence we must take the noncompact dual.

\begin{equation}
  \boxed{\AdS_2 \times H^5 \times S^4 \qquad \AdS_2 \times (\SL(3,\RR)/\SO(3)) \times S^4 \qquad F = f \sigma_4~.}
\end{equation}

\subsubsection{$\AdS_2 \times G_\CC(2,4) \times S^1$ backgrounds}
\label{sec:ads_2-times-g_cc2}

In the geometry $\AdS_2 \times G_\CC(2,4) \times S^1$ the most general invariant $4$-form is given by $F = f_0 \nu \wedge \omega + f_1 \Omega^{(1)} + f_2 \Omega^{(2)}$, where $\omega$, $\Omega^{(i)}$ are defined in Appendix \ref{sec:g_cc2-4}.  Using the equations \eqref{eq:GC24identities}, the field equation $F \wedge F = 0$ becomes $f_0(f_1+f_2)=0$.  Similarly, we compute $|F|^2 = -4f_0^2 + 3 (f_1^2 + f_2^2)$.  Since $F$ has no legs along $S^1$, that component of the Einstein equation just sets $|F|^2 = 0$, whence $f_0^2 =\tfrac34 (f_1^2 + f_2^2)$.  This is consistent with $f_0(f_1+f_2)=0$ (and $F\neq 0$) if and only if $f_2 = - f_1$, whence $f_0^2 = \tfrac32 f_1^2$.  The Einstein equation for $\AdS_2$ becomes $\Ric = - 2 f_0^2 g$, whereas the one for $G_\CC(2,4)$ becomes $\Ric = f_0^2 g$, so it is the compact $G_\CC(2,4)$ which appears.

\begin{equation}
  \boxed{\AdS_2 \times G_\CC(2,4) \times S^1 \qquad F = f \left( \sqrt{\tfrac32} \nu \wedge \omega \pm (\Omega^{(1)} - \Omega^{(2)})\right)~.}
\end{equation}

\subsubsection{$\AdS_2 \times \CP^3 \times S^3$ and $\AdS_2 \times G_\RR(2,5) \times S^3$ backgrounds}
\label{sec:ads_2-times-cp3}

In the geometry $\AdS_2 \times \CP^3 \times S^3$, the most general $F$ is now $F = f_0 \nu\wedge \omega + f_1 \half \omega^2$, with $\omega$ the Kähler form of $\CP^3$.  The equation $F\wedge F =0$ implies $f_0f_1=0$, whereas the norm $|F|^2 = 3 (f_1^2 - f_0^2)$.  The Einstein equations for $\CP^3$ say that $\Ric = \half f_1^2 g$, whence $f_1 \neq 0$.  This means that $f_0=0$.  The Einstein equations for $S^3$ are $\Ric = -\half f_1^2 g$, whence we must take $H^3$.  Finally, those of $\AdS_2$ are $\Ric = -\half f_1^2 g$.  The same calculation applies to the geometry $\AdS_2 \times G_\RR(2,5) \times S^3$.

\begin{equation}
  \boxed{\AdS_2 \times H^3 \times \CP^3 \qquad \AdS_2 \times H^3 \times G_\RR(2,5) \qquad F = f \half \omega^2~.}
\end{equation}

\subsubsection{$\AdS_2 \times \CP^3 \times T^3$ and $\AdS_2 \times G_\RR^+(2,5) \times  T^3$ backgrounds}
\label{sec:ads_2-times-cp3-1}

In these geometries the most general invariant $4$-form can be brought to the following form
  \begin{equation}
    F = f_0 \nu \wedge d\vartheta^{12} + f_1 \nu \wedge \omega + f_2 \omega \wedge d\vartheta^{12}
 + f_3 \half \omega^2 + f_4 \omega \wedge d\vartheta^{23}~,
  \end{equation}
where $\omega$ is the Kähler form and $\nu$ the $\AdS_2$ area form.  The equation $F\wedge F = 0$ translates into the following relations
\begin{equation}\label{eq:k6t3rels}
  f_1f_3 = 0 \qquad f_2f_3 = 0 \qquad f_3f_4 = 0 \qquad f_1f_4 = 0 \qquad\text{and}\qquad f_0 f_3 + 2 f_1 f_2 = 0~.
\end{equation}
The $T^3$ components of the Einstein equation impose the relation $f_2 f_4=0$ and in addition the following relations:
\begin{equation}
  -f_0^2 + 3 f_2^2 = -f_0^2 + 3 f_2^2 + 3 f_4^2 = 3 f_4^2 = \tfrac13 |F|^2~,
\end{equation}
where $|F|^2 = -f_0^2 - 3f_1^2 + 3 f_2^2 + 3 f_3^2 + 3 f_4^2$.  These relations imply $f_4=0$, $f_0^2 = 3 f_2^2$ and $f_1^2 = f_3^2$.  This last equation, together with the first relation in \eqref{eq:k6t3rels}, says that $f_1=f_3=0$, whence $F = f_0 \nu \wedge d\vartheta^{12} + f_2 \omega \wedge d\vartheta^{12}$.  The $\AdS_2$ Einstein equation is $\Ric = -\tfrac32 f_2^2 g$ and the $\CP^3$ Einstein equation is $\Ric = \half f_2^2 g$.

\begin{equation}
  \boxed{\AdS_2 \times \CP^3 \times T^3\qquad \AdS_2 \times G_\RR^+(2,5) \times  T^3 \qquad F = f (\omega \pm \sqrt{3} \nu) \wedge d\vartheta^{12}~.}
\end{equation}

\subsubsection{$\AdS_2 \times S^3 \times S^3 \times T^3$ backgrounds}
\label{sec:ads_2-times-s3}

In this geometry, the most general $F$ is given by
  \begin{equation}
    F = f_0 \nu \wedge d\vartheta^{12} + f_1 \sigma_3 \wedge d\vartheta^3 + f_2 \sigma_3 \wedge d\vartheta^1 + f_3 \sigma'_3 \wedge d\vartheta^1 + f_4 \sigma'_3 \wedge d\vartheta^2 + f_5 \sigma'_3 \wedge d\vartheta^3~.
  \end{equation}
The equation $F \wedge F =0$ imposes the following relations:
\begin{equation}\label{eq:s3s3t3rels}
  f_0 f_1 = 0,\quad f_0 f_5 =0,\quad f_1 f_4 =0,\quad f_2 f_4 = 0,\quad f_1 f_3 = f_2 f_5~.
\end{equation}
The Einstein equations along $T^3$ give the following equalities:
\begin{equation}\label{eq:f_3-f_4-=}
  f_3 f_4 = f_4 f_5 = f_1 f_2 + f_3 f_5 = 0~,
\end{equation}
and
\begin{equation}
  -f_0^2 + f_2^2 + f_3^2  = -f_0^2 + f_4^2 = -f_1^2 + f_5^2 = \tfrac13 (-f_0^2 -f_1^2 + f_2^2 + f_3^2 + f_4^2 + f_5^2)~,
\end{equation}
from where we immediately read that $f_4^2 = f_2^2 + f_3^2$ and that $f_5^2 = - f_0^2 + f_1^2 + f_2^2 + f_3^2$.  Inserting this into the remaining equality, we find that $f_0 = 0$.  Inserting into the relations \eqref{eq:s3s3t3rels}, only the last three remain.  If $f_4 \neq 0$, then those relations and also the first two in \eqref{eq:f_3-f_4-=} say that $f_1=f_2=f_3=f_5=0$, whence $f_4=0$ as well.  Therefore $f_4=0$ and hence $f_2=f_3=0$.  This means that $f_5^2 = f_1^2$.  In other words, $F =f (\sigma_3 + \sigma'_3) \wedge d\vartheta^3$.  The Einstein equations for the 3-spheres are of the form $\Ric = \tfrac16 \lambda g$, where $\lambda = f^2$ and $\lambda' = f^2$, whereas for $\AdS_2$ we find $\Ric = -\tfrac16 f^2 g$.  In summary,
we have

\begin{equation}
  \boxed{\AdS_2 \times S^3 \times S^3 \times T^3 \qquad F = f (\sigma_3  \pm \sigma'_3) \wedge d\vartheta^3}
\end{equation}

\subsubsection{$\AdS_2 \times S^4 \times S^3 \times S^2$ backgrounds}
\label{sec:ads_2-times-s4}

In this geometry, the most general invariant $4$-form is
\begin{equation}
  F=f_0 \nu \wedge \sigma_2 + f \sigma_4~,
\end{equation}
with $\sigma_k$ the volume form on $S^k$.  The equation $F \wedge F = 0$ says that $f_0 f_1 = 0$.  The Einstein equations for $\AdS_2$ and $S^2$ are formally the same, whence we must take $H^2$ instead of $S^2$.  The equation is then $\Ric = -\tfrac16 (2f_0^2 + f_1^2) g$.  The Einstein equation for $S^4$ is $\Ric = \tfrac16 (2f_1^2 + f_0^2) g$, whence we do have $S^4$.  Finally, the Einstein equation for $S^3$ is $\Ric = \tfrac16 (f_0^2 - f_1^2) g$.  We have two kinds of geometries, depending on whether $f_0=0$ or $f_1=0$.
\begin{equation}
  \boxed{\AdS_2 \times S^4 \times S^3 \times H^2 \qquad F = f \nu \wedge \sigma_2~,}
\end{equation}
and
\begin{equation}
  \boxed{\AdS_2 \times S^4 \times H^3 \times H^2 \qquad F = f \sigma_4~.}
\end{equation}
One might suspect the existence of a background $\AdS_2 \times S^4 \times H^2 \times T^3$ interpolating between them, but the form of $F$ would imply that $F=0$ on the boundary.  In fact, as shown in Appendix \ref{sec:geometries-with-no}, no such background exists.

\subsubsection{$\AdS_2 \times S^3 \times T^6$ backgrounds}
\label{sec:ads2s3t6}

In this geometry, the most general $F$ takes the form
  \begin{equation}
    F = \nu \wedge \alpha + \sigma_3 \wedge \beta + \gamma
  \end{equation}
where $\alpha \in \Omega^2(T^6)$, $\beta \in \Omega^1(T^6)$ and $\gamma \in \Omega^4(T^6)$ are parallel forms on $T^6$.  The equation $F \wedge F = 0$ is equivalent to $\alpha \wedge \beta = 0$, $\alpha \wedge \gamma =0$ and $\beta\wedge \gamma =0$.  We must distinguish two cases, depending on whether or not $\beta =0$.

\begin{enumerate}
\item If $\beta \neq 0$, the first and last equations have as solutions $\alpha = \beta \wedge \zeta$ and $\gamma = \beta \wedge \varphi$, for some $\zeta \in \Omega^1(T^6)$ and $\varphi \in \Omega^3(T^6)$.  Without loss of generality we can choose $\beta \propto d\vartheta^1$ and hence $\zeta \propto d\vartheta^2$.  We can then write $\varphi = d\vartheta^2 \wedge \psi + \chi$ for some $\psi\in \Omega^2(T^6)$ and $\chi \in \Omega^3(T^6)$, with $d\vartheta^i \wedge \psi = d\vartheta^i \wedge \chi = 0$ for $i=1,2$.  Without loss of generality can take $\psi = f_2 d\vartheta^{34} + f_3 d\vartheta^{56}$ and hence $\chi = f_4 d\vartheta^{1456} + f_5 d\vartheta^{1346}$.  Since $F = d\vartheta^1 \wedge H$ for some $3$-form $H$, it follows that the $T^6$ Einstein equations imply that $|F_i|^2 = 0$ (and hence $|F|^2=0$) for all $i=1,\dots,6$.  These equations are given by
\begin{equation}
  -f_0^2+f_1^2+f_2^2+f_3^2+f_4^2+f_5^2 = -f_0^2 + f_2^2 + f_3^2 = f_2^2 + f_5^2 = f_2^2 + f_4^2 + f_5^2 = f_3^2 + f_4^2 = f_3^2 + f_4^2 + f_5^2 = 0~,
\end{equation}
whose only solution is $f_i = 0$ for all $i$.

\item If $\beta = 0$, then $|F|^2 = -|\alpha|^2 + |\gamma|^2$ and the Einstein equations for the $\AdS_2$ and $S^3$ factors are of the form $\Ric = \tfrac16 \lambda g$, where
  \begin{equation}
    \lambda_{\AdS_2} = -(2|\alpha|^2 + |\gamma|^2) \qquad\text{and}\qquad \lambda_{S^3} = |\alpha|^2 - |\gamma|^2~.
  \end{equation}
The Einstein equations along the flat directions are
\begin{equation}
  -\left<\alpha_a,\alpha_b\right> + \left<\gamma_a,\gamma_b\right> = -\tfrac13 (|\alpha|^2-|\gamma|^2)g_{ab}~,
\end{equation}
which upon tracing with $g^{ab}$ becomes $\gamma =0$.  We can therefore bring $\beta$ to the form $\beta = f_1 d\vartheta^{12} + f_2 d\vartheta^{34} + f_3 d\vartheta^{56}$ where $f_1^2 = f_2^2 = f_3^2$ by virtue of the Einstein equations.

\begin{equation}
  \label{eq:ads2s3t6}
  \boxed{\AdS_2 \times S^3 \times T^6 \qquad F = f \nu \wedge (d\vartheta^{12} \pm d\vartheta^{34} \pm d\vartheta^{56})~.}
\end{equation}

\end{enumerate}
\subsubsection{$\AdS_2 \times \CP^2 \times T^5$ backgrounds}
\label{sec:ads_2-times-cp2}

In the geometry $\AdS_2 \times \CP^2 \times T^5$, the most general $F$ takes the form
\begin{equation}
  F = f_0 \nu \wedge \omega + f_1 \half \omega^2 + \nu \wedge \alpha + \omega \wedge \beta + \gamma~,
\end{equation}
where $\alpha, \beta \in \Omega^2(T^5)$ and $\gamma \in \Omega^4(T^5)$.  The equation $F \wedge F = 0$ gives rise to the following identities:
\begin{equation}
  2 f_0 \beta + f_1 \alpha = 0, \qquad f_1 \gamma + \beta^2 =0,\qquad f_0 \gamma + \alpha \wedge \beta = 0~.
\end{equation}
We can distinguish four cases:
\begin{enumerate}
\item $f_0\neq 0, f_1 =0$.  In this case, $\beta = \gamma = 0$ and hence $F$ can be brought to the form $F = \nu \wedge (f_0 \omega + f_2 d\vartheta^{12} + f_3 d\vartheta^{34})$.  Since $F$ has no legs along $\theta^5$, that component of the Einstein equations says that $|F|^2=0$, but since $|F|^2\leq 0$ here, it means that $F=0$.

\item $f_0=0, f_1 \neq 0$.  In this case, $\alpha = 0$ and $\gamma = -\beta^2/f_1$.  This means that $F$ can be written as $F = f_1 \half \omega^2 + \omega \wedge \beta - \beta^2/f_1$.  Without loss of generality we can let $\beta = f_2 d\vartheta^{12} + f_3 d\vartheta^{34}$, whence $\beta^2 = 2 f_2f_3 d\vartheta^{1234}$.  Again there is no $\theta^5$ leg, whence its Einstein equation sets $|F|^2=0$, but $|F|^2 \geq 0$ here, whence again $F=0$.

\item $f_0\neq0, f_1\neq 0$.  In this case, $\gamma =0$, $\beta^2=0$ and $\alpha = - 2f_0 \beta^2/f_1$.  Since $\beta^2 =0$, it is decomposable, whence we can bring it to the form $\beta = f_2 d\vartheta^{12}$ without loss of generality.  Then we have
  \begin{equation}
    F = f_0 \nu \wedge \omega + f_1\half \omega^2 - \frac{2f_0f_2}{f_1}\nu \wedge d\vartheta^{12} + f_2 \omega \wedge d\vartheta^{12}~.
 \end{equation}
Since $F$ has no legs along the $\theta^{3,4,5}$ directions, any of these Einstein equations sets $|F|^2=0$, whence $f_1^2 + 2f_2^2 = 2f_0^2 + 4 f_0^2 f_2^2/f_1^2$, which is equivalent to $f_1^2 = 2f_0^2$.  The Einstein equations for $\AdS_2$ and $\CP^2$ are of the form $\Ric = \lambda g$, with
\begin{equation}
  - \lambda_{\AdS_2} = f_0^2 + f_2^2 \qquad\text{and}\qquad
  \lambda_{\CP^2} = \half (f_0^2 + f_2^2)~.
\end{equation}
Since $f_0\neq 0$, we have $\AdS_2 \times \CP^2 \times T^5$ for any allowed values of $f_0,f_2$.

\item The final case is when $f_0=f_1=0$.  Here $\beta^2=0$ and $\alpha\wedge \beta = 0$.  The equation $\beta^2=0$ says that $\beta$ is decomposable, whence without loss of generality we can let $\beta = f_2 d\vartheta^{12}$.  Then $\alpha \wedge \beta = 0$ forces $\alpha$ to be a linear combination of $d\vartheta^{12},d\vartheta^{13},d\vartheta^{14},d\vartheta^{15},d\vartheta^{23},d\vartheta^{24},d\vartheta^{25}$.  Using the freedom to rotate in the $(345)$ directions, we can assume that $d\vartheta^{24}$ and $d\vartheta^{25}$ do not appear.  This leaves the possibility to rotate separately in the $(12)$ and $(45)$ planes.  Rotating in the $(12)$ plane we get rid of $d\vartheta^{23}$ and finally rotating in the $(45)$ plane gets rid of $d\vartheta^{15}$, whence $\alpha = d\vartheta^1\wedge (f_3 d\vartheta^2 + f_4 d\vartheta^3 + f_5 d\vartheta^4)$.  We have now used all the symmetry, so we are forced to take $\gamma$ to be the most general 4-form, namely
  \begin{equation}
    \gamma = f_6 d\vartheta^{2345} + f_7 d\vartheta^{1345} +f_8 d\vartheta^{1245} + f_9 d\vartheta^{1235} + f_{10} d\vartheta^{1234}~.
  \end{equation}
The Einstein equations along the $T^5$ directions become the following system of equalities:
\begin{equation}
  f_6 f_7 = f_6 f_8 = f_6 f_9 = f_6 f_{10} = f_7 f_{10} = f_8 f_{10} = f_9 f_{10} = f_7 f_8 - f_3 f_4 = f_4 f_5 - f_8 f_9 = f_3 f_5 - f_7 f_9 = 0~,
\end{equation}
and
\begin{multline}
  f_{6}^2 + f_{7}^2 + f_{8}^2 + f_{9}^2 = f_{10}^2 - f_{5}^2 + f_{6}^2  + f_{7}^2 + f_{8}^2 = f_{10}^2 - f_{4}^2 + f_{6}^2 + f_{7}^2 + f_{9}^2 = f_{10}^2 + 2f_{2}^2 - f_{3}^2 + f_{6}^2 + f_{8}^2 + f_{9}^2\\
  = f_{10}^2 + 2f_{2}^2 - f_{3}^2 - f_{4}^2 - f_{5}^2 + f_{7}^2 + f_{8}^2 + f_{9}^2 = \tfrac{1}{3} \left(f_{10}^2 + 2f_{2}^2 - f_{3}^2 - f_{4}^2 - f_{5}^2 + f_{6}^2 + f_{7}^2 + f_{8}^2 + f_{9}^2\right)~.
\end{multline}
The only solutions are those for which all $f_i=0$ except for $f_2$ and $f_3$, which satisfy $2f_2^2 = f_3^2$.  This branch is obtained formally from the previous one by setting $f_0 = f_1 = 0$.
\end{enumerate}

In summary, we have
\begin{equation}\label{eq:ads2cp2t5}
  \boxed{\AdS_2 \times \CP^2 \times T^5 \qquad F = \frac{f_0}{\sqrt{2}} \left(\sqrt{2}\nu \pm \omega\right) \wedge \omega \mp f_2 \left( \sqrt{2} \nu \mp \omega \right) \wedge d\vartheta^{12}~,}
  \end{equation}
where the signs are correlated.

\subsubsection{$\AdS_2 \times \SLAG_3 \times \CP^2$, $\AdS_2 \times S^5 \times \CP^2$, $\AdS_2 \times \SLAG_3 \times T^4$ and $\AdS_2 \times S^5 \times T^4$ backgrounds}
\label{sec:ads_2-times-slag_3-2}

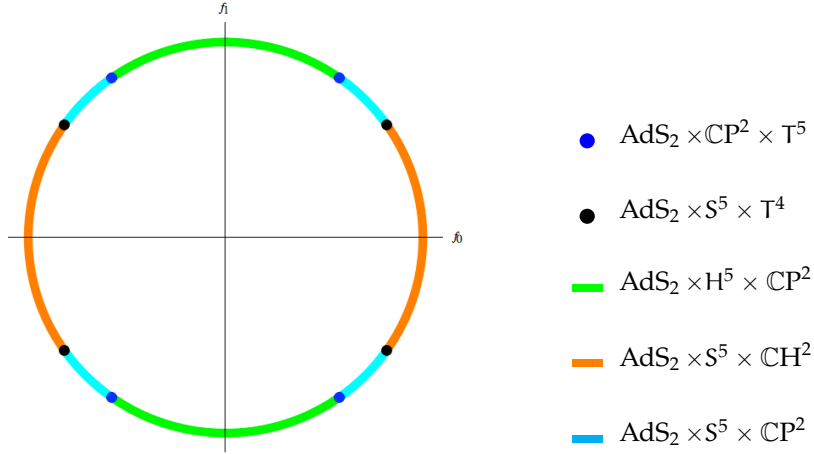
\begin{figure}[h!]
  \centering
  \includegraphics[width=6cm]{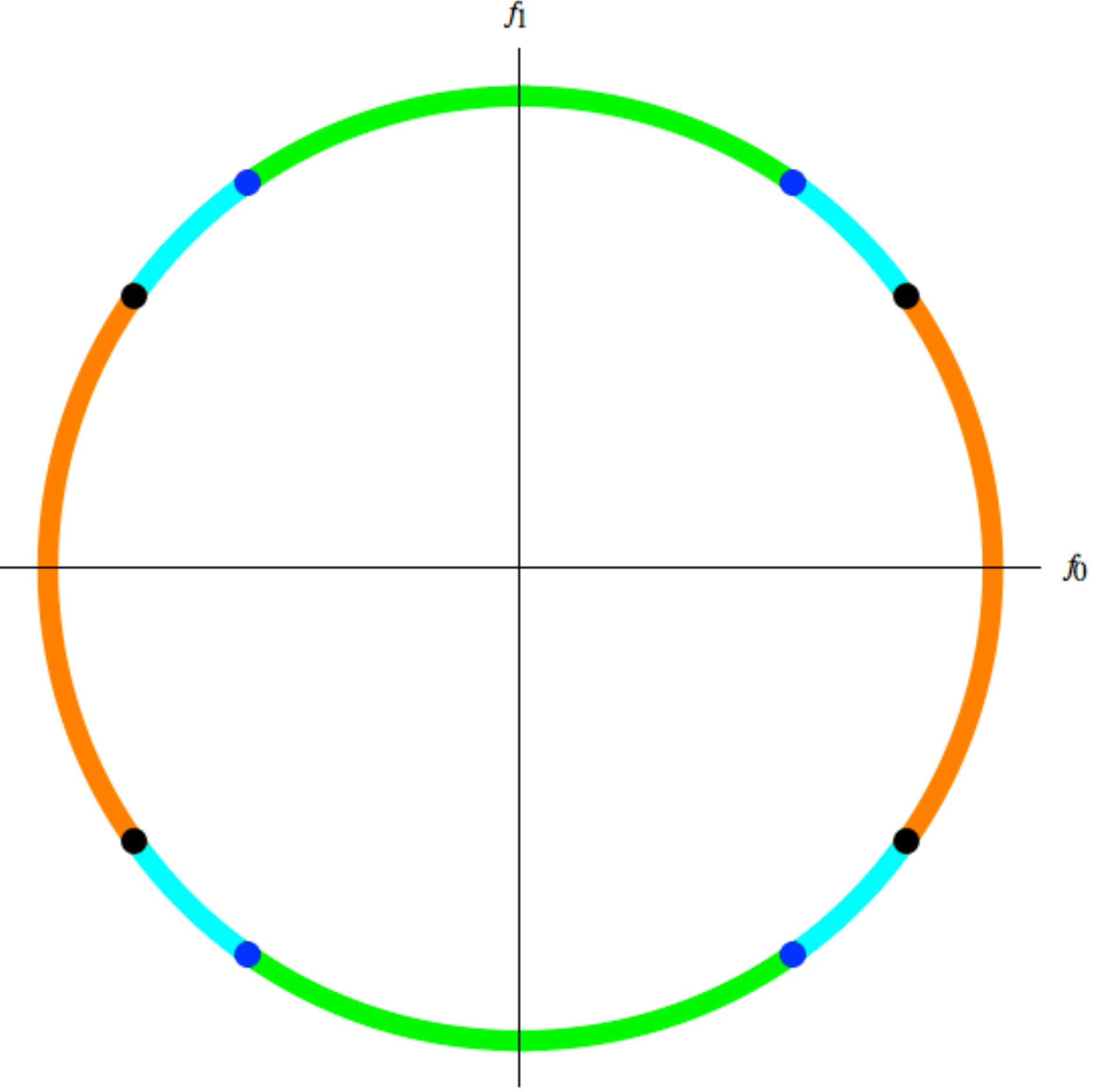}\qquad\qquad
  \begin{tikzpicture} [scale=1]
    \def\thicklinepath{-- ++(4mm,0mm) -- ++(0mm,1mm) -- ++(-4mm,0mm) -- cycle}
    \fill [cyan] (0,0) \thicklinepath;
    \draw (0,0.15) node [black,right=5mm] {$\AdS_2 \times S^5 \times \CP^2$};
    \fill [orange] (0,1) \thicklinepath;
    \draw (0,1.15) node [black,right=5mm] {$\AdS_2 \times S^5 \times \CH^2$};
    \fill [green] (0,2) \thicklinepath;
    \draw (0,2.15) node [black,right=5mm] {$\AdS_2 \times H^5 \times \CP^2$};
    \fill [black] (0.2,3) circle (1mm);
    \draw (0,3.1) node [black,right=5mm] {$\AdS_2 \times S^5 \times T^4$};
    \fill [blue] (0.2,4) circle (1mm);
    \draw (0,4.1) node [black,right=5mm] {$\AdS_2 \times \CP^2 \times T^5$};
\end{tikzpicture}
  \caption{$F$-moduli for $\AdS_2 \times S^5 \times \CP^2$ geometries}
  \label{fig:FmoduliAdS2S5CP2}
\end{figure}

The resulting moduli space is illustrated in Figure~\ref{fig:FmoduliAdS2S5CP2}, which although labelled for $S^5/H^5$ applies as well to $\SLAG_3$ and its noncompact dual.  There are boundaries between the regions corresponding to backgrounds of the form $\AdS_3 \times \CP^2 \times T^5$ discussed in Section~\ref{sec:ads_2-times-cp2}.  They correspond to the branch of those backgrounds where $f_0=0$ in equation~\eqref{eq:ads2cp2t5}.

In the geometries $\AdS_2 \times \SLAG_3 \times \CP^2$ and $\AdS_2 \times S^5 \times \CP^2$, the most general $4$-form is
\begin{equation}
  F= f_0 \nu \wedge \omega + f_1 \half \omega^2~,
\end{equation}
which automatically satisfies $F \wedge F =0$ and has norm $|F|^2 = -2f_0^2 + f_1^2$.  The Einstein equations along $S^5$ (or $\SLAG_3$) are $\Ric = \tfrac16 (2f_0^2 - f_1^2) g$, whereas those along $\CP^2$ are $\Ric = \tfrac16 (2 f_1^2 - f_0^2) g$ and along $\AdS_2$ are $\Ric = -\tfrac16 (4f_0^2 + f_1^2) g$.  In summary, we have
\begin{equation}
  \boxed{\AdS_2 \times \SLAG_3 \times \CP^2 \qquad \AdS_2 \times S^5 \times \CP^2 \qquad 4f_1^2 > 2f_0^2 > f_1^2~,}
\end{equation}
\begin{equation}
  \boxed{\AdS_2 \times \SLAG_3 \times \CH^2 \qquad \AdS_2 \times S^5 \times \CH^2 \qquad f_0^2 > 2 f_1^2~,}
\end{equation}
and
\begin{equation}
  \boxed{\AdS_2 \times (\SL(3,\RR)/\SO(3)) \times \CP^2 \qquad \AdS_2 \times H^5 \times \CP^2 \qquad f_1^2 > 2 f_0^2~.}
 \end{equation}

If we now let the $\CP^2$ curvature go to zero, we obtain backgrounds with geometries $\AdS_2 \times \SLAG_3 \times T^4$ and $\AdS_2 \times S^5 \times T^4$, respectively.  In principle, on these geometries we could have a more general $F$, namely
\begin{equation}
  F = f_0 \nu \wedge d\vartheta^{12} + f_1 d\vartheta^{1234} + f_2 \nu \wedge d\vartheta^{34}~,
\end{equation}
which obeys $F \wedge F = 0$ and $|F|^2 = -f_0^2 + f_1^2 - f_2^2$.  The $T^4$ components of the Einstein equation say $|F_i|^2 = \tfrac13 |F|^2$.  This translates into the equalities $-f_0^2 + f_1^2 = -f_2^2 + f_1^2 = \tfrac13 (-f_0^2 + f_1^2 - f_2^2)$.  The first equality says $f_0^2 = f_2^2$, whereas the second says $f_2^2 = 2 f_1^2$.  This means $|F|^2=-3f_1^2$.  The Einstein equation for $\AdS_2$ says $\Ric = -\tfrac32 f_1^2 g$, whereas that of the $5$-dimensional space is $\Ric = \half f_1^2 g$.  In summary, we have
\begin{equation}\label{eq:ads2s5t4}
  \boxed{\AdS_2 \times \SLAG_3 \times T^4 \qquad \AdS_2 \times S^5 \times T^4 \qquad F = f \left(d\vartheta^{1234} \pm \sqrt{2} \nu \wedge (d\vartheta^{12} \pm d\vartheta^{34})\right)~,}
\end{equation}
where the signs are uncorrelated.  These is precisely (up to change of local orthonormal frame) the backgrounds obtained by letting the $\CP^2$ curvature above vanish.

If on the contrary we let the curvature of $S^5$ or $\SLAG_3$ vanish, we obtain the branch of $\AdS_3 \times \CP^2 \times T^5$ backgrounds with $f_0=0$ in equation~\eqref{eq:ads2cp2t5}.

\subsubsection{$\AdS_2 \times \SLAG_3 \times S^2 \times S^2$, $\AdS_2 \times S^5 \times S^2 \times S^2$, $\AdS_2 \times \SLAG_3 \times S^2 \times T^2$ and $\AdS_2 \times S^5 \times S^2 \times T^2$ backgrounds}
\label{sec:ads_2-times-slag_3-3}

\begin{figure}[h!]
  \centering
  \includegraphics[width=6cm]{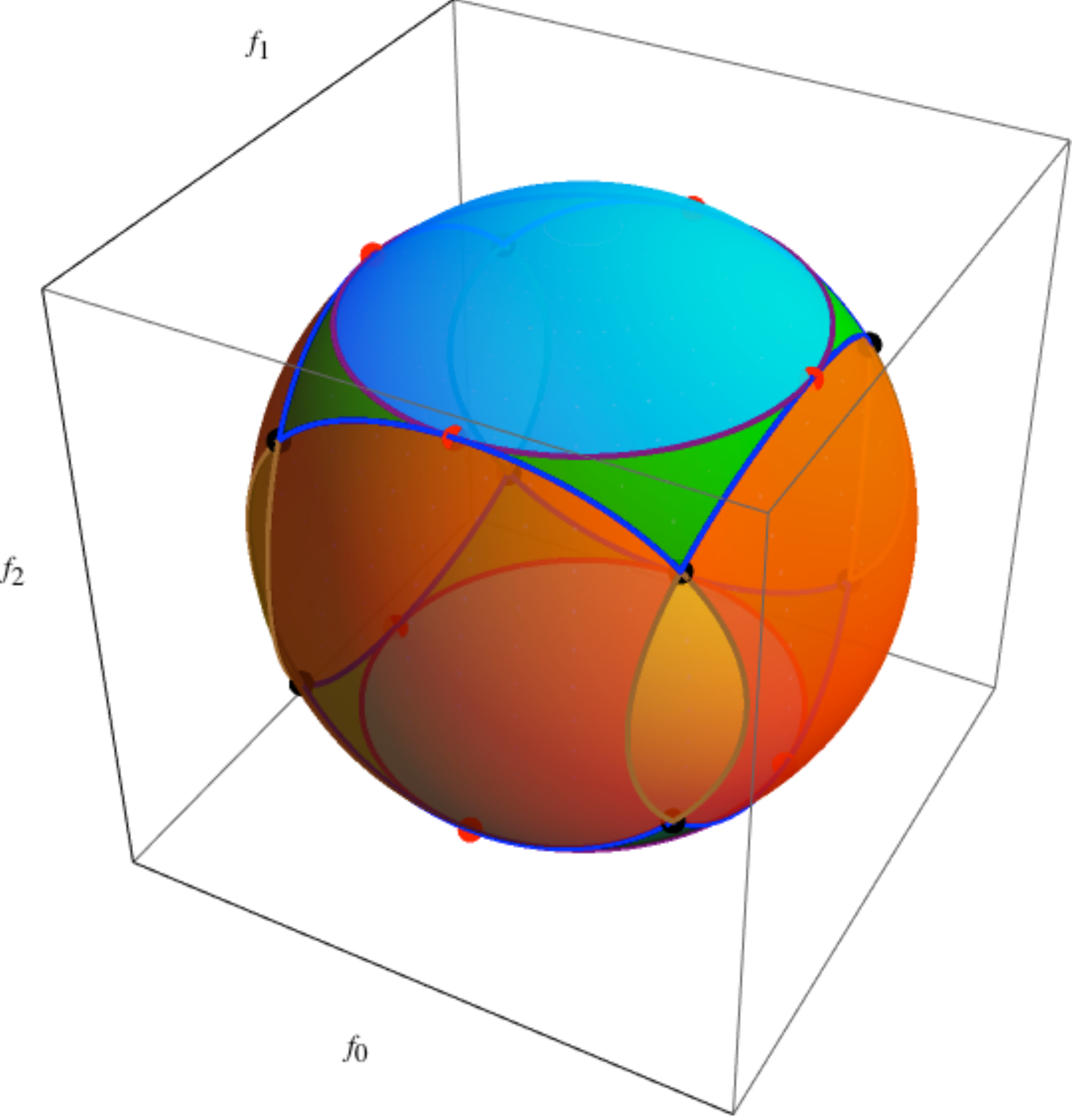}\qquad\qquad
  \begin{tikzpicture} [scale=0.75]
    \def\rectanglepath{-- ++(4mm,0mm) -- ++(0mm,4mm) -- ++(-4mm,0mm) -- cycle}
    \def\thicklinepath{-- ++(4mm,0mm) -- ++(0mm,1mm) -- ++(-4mm,0mm) -- cycle}
    \fill [green] (0,0) \rectanglepath;
    \draw (0,0.2) node [black,right=5mm] {$\AdS_2 \times S^5 \times S^2 \times S^2$};
    \fill [orange] (0,1) \rectanglepath;
    \draw (0,1.2) node [black,right=5mm] {$\AdS_2 \times S^5 \times S^2 \times H^2$};
    \fill [yellow] (0,2) \rectanglepath;
    \draw (0,2.2) node [black,right=5mm] {$\AdS_2 \times S^5 \times H^2 \times H^2$};
    \fill [cyan] (0,3) \rectanglepath;
    \draw (0,3.2) node [black,right=5mm] {$\AdS_2 \times H^5 \times S^2 \times S^2$};
    \fill [blue] (0,4) \thicklinepath;
    \draw (0,4.15) node [black,right=5mm] {$\AdS_2 \times S^5 \times S^2 \times T^2$};
    \fill [brown] (0,5) \thicklinepath;
    \draw (0,5.15) node [black,right=5mm] {$\AdS_2 \times S^5 \times H^2 \times T^2$};
    \fill [purple] (0,6) \thicklinepath;
    \draw (0,6.15) node [black,right=5mm] {$\AdS_2 \times T^5 \times S^2 \times S^2$};
    \fill [black] (0.2,7) circle (1mm);
    \draw (0,7.1) node [black,right=5mm] {$\AdS_2 \times S^5 \times T^4$};
    \fill [red] (0.2,8) circle (1mm);
    \draw (0,8.1) node [black,right=5mm] {$\AdS_2 \times S^2 \times T^7$};
\end{tikzpicture}
  \caption{$F$-moduli for $\AdS_2 \times S^5 \times S^2 \times S^2$ geometries}
  \label{fig:FmoduliAdS2S5S2S2}
\end{figure}
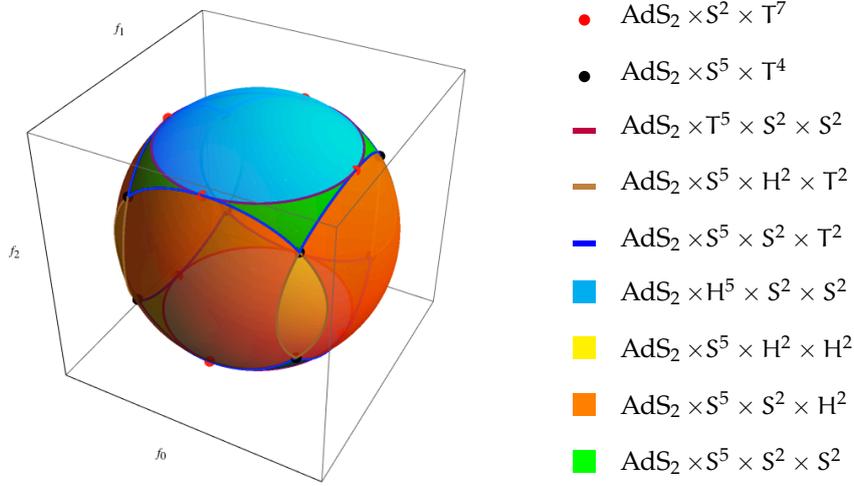

The $F$-moduli space is illustrated for the case of $S^5$ in Figure~\ref{fig:FmoduliAdS2S5S2S2}.  There is a similar diagram for $\SLAG_3$ replacing the $S^5$.

The most general invariant $4$-form is given by
\begin{equation}
  F= f_0 \nu \wedge \sigma_1 + f_1 \nu \wedge \sigma_2 + f_2 \sigma_1 \wedge \sigma_2~,
\end{equation}
where $\sigma_i$ are the area forms on the $S^2$s.  The norm is given by $|F|^2 = -f_0^2 -f_1^2 + f_2^2$.  The equation $F\wedge F =0$ holds for dimensional reasons, since $F$ has legs along a six-dimensional space.  The Einstein equations for $S^5$ (or $\SLAG_3$) are $\Ric = \tfrac16 (f_0^2+f_1^2 - f_2^2) g$, whereas the ones for the two $S^2$ are $\Ric = \tfrac16 (f_1^2 + 2f_2^2 - 2f_0^2) g$ and $\Ric = \tfrac16 (f_0^2 + 2f_2^2 - 2f_1^2) g$, respectively.  The one for $\AdS_2$ says that $\Ric = -\tfrac16 (2f_0^2 + 2f_1^2 + f_2^2) g$.  It is not hard to show that if the scalar curvatures of the $S^5$ is negative (so we have $H^5$ instead) then the scalar curvatures of the two $S^2$s must be positive.  On the other hand, if the scalar curvature of the $S^5$ is positive, then the scalar curvatures of the two $S^2$s are not constrained.  Hence we get four types of geometries, depending on the values of $f_0,f_1,f_2$.

\begin{equation}
  \boxed{\AdS_2 \times \SLAG_3 \times S^2 \times S^2 \qquad \AdS_2 \times S^5 \times S^2 \times S^2 \qquad 
  \begin{cases}
    f_0^2 + f_1^2 > f_2^2\\
    f_1^2 + 2 f_2^2 > 2 f_0^2\\
    f_0^2 + 2 f_2^2 > 2 f_1^2
  \end{cases}}
\end{equation}

\begin{equation}
  \boxed{\AdS_2 \times \SLAG_3 \times S^2 \times H^2 \qquad \AdS_2 \times S^5 \times S^2 \times H^2 \qquad
  \begin{cases}
    2 f_1^2 > f_0^2 + 2 f_2^2\\
    2 f_0^2 < f_1^2 + 2 f_2^2
  \end{cases}}
\end{equation}

\begin{equation}
  \boxed{\AdS_2 \times \SLAG_3 \times H^2 \times H^2 \qquad \AdS_2 \times S^5 \times H^2 \times H^2 \qquad
  \begin{cases}
    2 f_1^2 > f_0^2 + 2 f_2^2\\
    2 f_0^2 > f_1^2 + 2 f_2^2
  \end{cases}}
\end{equation}

\begin{equation}
  \boxed{\AdS_2 \times (\SL(3,\RR)/\SO(3)) \times S^2 \times S^2 \qquad \AdS_2 \times H^5 \times S^2 \times S^2 \qquad f_2^2 > f_0^2 + f_1^2}
\end{equation}

Letting the curvature of one of $S^2$ (or $H^2$) vanish, we obtain backgrounds of the type $\AdS_2 \times \SLAG_3 \times S^2 \times T^2$ and $\AdS_2 \times S^5 \times S^2 \times T^2$.  The most general $F$ in these geometries is not more general than the one just considered, hence the backgrounds are obtained from the ones above simply by letting one of $S^2$ or $H^2$ in the first two of the above backgrounds become flat:

\begin{equation}
  \boxed{\AdS_2 \times \SLAG_3 \times S^2 \times T^2 \qquad \AdS_2 \times S^5 \times S^2 \times T^2 \qquad
  \begin{cases}
    2 f_0^2 = 2 f_2^2 + f_1^2\\
    2 f_2^2 > f_1^2
  \end{cases}}
\end{equation}
  
\begin{equation}
  \boxed{\AdS_2 \times \SLAG_3 \times H^2 \times T^2 \qquad \AdS_2 \times S^5 \times H^2 \times T^2 \qquad
  \begin{cases}
    2 f_0^2 = 2 f_2^2 + f_1^2\\
    2 f_2^2 < f_1^2
  \end{cases}}
\end{equation}

If we let both $S^2$ become flat, we obtain the backgrounds with geometry $\AdS_2 \times S^5 \times T^4$ or $\AdS_2 \times \SLAG_3 \times T^4$ in equation \eqref{eq:ads2s5t4}, whereas if we let the curvature of the $S^5$ or $\SLAG_3$ vanish, we obtain backgrounds with geometries $\AdS_2 \times S^2 \times S^2 \times T^5$ and $\AdS_2 \times S^2 \times T^7$ to be discussed below in more generality.

\subsubsection{$\AdS_2 \times \CP^2 \times S^3 \times S^2$ and $\AdS_2 \times \CP^2 \times S^3 \times T^2$ backgrounds}
\label{sec:ads_2-times-cp2-1}

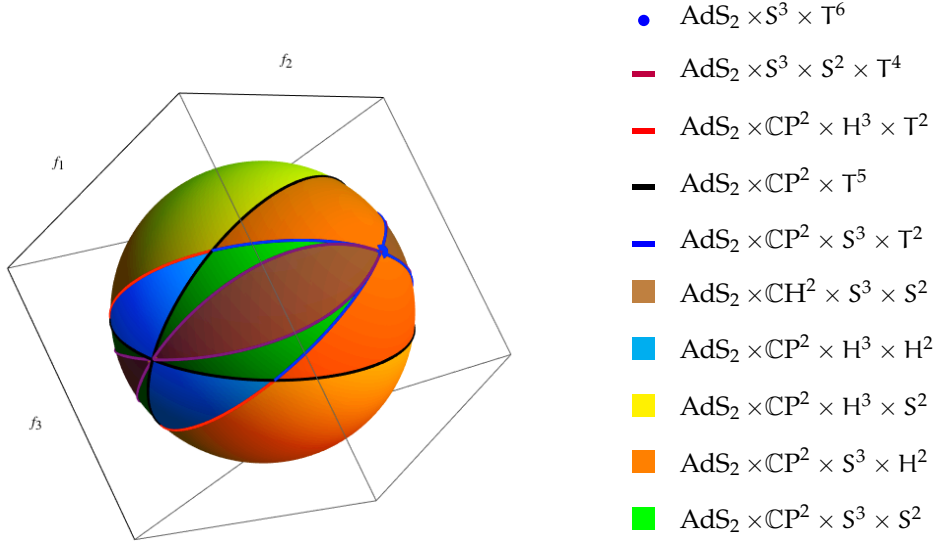
\begin{figure}[h!]
  \centering
  \includegraphics[width=7cm]{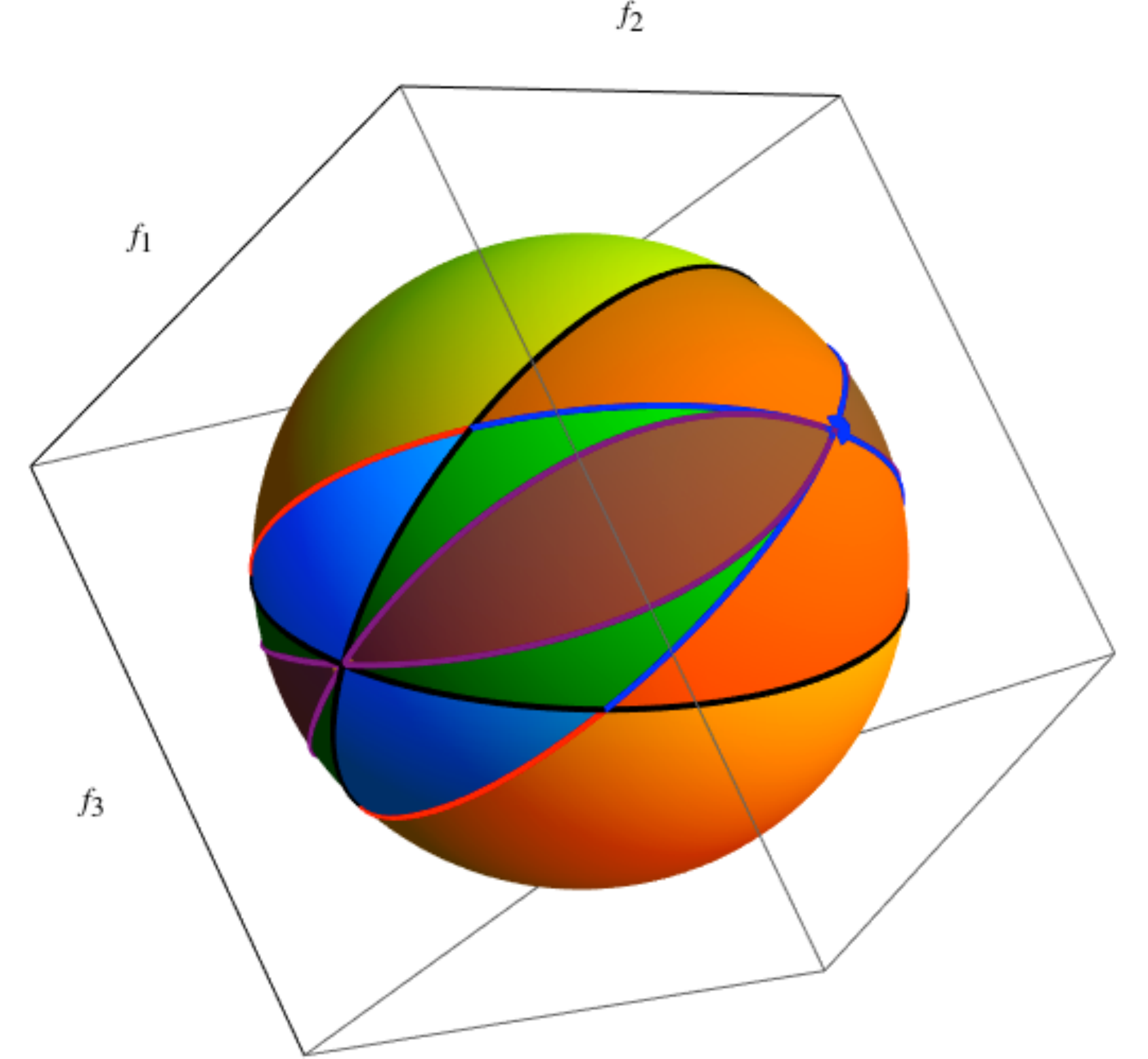}\qquad\qquad
  \begin{tikzpicture} [scale=0.75]
    \def\rectanglepath{-- ++(4mm,0mm) -- ++(0mm,4mm) -- ++(-4mm,0mm) -- cycle}
    \def\thicklinepath{-- ++(4mm,0mm) -- ++(0mm,1mm) -- ++(-4mm,0mm) -- cycle}
    \fill [green] (0,0) \rectanglepath;
    \draw (0,0.2) node [black,right=5mm] {$\AdS_2 \times \CP^2 \times S^3 \times S^2$};
    \fill [orange] (0,1) \rectanglepath;
    \draw (0,1.2) node [black,right=5mm] {$\AdS_2 \times \CP^2 \times S^3 \times H^2$};
    \fill [yellow] (0,2) \rectanglepath;
    \draw (0,2.2) node [black,right=5mm] {$\AdS_2 \times \CP^2 \times H^3 \times S^2$};
    \fill [cyan] (0,3) \rectanglepath;
    \draw (0,3.2) node [black,right=5mm] {$\AdS_2 \times \CP^2 \times H^3 \times H^2$};
    \fill [brown] (0,4) \rectanglepath;
    \draw (0,4.2) node [black,right=5mm] {$\AdS_2 \times \CH^2 \times S^3 \times S^2$};
    \fill [blue] (0,5) \thicklinepath;
    \draw (0,5.15) node [black,right=5mm] {$\AdS_2 \times \CP^2 \times S^3 \times T^2$};
    \fill [black] (0,6) \thicklinepath;
    \draw (0,6.15) node [black,right=5mm] {$\AdS_2 \times \CP^2 \times T^5$};
    \fill [red] (0,7) \thicklinepath;
    \draw (0,7.15) node [black,right=5mm] {$\AdS_2 \times \CP^2 \times H^3 \times T^2$};
    \fill [purple] (0,8) \thicklinepath;
    \draw (0,8.15) node [black,right=5mm] {$\AdS_2 \times S^3 \times S^2 \times T^4$};
    \fill [blue] (0.2,9) circle (1mm);
    \draw (0,9.1) node [black,right=5mm] {$\AdS_2 \times S^3 \times T^6$};
\end{tikzpicture}
\caption{$F$-moduli for $\AdS_2 \times \CP^2 \times S^3 \times S^2$ backgrounds}
\label{fig:FmoduliAdS2CP2S3S2}
\end{figure}

The F-moduli for these backgrounds is illustrated in Figure~\ref{fig:FmoduliAdS2CP2S3S2}.

In this geometry, the most general invariant $4$-form is
\begin{equation}
  \label{eq:Fads2cp2s3s2}
  F = f_0 \nu \wedge \omega + f_1 \nu \wedge \sigma_2 + f_2 \half \omega^2 + f_3 \omega \wedge \sigma_2
\end{equation}
The equation $F\wedge F = 0$ imposes the condition $2 f_0f_3 + f_1f_2 = 0$.  The norm is given by $|F|^2 = -2 f_0^2 -f_1^2 + f_2^2 + 2 f_3^2$.  The Einstein equations for $\AdS_2$, $\CP^2$, $S^3$ and $S^2$ are of the form $\Ric = \tfrac16  \lambda g$, where the constant $\lambda$ takes the following values:
  \begin{equation}
    \begin{aligned}[m]
      \lambda_{\AdS_2} &= - 4 f_0^2 - 2 f_1^2 - f_2^2 - 2 f_3^2\\
      \lambda_{\CP^2} &= - f_0^2 + f_1^2 + 2 f_2^2 + f_3^2
    \end{aligned}\qquad\qquad
    \begin{aligned}[m]
      \lambda_{S^3} &= 2 f_0^2 + f_1^2 - f_2^2 - 2 f_3^2 \\
      \lambda_{S^2} &= 2 f_0^2 - 2 f_1^2 - f_2^2 + 4 f_3^2~.
    \end{aligned}
  \end{equation}
The sum of all but $\lambda_{\AdS_2}$ is non-negative, whence they cannot all be negative.  Similarly, if $\lambda_{\CP^2}$ is negative then $\lambda_{S^3}$ and $\lambda_{S^2}$ are both positive.  Any other combination of signs is possible.  In summary, we have the following backgrounds:
\begin{equation}
  \boxed{\AdS_2 \times \CP^2 \times S^3 \times S^2 \qquad f_1^2 + 2 f_2^2 + f_3^2 > f_0^2\qquad 2 f_0^2 + f_1^2 > f_2^2 + 2 f_3^2\qquad 2 f_0^2 + 4 f_3^2 > 2 f_1^2 + f_2^2~,}
\end{equation}
\begin{equation}
  \boxed{\AdS_2 \times \CP^2 \times S^3 \times H^2 \qquad f_1^2 + 2 f_2^2 + f_3^2 >  f_0^2 \qquad 2 f_0^2 + f_1^2 > f_2^2 + 2 f_3^2 \qquad 2 f_0^2 + 4 f_3^2 < 2 f_1^2 + f_2^2~,}
\end{equation}
\begin{equation}
  \boxed{\AdS_2 \times \CP^2 \times H^3 \times S^2 \qquad f_1^2 + 2 f_2^2 + f_3^2 > f_0^2\qquad 2 f_0^2 + f_1^2 < f_2^2 + 2 f_3^2\qquad 2 f_0^2 + 4 f_3^2 > 2 f_1^2 + f_2^2~,}
\end{equation}
\begin{equation}
  \boxed{\AdS_2 \times \CP^2 \times H^3 \times H^2 \qquad f_1^2 + 2 f_2^2 + f_3^2 > f_0^2\qquad 2 f_0^2 + f_1^2 < f_2^2 + 2 f_3^2\qquad 2 f_0^2 + 4 f_3^2 < 2 f_1^2 + f_2^2~,}
\end{equation}
and
\begin{equation}
  \boxed{\AdS_2 \times \CH^2 \times S^3 \times S^2 \qquad f_1^2 + 2 f_2^2 + f_3^2 < f_0^2\qquad 2 f_0^2 + f_1^2 > f_2^2 + 2 f_3^2\qquad 2 f_0^2 + 4 f_3^2 > 2 f_1^2 + f_2^2~.}
\end{equation}

Letting the curvature of the $S^2$ (or $H^2$) vanish we arrive at backgrounds with geometry $\AdS_2 \times \CP^2 \times S^3 \times T^2$. However in principle such a geometry admits more general $F$: namely,
\begin{equation}
  F = f_0 \nu \wedge \omega + f_1 \nu \wedge d\vartheta^{12} + f_2 \half \omega^2 + f_3 \omega \wedge d\vartheta^{12} + f_4 \sigma_3 \wedge d\vartheta^1~,
\end{equation}
without loss of generality.  The equation $F \wedge F = 0$ imposes the relations $f_0 f_4 =0$, $f_2 f_4 =0$ and $f_1 f_2 + 2 f_0 f_3 = 0$, whereas the Einstein equations along the flat directions become $-f_1^2 + 2 f_3^2 = -f_1^2 + 2 f_3^2 + f_4^2 = \tfrac13 |F|^2$, where $|F|^2 = -f_1^2 -2 f_0^2 + 2 f_3^2 + f_2^2 + f_4 ^2$.  But now the first equality translates into $f_4=0$, whence we are back to the same Ansatz for $F$ as in equation \eqref{eq:Fads2cp2s3s2}, except that we impose the additional Einstein equation: $2f_0^2 + 4 f_3^2 = 2f_1^2 + f_2^2$, which has the effect of setting the curvature of the $S^2$ to zero.  In summary, we have

\begin{equation}
  \boxed{\AdS_2 \times \CP^2 \times S^3 \times T^2  \qquad f_0^2 + 3 f_3^2 > f_1^2 > 2 f_3^2~,}
\end{equation}
\begin{equation}
  \boxed{\AdS_2 \times \CH^2 \times S^3 \times T^2  \qquad f_1^2 > 3 f_3^2 + f_0^2 ~,}
\end{equation}
and
\begin{equation}
  \boxed{\AdS_2 \times \CP^2 \times H^3 \times T^2  \qquad 2 f_3^2 >  f_1^2~.}
\end{equation}

If we let instead the curvature of $S^3$ vanish we do not obtain any backgrounds within this Ansatz; although as we shall see in Section \ref{sec:ads2cp2s2t3} below, there are symmetric backgrounds with underlying geometry $\AdS_2 \times \CP^2 \times H^2 \times T^3$.

If we let the curvature of $\CP^2$ vanish, we obtain a branch of $\AdS_2 \times S^3 \times S^2 \times T^4$ backgrounds to be discussed in more generality below in Section \ref{sec:ads_2-times-s3-2}.  In addition we can let the curvature of $S^3$ or of $S^2$ (but not both) vanish to obtain $\AdS_2 \times S^2 \times T^7$ and $\AdS_2 \times S^2 \times T^6$ backgrounds.  These latter backgrounds were discussed in Section \ref{sec:ads2s3t6}.

\subsubsection{$\AdS_2 \times S^3 \times S^2 \times S^2 \times S^2$, $\AdS_2 \times S^3 \times S^2 \times S^2 \times T^2$ and $\AdS_2 \times S^3 \times S^2 \times T^4$ backgrounds}
\label{sec:ads_2-times-s3-2}

In the first of these geometries, the most general $F$ is given by
  \begin{equation}\label{eq:Fads2s3s2s2s2}
    F = f_1 \nu \wedge \sigma_1 + f_2 \nu \wedge \sigma_2 + f_3 \nu \wedge \sigma_3 + f_4 \sigma_1 \wedge \sigma_2 + f_5 \sigma_1 \wedge \sigma_3 + f_6 \sigma_2 \wedge \sigma_3~,
  \end{equation}
whereas the equation $F \wedge F = 0$ only imposes the condition $f_1 f_6 + f_2 f_5 + f_3 f_4 = 0$.  The norm of $F$ is given by $|F|^2 = -f_1^2 - f_2^2 - f_3^2 + f_4^2 + f_5^2 + f_6^2$.  The Einstein equations for the different factors are $\Ric = \tfrac16 \lambda g$, where
\begin{equation}
  \begin{aligned}[m]
    \lambda_{S^3} &= f_1^2 + f_2^2 + f_3^2 - f_4^2 - f_5^2 - f_6^2\\
    \lambda_{S_{(1)}^2} &= - 2 f_1^2 + f_2^2 + f_3^2 + 2 f_4^2 + 2 f_5^2 - f_6^2\\
    \lambda_{S_{(2)}^2} &= f_1^2 - 2 f_2^2 + f_3^2 + 2 f_4^2 - f_5^2 + 2 f_6^2\\
    \lambda_{S_{(3)}^2} &= f_1^2 + f_2^2 - 2 f_3^2 - f_4^2 + 2 f_5^2 + 2 f_6^2
 \end{aligned}
\end{equation}
It is clear that the sum of any three is non-negative, whence at most two of the $\lambda$s can be negative.  This gives rise to the following backgrounds.

\begin{equation}
  \boxed{\AdS_2 \times S^3 \times S^2 \times S^2 \times S^2  \qquad     \begin{cases}
     f_1^2 + f_2^2 + f_3^2 > f_4^2 + f_5^2 + f_6^2 \\
    f_2^2 + f_3^2 + 2 f_4^2 + 2 f_5^2 > 2 f_1^2 + f_6^2 \\
    f_1^2 + f_3^2 + 2 f_4^2 + 2 f_6^2 > 2 f_2^2 + f_5^2 \\
    f_1^2 + f_2^2 + 2 f_5^2 + 2 f_6^2 > 2 f_3^2 + f_4^2
   \end{cases}}
\end{equation}

\begin{equation}
  \boxed{\AdS_2 \times H^3 \times S^2 \times S^2 \times S^2 \qquad     \begin{cases}
     f_1^2 + f_2^2 + f_3^2 < f_4^2 + f_5^2 + f_6^2 \\
    f_2^2 + f_3^2 + 2 f_4^2 + 2 f_5^2 > 2 f_1^2 + f_6^2 \\
    f_1^2 + f_3^2 + 2 f_4^2 + 2 f_6^2 > 2 f_2^2 + f_5^2 \\
    f_1^2 + f_2^2 + 2 f_5^2 + 2 f_6^2 > 2 f_3^2 + f_4^2
   \end{cases}}
\end{equation}

\begin{equation}
  \boxed{\AdS_2 \times S^3 \times H^2 \times S^2 \times S^2 \qquad     \begin{cases}
     f_1^2 + f_2^2 + f_3^2 > f_4^2 + f_5^2 + f_6^2 \\
    f_2^2 + f_3^2 + 2 f_4^2 + 2 f_5^2 < 2 f_1^2 + f_6^2 \\
    f_1^2 + f_3^2 + 2 f_4^2 + 2 f_6^2 > 2 f_2^2 + f_5^2 \\
    f_1^2 + f_2^2 + 2 f_5^2 + 2 f_6^2 > 2 f_3^2 + f_4^2
   \end{cases}}
\end{equation}

\begin{equation}
  \boxed{\AdS_2 \times H^3 \times H^2 \times S^2 \times S^2 \qquad     \begin{cases}
     f_1^2 + f_2^2 + f_3^2 < f_4^2 + f_5^2 + f_6^2 \\
    f_2^2 + f_3^2 + 2 f_4^2 + 2 f_5^2 < 2 f_1^2 + f_6^2 \\
    f_1^2 + f_3^2 + 2 f_4^2 + 2 f_6^2 > 2 f_2^2 + f_5^2 \\
    f_1^2 + f_2^2 + 2 f_5^2 + 2 f_6^2 > 2 f_3^2 + f_4^2
   \end{cases}}
\end{equation}

\begin{equation}
  \boxed{\AdS_2 \times S^3 \times H^2 \times H^2 \times S^2 \qquad     \begin{cases}
     f_1^2 + f_2^2 + f_3^2 > f_4^2 + f_5^2 + f_6^2 \\
    f_2^2 + f_3^2 + 2 f_4^2 + 2 f_5^2 < 2 f_1^2 + f_6^2 \\
    f_1^2 + f_3^2 + 2 f_4^2 + 2 f_6^2 < 2 f_2^2 + f_5^2 \\
    f_1^2 + f_2^2 + 2 f_5^2 + 2 f_6^2 > 2 f_3^2 + f_4^2
   \end{cases}}
\end{equation}

Letting the curvature of one the $S^2$ vanish, we arrive at backgrounds with underlying geometry $\AdS_2 \times S^3 \times S^2 \times S^2 \times T^2$.  However such a geometry admits, in principle, a more general $F$: namely, one which can be brought to the form
  \begin{equation}
    F = f_1 \nu \wedge \sigma_1 + f_2 \nu \wedge \sigma_2 + f_3 \nu \wedge d\vartheta^{12} + f_4 \sigma_1 \wedge \sigma_2 + f_5 \sigma_1 \wedge d\vartheta^{12} + f_6 \sigma_2 \wedge d\vartheta^{12} + f_7 \upsilon \wedge d\vartheta^1
  \end{equation}
without loss of generality, where $\upsilon$ is the volume form on the $S^3$.  The equation $F \wedge F = 0$ results in the relations
\begin{equation}
  f_3 f_4 + f_1 f_6 + f_2 f_5 = 0, \qquad f_7 f_4 = 0, \qquad f_1 f_7 = 0,\qquad f_2 f_7 = 0~,
\end{equation}
whereas the $T^2$ Einstein equations become
\begin{equation}
  -f_3^2 + f_7^2 + f_5^2 + f_6^2 = -f_3^2 + f_5^2 + f_6^2 = \tfrac13 (f_7^2 - f_1^2 - f_2^2 - f_3^2 + f_4^2 + f_5^2 + f_6^2)~.
\end{equation}
These equations are equivalent to $f_7=0$ and
\begin{equation}\label{eq:AdS2S3S2S2T2rels}
  f_1^2 + f_2^2 + 2 f_5^2 + 2 f_6^2 = 2 f_3^2 + f_4^2~.
\end{equation}
In particular, since $f_7 =0$ we are back in the Ansatz for $F$ given in equation \eqref{eq:Fads2s3s2s2s2}, but with a particular choice of which one of the three $S^2$ has become flat.  This means that we can obtain these backgrounds simply by setting $\lambda_{S^2_{(3)}} = 0$.  Clearly there are similar classes of backgrounds obtained by setting any of the other $\lambda_{S^2}$ to zero instead.

In summary, we have -- up to relabeling -- the following backgrounds:
\begin{equation}
  \boxed{\AdS_2 \times S^3 \times S^2 \times S^2 \times T^2 \qquad
    f_3^2 > f_5^2 + f_6^2 \qquad
    f_3^2 + f_4^2 > f_1^2 + f_6^2 \qquad
    f_3^2 + f_4^2 > f_2^2 + f_5^2~,}
\end{equation}
\begin{equation}
  \boxed{\AdS_2 \times H^3 \times S^2 \times S^2 \times T^2 \qquad
    f_3^2 < f_5^2 + f_6^2 \qquad
    f_3^2 + f_4^2 > f_1^2 + f_6^2 \qquad
    f_3^2 + f_4^2 > f_2^2 + f_5^2~,}
\end{equation}
and
\begin{equation}
  \boxed{\AdS_2 \times S^3 \times S^2 \times H^2 \times T^2 \qquad
    f_3^2 > f_5^2 + f_6^2 \qquad
    f_3^2 + f_4^2 > f_1^2 + f_6^2 \qquad
    f_3^2 + f_4^2 < f_2^2 + f_5^2~.}
\end{equation}

Letting the curvature of another of the $S^2$ vanish, we obtain backgrounds with underlying geometry $\AdS_2 \times S^3 \times S^2 \times T^4$.  Again, though, such a geometry admits a more general $F$: namely,
\begin{equation}
  F = f_1 \nu \wedge \sigma + \nu \wedge \alpha + \upsilon \wedge \beta + \sigma \wedge \gamma + f_6 d\vartheta^{1234}~,
\end{equation}
where $\alpha,\gamma \in \Omega^2(T^4)$ and $\beta \in \Omega^1(T^4)$ are parallel forms on $T^4$ and $\sigma$ and $\upsilon$ are the volume forms on $S^2$ and $S^3$, respectively.  The equation $F \wedge F = 0$ imposes the following relations
\begin{equation}
  \alpha \wedge \beta = 0,\qquad \alpha \wedge \gamma + f_1 f_6 d\vartheta^{1234} = 0~,\qquad f_1 \beta = 0, \qquad \beta \wedge \gamma = 0~.
\end{equation}
We must distinguish two cases, depending on whether or not $\beta=0$.
\begin{enumerate}
\item If $\beta$ is a nonzero $1$-form, then $f_1=0$ and the equations $\alpha \wedge \beta = 0$ and $\beta \wedge \gamma =0$ imply the existence of $1$-forms $\zeta$ and $\xi$ such that $\alpha = \zeta \wedge \beta$ and $\gamma = \xi \wedge \beta$.  The most general $F$ can now be brought to the form
\begin{equation}
  F = f_0 \upsilon_3 \wedge d\vartheta^1 + f_2 \sigma \wedge d\vartheta^{12} + f_6 d\vartheta^{1234} + f_4 \nu \wedge d\vartheta^{12} + f_5 \nu \wedge d\vartheta^{13}~.
\end{equation}
Indeed, we can choose $\beta = f_0 d\vartheta^1$ without loss of generality, and hence $\zeta$ to be any linear combination of $d\vartheta^2$ and $d\vartheta^3$.  But now acting by isometries fixing $\beta$ we can assume that $\zeta$ has no $d\vartheta^3$ component.  Finally this leaves $\xi$ to be an arbitrary linear combination of $d\vartheta^2$ and $d\vartheta^3$.  The equation $F\wedge F =0$ is now automatically satisfied, whereas the $T^4$ Einstein equations become the following equalities:
\begin{equation}
  f_0^2 + f_2^2 + f_6^2 - f_4^2 - f_5^2 = f_2^2 + f_6^2 - f_4^2 = f_6^2 - f_5^2 = f_6^2 = 0~,
\end{equation}
and also $f_4 f_5 = 0$.  This implies that $f_6=f_5=f_0=0$ and $f_2^2 = f_4^2$.  In particular, it follows that $|F|=0$ and since $F$ has remained without legs along $S^3$, the Einstein equations would say $S^3$ is Ricci-flat, which is absurd.

\item On the other hand, if $\beta = 0$, then $F$ can be brought to the following form
\begin{equation}
  F = \nu \wedge (f_1 \sigma + f_2 d\vartheta^{12} + f_3 d\vartheta^{34}) + f_6 d\vartheta^{1234} + \sigma \wedge (f_4 d\vartheta^{12} + f_5 d\vartheta^{34} + f_7 d\vartheta^{13} + f_8 d\vartheta^{14})~,
\end{equation}
and the only relation left from $F \wedge F =0$ is $f_2 f_5 + f_1 f_6 + f_3 f_4 = 0$.  The $T^4$ Einstein equations translate into the following equalities:
\begin{equation}
  f_5 f_8 = f_5 f_7 = f_4 f_7 = f_4 f_8 = f_7 f_8 = 0~,
\end{equation}
and
\begin{equation}
  -f_2^2 + f_6^2 + f_4^2 + f_7^2 + f_8^2 = -f_2^2 + f_6^2 + f_4^2 = - f_3^2 + f_6^2 + f_5^2 + f_7^2 = -f_3^2 f_6^2 + f_5^2 + f_8^2 = \tfrac13 |F|^2
\end{equation}
where $|F|^2 = -f_1^2 - f_2^2 - f_3^2 + f_6^2 + f_4^2 + f_5^2 + f_7^2 + f_8^2$.  The first of the above equalities says that $f_7=f_8=0$, which means that $F$ now conforms to the Ansatz in equation \eqref{eq:Fads2s3s2s2s2}.  This means that this family of backgrounds is obtained by setting, say, $\lambda_{S^2_{(2)}} =0$ in addition.
\end{enumerate}

In summary, we have $\lambda_{S^2_{(2)}} = \lambda_{S^2_{(3)}} = 0$ and the following backgrounds:
\begin{equation}
  \boxed{\AdS_2 \times S^3 \times S^2 \times T^4\qquad 2f_3^2 + 3f_4^2 > f_1^2 + f_2^2 > f_4^2~,}
\end{equation}
and
\begin{equation}
  \boxed{\AdS_2 \times S^3 \times H^2 \times T^4 \qquad f_1^2 + f_2^2 > 2 f_3^2 + 3f_4^2~.}
\end{equation}

\subsubsection{$\AdS_2 \times \CP^2 \times S^2 \times T^3$ backgrounds}
\label{sec:ads2cp2s2t3}

In this geometry, the most general $F$ is given by
  \begin{equation}
    F = f_0 \nu \wedge \omega + f_1 \nu \wedge \sigma_2 + f_2 \omega \wedge \sigma_2 + f_3 \half \omega^2 + \nu \wedge \alpha + \omega \wedge \beta + \sigma_2 \wedge \gamma~,
  \end{equation}
where $\alpha,\beta,\gamma \in \Omega^2(T^3)$.  The equation $F \wedge F = 0$ implies the following relations:
\begin{equation}
  2 f_0 f_2 + f_1 f_3 = 0, \qquad 2 f_0 \beta + f_3 \alpha = 0, \qquad f_0 \gamma + f_1 \beta + f_2 \alpha = 0,\qquad 2 f_2 \beta + f_3 \gamma = 0~.
\end{equation}
Without loss of generality with can choose the one-forms $\star\alpha = f_4 d\vartheta^1$, $\star\beta = f_5 d\vartheta^1 + f_6 d\vartheta^2$ and $\star\gamma = f_7 d\vartheta^1 + f_8 d\vartheta^2 + f_9 d\vartheta^3$,whence
\begin{multline}
  F = f_0 \nu \wedge \omega + f_1 \nu \wedge \sigma_2 + f_2 \omega \wedge \sigma_2 + f_3 \half \omega^2 + f_4 \nu \wedge d\vartheta^{23}\\
  + f_5 \omega \wedge d\vartheta^{23} - f_6 \omega \wedge d\vartheta^{13} + f_7  \sigma_2 \wedge d\vartheta^{23} - f_8  \sigma_2 \wedge d\vartheta^{13} + f_9  \sigma_2 \wedge d\vartheta^{12}~,
\end{multline}
with relations
\begin{equation}
  \begin{split}\label{eq:AdS2CP2S2T3rels}
    2 f_0 f_2 + f_1 f_3 = 0,\quad 2 f_0 f_5 + f_3 f_4 = 0,\quad 2 f_2 f_5 + f_3 f_7 =0, \quad 2 f_2 f_6 + f_3 f_8 =0,\\
    f_0 f_6 = 0,\quad f_3 f_9 = 0,\quad f_0 f_9 = 0,\quad f_0 f_8 + f_1 f_6 = 0, \quad
    f_0 f_7 + f_1 f_5 + f_2 f_4 = 0~.
  \end{split}
\end{equation}
The $T^3$ Einstein equations impose the further equalities
\begin{equation}\label{eq:f_7-f_9-=}
  f_7 f_9 = f_8 f_9 = 2 f_5 f_6 + f_7 f_8 = 0~,
\end{equation}
and
\begin{equation}
  2 f_6^2 + f_8^2 + f_9^2 = - f_4^2 + 2 f_5^2 + f_7^2 + f_9^2 = - f_4^2 + 2 f_5^2 + 2 f_6^2 + f_7^2 + f_8^2 = \tfrac13 (-2 f_0^2 - f_1^2 + 2 f_2^2 + f_3^2 - f_4^2 + 2f_5^2 + 2 f_6^2 + f_7^2 + f_8^2 + f_9^2)~,
\end{equation}
which can be rewritten as
\begin{equation}\label{eq:T3EEAdS2}
  f_4^2 + 2 f_6^2 + f_8^2 = 2 f_5^2 + f_7^2 = f_4^2 + f_9^2 \qquad\text{and}\qquad
  2 f_0^2 + f_1^2 + 3 f_9^2  = 2 f_2^2 + f_3^2~.
\end{equation}

The above system of equations is not hard to reduce.  One can see fairly quickly that $f_6=f_8=f_9=0$.  This means that $d\vartheta^1$ does not appear in $F$, whence the Einstein equation along $\vartheta^1$ will force $|F|=0$.  In all cases, the Einstein equations for the different factors are of the form $\Ric = \half \lambda g$ for some $\lambda$ to be described below on a case-by-case analysis.

There are three different branches of solutions to equations \eqref{eq:AdS2CP2S2T3rels} and \eqref{eq:T3EEAdS2}.
\begin{enumerate}
\item In the first branch, $f_0=f_1=f_2=f_3=f_6=f_8=f_9=0$, whence
\begin{equation}
  F = (f_4 \nu + f_5 \omega + f_7 \sigma_2)\wedge d\vartheta^{23}~,
\end{equation}
subject to $|F|^2 =0$, i.e., $f_4^2 = 2f_5^2 + f_7^2$.  This implies that
\begin{equation}
   \lambda_{\CP^2} = f_5^2,\qquad \lambda_{S^2} = f_7^2,\qquad\text{and}\qquad \lambda_{\AdS_2} = - f_4^2
\end{equation}
whence we have the following background:

\begin{equation}
  \boxed{\AdS_2 \times \CP^2 \times S^2 \times T^3 \qquad  F = (f_4 \nu + f_5 \omega + f_7 \sigma_2)\wedge d\vartheta^{23} \qquad f_4^2 = 2f_5^2 + f_7^2~.}
 \end{equation}

If we set $f_5=0$, we get

\begin{equation}
  \boxed{\AdS_2 \times S^2 \times T^7 \qquad  F = (f_4 \nu + f_7 \sigma_2)\wedge d\vartheta^{23} \qquad f_4^2 = f_7^2~.}
 \end{equation}

\item In the second branch, $f_1=f_2=f_6=f_7=f_8=f_9=0$, whence
  \begin{equation}
  F = f_0 \nu \wedge \omega + f_3 \half \omega^2 + f_4 \nu \wedge d\vartheta^{23} + f_5 \omega \wedge d\vartheta^{23}~,
 \end{equation}
where $f_3 = \pm \sqrt{2} f_0$ and $f_4 = \mp \sqrt{2} f_5$, with correlated signs.  It follows that
\begin{equation}
   \lambda_{\CP^2} = f_0^2 + f_5^2,\qquad \lambda_{S^2} = 0,\qquad\text{and}\qquad \lambda_{\AdS_2} = - (f_0^2 + f_5^2)~,
\end{equation}
whence we have an $\AdS_2 \times \CP^2 \times T^5$ background, of the type already discussed in Section \ref{sec:ads_2-times-cp2}.

\item The final branch is very similar to the previous one.  In this one we have $f_4=f_5=f_6=f_7=f_8=f_9=0$, whence
  \begin{equation}
  F = f_0 \nu \wedge \omega + f_1 \nu \wedge \sigma_2 + f_2 \omega \wedge \sigma_2 + f_3 \half \omega^2~,
 \end{equation}
with $f_3 = \pm \sqrt{2} f_0$ and $f_1 = \mp \sqrt{2} f_2$, with correlated signs.  It follows that
\end{enumerate}
\begin{equation}
   \lambda_{\CP^2} = f_0^2 + f_2^2,\qquad \lambda_{S^2} = 0,\qquad\text{and}\qquad \lambda_{\AdS_2} = - (f_0^2 + f_2^2)~,
\end{equation}
whence we have an $\AdS_2 \times \CP^2 \times T^5$ background, of the type already discussed in Section \ref{sec:ads_2-times-cp2}.  Indeed, notice that since $\lambda_{S^2} =0$, $\sigma_2$ can be understood as $d\vartheta^{45}$, say.

\subsubsection{$\AdS_2 \times S^2 \times S^2 \times S^2 \times S^2 \times S^1$ backgrounds}
\label{sec:ads_2-times-s2}

In this geometry, the most general $F$ is given by
  \begin{equation}
    F = \sum_{i=1}^4 f_i \nu \wedge \sigma_i + \sum_{1\leq i<j\leq 4} f_{ij} \sigma_i \wedge \sigma_j~,
  \end{equation}
and the equation $F\wedge F =0$ results in the following identities:
\begin{equation}
  \begin{aligned}[t]
    f_1 f_{23} + f_2 f_{13} + f_3 f_{12} &= 0\\
    f_1 f_{24} + f_2 f_{14} + f_4 f_{12} &= 0\\
    f_{12} f_{34} + f_{13} f_{24} + f_{14} f_{23} &= 0
  \end{aligned}
  \qquad\qquad
  \begin{aligned}[t]
   f_1 f_{34} + f_3 f_{14} + f_4 f_{13} &= 0\\
    f_2 f_{34} + f_3 f_{24} + f_4 f_{23} &= 0
 \end{aligned}
\end{equation}
Since $F$ has no legs along the $S^1$, the $S^1$ Einstein equation sets $|F|^2=0$, which translates into $\sum_i f_i^2 = \sum_{i<j} f_{ij}^2$.  The Einstein equations for the $S^2$ factors are of the form $\Ric = \half \lambda g$, where
\begin{equation}
  \begin{aligned}[m]
    \lambda_1 &= -f_1^2 + f_{12}^2 + f_{13}^2 + f_{14}^2\\
    \lambda_2 &= -f_2^2 + f_{12}^2 + f_{23}^2 + f_{24}^2
\end{aligned}
  \qquad\qquad
  \begin{aligned}[m]
    \lambda_3 &= -f_3^2 + f_{13}^2 + f_{23}^2 + f_{34}^2\\
    \lambda_4 &= -f_4^2 + f_{14}^2 + f_{24}^2 + f_{34}^2
\end{aligned}
\end{equation}
The sum of any three is non-negative, whence at most two can be negative.  This gives rise to three possible geometries summarised below.  It is easy to show that the first two cases can indeed occur, by considering special solutions with two of the $f_i$ vanishing.

\begin{equation}
  \boxed{\AdS_2 \times S^2 \times S^2 \times S^2 \times S^2 \times S^1 \qquad
    \begin{cases}
    f_{12}^2 + f_{13}^2 + f_{14}^2 > f_1^2 \\
    f_{12}^2 + f_{23}^2 + f_{24}^2 > f_2^2 \\
    f_{13}^2 + f_{23}^2 + f_{34}^2 > f_3^2 \\
    f_{14}^2 + f_{24}^2 + f_{34}^2 > f_4^2 
   \end{cases}~,}
\end{equation}
\begin{equation}
  \boxed{\AdS_2 \times H^2 \times S^2 \times S^2 \times S^2 \times S^1 \qquad
    \begin{cases}
    f_{12}^2 + f_{13}^2 + f_{14}^2 < f_1^2 \\
    f_{12}^2 + f_{23}^2 + f_{24}^2 > f_2^2 \\
    f_{13}^2 + f_{23}^2 + f_{34}^2 > f_3^2 \\
    f_{14}^2 + f_{24}^2 + f_{34}^2 > f_4^2 
   \end{cases}~,}
\end{equation}
and
\begin{equation}
  \boxed{\AdS_2 \times H^2 \times H^2 \times S^2 \times S^2 \times S^1 \qquad
    \begin{cases}
    f_{12}^2 + f_{13}^2 + f_{14}^2 < f_1^2 \\
    f_{12}^2 + f_{23}^2 + f_{24}^2 < f_2^2 \\
    f_{13}^2 + f_{23}^2 + f_{34}^2 > f_3^2 \\
    f_{14}^2 + f_{24}^2 + f_{34}^2 > f_4^2 
   \end{cases}
    ~.}
\end{equation}

I have yet to show that the last system can indeed have solutions.

\subsubsection{$\AdS_2 \times S^2 \times S^2 \times S^2 \times T^3$, $\AdS_2 \times S^2 \times S^2 \times T^5$ and $\AdS_2 \times S^2 \times T^7$  backgrounds}
\label{sec:remainingbackgrounds}

The determination of the $F$-moduli space for these families of backgrounds is computationally very complex and I have not been able to determine the moduli space.  Backgrounds with such geometries do exist: indeed, they have appeared already as limiting cases of other backgrounds.  The question remains whether some of the $S^2$ in the geometries can be replaced by hyperbolic planes.

\section*{Acknowledgments}

This paper has had a very long gestation period and there are many people I'd like to thank.  During this time I have benefited from conversations with a number of people, originally Dmitri Alekseevsky and Elmer Rees and more recently Noel Hustler.  The bulk of the work was done while I was visiting the High-Energy Physics department of the University of Tōkyō during the month of April 2011, whereas it was during a second visit in December 2011 that I put the finishing touches to the paper.  I am grateful to Teruhiko Kawano for arranging both visits and providing a very pleasant environment in which to do research.  In particular it was thanks to a question he posed to me about my earlier results that I realised I had incorrectly ruled out a couple of cases.  Some of the final calculations were done while visiting the University of Murcia in May 2011 and I am grateful to Ángel Ferrández for the invitation to visit and for his hospitality.  I am happy to acknowledge a conversation with Paul Reynolds during which a small piece of the puzzle finally fell into place.  I am grateful to Roger Bielawski and Martin Speight for the invitation to participate in the workshop \emph{Gauge Theory and Complex Geometry} at the University of Leeds and for the opportunity to talk about this work while still not quite finished.  In particular, I am grateful to Jacques Hurtubise for a penetrating question about the nature of topological transitions in the F-moduli spaces.  I also had the chance to talk about this work at the Lorentzian Geometry meeting GELOGRA'11 which took place at the University of Granada in September 2011.  I am grateful to Miguel Sánchez Caja for the invitation to participate.  Finally, I would like to thank user Sasha at StackOverflow for answering a question  \cite{SO5788842} which helped me in producing some of the figures.

Last, but not least, this work was supported in part by grants ST/G000514/1 ``String Theory Scotland'' and ST/J000329/1 ``Particle Theory at the Tait Institute'' from the UK Science and Technology Facilities Council. 

\appendix

\section{Geometry of some symmetric spaces}
\label{sec:geom-some-symm}

In this appendix we collect some information relevant to the calculations with symmetric spaces.  We use the following notation: $E_{ij}$ is the elementary matrix with a $1$ in the $(i,j)$ entry and $0$ otherwise.  We let $A_{ij} = E_{ij} - E_{ji}$ and $S_{ij} = E_{ij} + E_{ji}$ denote the corresponding elementary skew- and symmetric matrices.

\subsection{$\CP^3$}
\label{sec:cp3}

This is the symmetric space $\U(4)/\U(3)\times\U(1)$.  The Lie algebra $\fu(4)$ is the subalgebra of $4\times 4$ skew-hermitian complex matrices, whereas the subalgebra $\fk = \fu(3)\oplus \fu(1)$ is the subalgebra of matrices of the form
\begin{equation}
  \begin{pmatrix}
    A & 0 \\ 0 & a
  \end{pmatrix}\qquad\text{where}\qquad A \in \Mat(3,\CC),~\overline A^T = -A,~a \in i\RR~.
\end{equation}
The complement $\fp$ of $\fk$ can be taken to be the $\RR$-vector space of matrices of the form
\begin{equation}
  \begin{pmatrix}
    0 & v \\ -\overline v^T & 0
  \end{pmatrix}\qquad\text{where}\qquad v\in \CC^3~.
\end{equation}
An explicit $\RR$-basis for $\fk$ is given by the following 10 matrices:
\begin{equation}
  X_j = i E_{jj},~ j=1,\dots,4,~ X_5 = A_{12},~ X_6 = A_{13},~ X_7 = A_{23},~ X_8 = i S_{12},~ X_9 = i S_{13},~ X_{10} = i S_{23}~,
\end{equation}
whereas an explicit $\RR$-basis for $\fp$ is given by the following 6 matrices:
\begin{equation}
  Y_1 = A_{14},~ Y_2 = iS_{14},~ Y_3 = A_{24},~ Y_4 = iS_{24},~ Y_5 = A_{34},~ Y_6 = iS_{34}~.
\end{equation}
We define an inner product on $\fp$ by
\begin{equation}
  \left<Y,Y'\right> = -\half \tr Y Y' \qquad\text{for all $Y,Y'\in\fp$,}
\end{equation}
relative to which the above basis for $\fp$ is orthonormal.  The linear isotropy representation of $\fk$ on $\fp$ preserves this inner product.  It also preserves the symplectic structure
\begin{equation}
  \omega = \theta^{12} + \theta^{34} + \theta^{56},
\end{equation}
where $\theta^a$ is the canonically dual basis to $Y_a$ and $\theta^{ab\dots c}$ denotes their wedge product.  This means that the linear isotropy representation sends $\fk$ to $\fu(\fp)$, which just restates that $\CP^3$ is a hermitian symmetric space.  One sees that $|\omega|^2 = 3$ and that $\Omega = \half \omega^2 = \theta^{1234} + \theta^{1256} + \theta^{3456}$, whence $|\Omega|^2 = 3$.  It follows that $\left<\sigma_a,\sigma_b\right> = \delta_{ab}$ and that $\left<\Omega_a,\Omega_b\right> = 2 \delta_{ab}$.

\subsection{$G_\RR^+(2,5)$}
\label{sec:g_rr+2-5}

This is the symmetric space $\Sp(2)/\U(2)$.  The Lie algebra $\fg = \fsp(2)$ consists of $2\times 2$ skew-hermitian quaternionic matrices; that is, $X \in \Mat(2,\HH)$ with $\overline{X}^t = -X$.  Explicitly,
\begin{equation}
  \fsp(2) = \left\{ \begin{pmatrix} a & b \\ -\overline b & c \end{pmatrix} \middle | a,c \in \Im\HH,~ b \in \HH\right\}~.
\end{equation}
It is clear that $\dim_\RR\fsp(2) = 10$, as expected.  The subalgebra $\fk = \fu(2)$ consists of those matrices in $\fsp(2)$ which have complex entries.  Explicitly,
\begin{equation}
  \fu(2) = \left\{ \begin{pmatrix} a & b \\ -\overline b & c \end{pmatrix} \middle | a,c \in \Im\CC,~ b \in \CC\right\}~.
\end{equation}
Again, clearly $\dim_\RR\fu(2) = 4$, as expected.  A natural complement $\fp$ of $\fu(2)$ into $\fsp(2)$ consists of those matrices where $a,b,c$ are in $j\CC$; that is,
\begin{equation}
  \fp = \left\{ \begin{pmatrix} a & b \\ -\overline b & c \end{pmatrix} \middle | a,b,c \in j\CC\right\}~.
\end{equation}
Of course, $\dim_\RR\fp = 6$.  It is clearly a symmetric split, since the entries of the matrices in $\fu(2)$ are complex, whereas those in $\fp$ are in $j\CC$, and $j^2=-1$ shows that $[\fu(2),\fp] \subset \fp$ and $[\fp,\fp] \subset \fu(2)$.

It is perhaps convenient to write these quaternionic matrices in terms of complex matrices of twice the size.  We define a right $\CC$-vector space isomorphism $\HH^2 \to \CC^2 \oplus \CC^2$ by $x + j y \mapsto (x,y)$, where $x,y\in\CC^2$.  This induces an $\RR$-linear embedding $\Mat(2,\HH) \to \Mat(4,\CC)$, under which
\begin{equation}
  A + j B \mapsto
  \begin{pmatrix}
    A & - \overline B \\ B & \overline A
  \end{pmatrix} \qquad\text{where}\qquad A,B \in \Mat(2,\CC)~.
\end{equation}
The image of $\fsp(2)$ under this embedding is the unitary symplectic Lie algebra $\fusp(4)$.  The matrix $A + j B \in \fsp(2)$ if and only if $\overline A^t = -A$ and $B^t = B$.  The image of the $\fu(2)$ subalgebra of $\fsp(2)$ consists of matrices in $\fusp(4)$ with $B=0$, whence
\begin{equation}
  \left\{
    \begin{pmatrix}
      a & b & 0 & 0\\ -\overline b & c & 0 & 0 \\ 0 & 0 & - a & \overline b \\ 0 & 0 & -b & - c
    \end{pmatrix}
    \middle | a,c \in \Im\CC,~b \in \CC\right\}~,
\end{equation}
whereas the image of $\fp$ in $\fusp(4)$ consists of those matrices in $\fusp(4)$ which have $A=0$, whence
\begin{equation}
  \left\{
    \begin{pmatrix}
      0 & 0 & -\overline a & -\overline b \\  0 & 0 & -\overline b & -\overline c \\ a & b & 0 & 0 \\ b & c & 0 & 0
    \end{pmatrix}
    \middle | a,b,c \in \CC\right\}~.
\end{equation}
Henceforth we will identify $\fk$ and $\fp$ with their images in $\fusp(4)$.  In other words, we will identify $\fk$ with the $\RR$-span of the following four matrices:
\begin{equation}
  X_1 = i (E_{11}-E_{33}), \quad X_2 = i(E_{22}-E_{44}), \quad X_3 = A_{12} + A_{34}, \quad X_4 = i(S_{12}-S_{34})~.
\end{equation}
and we will identify $\fp$ with the $\RR$-span of the following six matrices:
\begin{equation}
  Y_1 = A_{13},\quad Y_2 = i S_{13},\quad Y_3 = A_{24},\quad Y_4 = i S_{24},\quad Y_5 = \tfrac1{\sqrt{2}}(A_{14} + A_{23}),\quad Y_6 = \tfrac{i}{\sqrt{2}}(S_{14}+S_{23})~,
\end{equation}
where the reason for the factors of $\sqrt{2}$ is to ensure that the $Y_i$ form an orthonormal basis with respect to the following inner product on $\fp$:
\begin{equation}
  \left<Y,Y'\right> = -\half \tr Y Y' \qquad\text{for all $Y,Y' \in \fp$.}
\end{equation}
It is easy to check that the linear isotropy representation $X_i \cdot Y_a := [X_i,Y_a]$ is antisymmetric relative to this inner product and hence defines a homomorphism $\fk \to \fso(\fp)$.  In addition, the linear isotropy representation preserves the symplectic structure $\omega$ on $\fp$ defined by
\begin{equation}
  \omega = \theta^{12} + \theta^{34} + \theta^{56}~,
\end{equation}
where $\theta^a$ is the canonically dual basis to $Y_a$.  This means that the image of the linear isotropy representation belongs to $\fso(\fp) \cap \fsp(\fp) = \fu(\fp)$, restating that $G_\RR^+(2,5)$ is a hermitian symmetric space.

\subsection{$\SLAG_4$}
\label{sec:slag_4}

The grassmannian of special lagrangian planes in $\CC^4$ is the symmetric space $\SU(4)/\SO(4)$.  The Lie algebra $\fsu(4)$ is the Lie algebra of skew-hermitian $4\times 4$ complex matrices, whereas $\fk = \fso(4)$ is the subalgebra consisting of real (and hence skew-symmetric) matrices.  The complementary subspace $\fp$ can be taken to be the subspace of traceless, imaginary skew-hermitian (hence symmetric) matrices.  An explicit $\RR$-basis for $\fk$ is given by the $6$ skew-symmetric matrices:
\begin{equation}
  X_1 = A_{12},\quad X_2 = A_{13},\quad X_3 = A_{14},\quad X_4 = A_{23},\quad X_5 = A_{24},\quad X_6 = A_{34}~,
\end{equation}
whereas an explicit $\RR$-basis for $\fp$ is given by the following $9$ matrices:
\begin{equation}
  \begin{aligned}[m]
    &Y_1 = i (E_{11}-E_{22}),\quad Y_2 = i (E_{33}-E_{44}),\quad Y_3 = \tfrac{i}{\sqrt{2}}(E_{11}+E_{22}-E_{33}-E_{44}),\\
    &Y_4 = i S_{12},\quad  Y_5 = i S_{13},\quad Y_6 = i S_{14},\quad Y_7 = i S_{23},\quad Y_8 = i S_{24},\quad Y_9 = i S_{34}~,
 \end{aligned}
\end{equation}
which has the virtue of being orthonormal relative to the $\SO(4)$-invariant inner product
\begin{equation}
  \left<Y,Y'\right> = -\tfrac12 \tr Y Y' \qquad\text{for all $Y,Y'\in\fp$.}
\end{equation}
Relative to the basis $\theta^a$, canonically dual to $Y_a$, the $\SO(4)$-invariant $4$-form is given by
\begin{equation}\label{eq:slag4form}
  \Omega = \sqrt{2} \left(\theta^{1249} - \theta^{1456} - \theta^{1478} - \theta^{2579} - \theta^{2689} \right) - \theta^{1358} + \theta^{1367} + \theta^{2358} + \theta^{2367} - \theta^{3456} + \theta^{3478} + \theta^{3579} - \theta^{3689}~.
\end{equation}
It follows easily that $|\Omega|^2 = 18$ and $\left<\Omega_a,\Omega_b\right> = 8 \delta_{ab}$.

\subsection{$G_\CC(2,4)$}
\label{sec:g_cc2-4}

The grassmannian of complex planes in $\CC^4$ embeds in $\CP^5$ as the Klein quadric.  It is the symmetric space $\U(4)/\U(2)\times\U(2)$.  The Lie algebra $\fu(4)$ consists of skew-hermitian $4\times 4$ complex matrices.  The Lie subalgebra $\fk = \fu(2)\oplus \fu(2)$ consists of those skew-hermitian matrices of the form
\begin{equation}
  \begin{pmatrix}
    X & 0 \\ 0 & Y
  \end{pmatrix} \qquad\text{where $X,Y$ are $2\times 2$ skew-hermitian matrices.}
\end{equation}
The complementary subspace $\fp$ is then the space of matrices of the form
\begin{equation}
  \begin{pmatrix}
    0 & Z \\ -\overline Z^T & 0
  \end{pmatrix}\qquad\text{where $Z$ is an arbitrary $2\times 2$ matrix.}
\end{equation}
It is clear that $\dim_\RR\fp = 8$.  An explicit $\RR$-basis for $\fk$ is given by the following $8$ matrices:
\begin{equation}
  X_1 = i E_{11},\quad X_2 = i E_{22},\quad X_3 = i E_{33},\quad X_4 = i E_{44},\quad X_5 = A_{12},\quad X_6= i S_{12},\quad X_7 = A_{34},\quad X_8 = i S_{34}~,
\end{equation}
whereas an explicit $\RR$-basis for $\fp$ is given by the following $8$ matrices:
\begin{equation}
  Y_1 = A_{13},\quad Y_2 = i S_{13},\quad Y_3 = A_{23},\quad Y_4 = i S_{23},\quad Y_5 = A_{14},\quad Y_6 = i S_{14},\quad Y_7 = A_{24},\quad Y_8 = i S_{24}~.
\end{equation}
This basis is orthonormal relative to the inner product
\begin{equation}
  \left<Y,Y'\right> = -\tfrac12 \tr Y Y' \qquad\text{for all $Y,Y'\in\fp$,}
\end{equation}
which is invariant under the linear isotropy representation of $\fk$.   Letting $\theta^a$ denote the canonically dual basis to the $Y_a$ and in the usual shorthand notation, there is an invariant symplectic structure given by
\begin{equation}
  \omega = \theta^{12} + \theta^{34} + \theta^{56} + \theta^{78}~.
\end{equation}
We have $\left<\sigma_a,\sigma_b\right> = \delta_{ab}$ and $|\omega|^2 = 4$.  In addition there are two linearly independent invariant $4$-forms:
\begin{equation}
  \begin{aligned}[m]
    \Omega^{(1)} &= \theta^{1256} + \theta^{3478} + \half ( \theta^{1278} + \theta^{3456} ) + \tfrac14 ( \theta^{1357}+ \theta^{1368} + \theta^{1458} - \theta^{1467} - \theta^{2358} +  \theta^{2367} + \theta^{2457} + \theta^{2468} ) \\
    \Omega^{(2)} &= \theta^{1234} +\theta^{5678} + \half (\theta^{1278} + \theta^{3456}) - \tfrac14 (\theta^{1357} + \theta^{1368} + \theta^{1458} - \theta^{1467} - \theta^{2358} + \theta^{2367} + \theta^{2457} + \theta^{2468})\\
  \end{aligned}
\end{equation}
Notice that
\begin{equation}\label{eq:GC24identities}
  \Omega^{(1)} + \Omega^{(2)} = \half \omega^2 \qquad\text{and}\qquad \omega \wedge \Omega^{(1)} = \omega \wedge \Omega^{(2)} = \tfrac14 \omega^3~.
\end{equation}
In addition, the forms $\Omega^{(i)}$ are self-dual and satisfy $\Omega^{(i)} \wedge \Omega^{(j)} = 3 \delta^{ij} \theta^{12345678}$, whence $|\Omega^{(i)}|^2 = 3$.  Finally, we also have $\left<\Omega^{(i)}_a, \Omega^{(j)}_b\right> = \tfrac32 \delta_{ab} \delta^{ij}$.

\section{Inadmissible anti~de~Sitter geometries}
\label{sec:geometries-which-do}

Here we list those geometries which do not admit symmetric supergravity backgrounds.  There are two main reasons: either no nonzero $F$ is available (and they are not Ricci-flat), or there are suitable $F$s but the equations of motion force $F=0$, either $F\wedge F \neq 0$ or the Einstein equation again contradicts the fact that the metrics are not Ricci-flat.

\subsection{Geometries with $F=0$}
\label{sec:geom-with-suit}

We now proceed to list those geometries which cannot appear due to not having nonzero invariant $4$-forms.  Only the compact forms are listed, but in fact the noncompact duals cannot appear either.

\begin{multicols}{3}
  \begin{itemize}
  \item $\AdS_6 \times \SLAG_3$
  \item $\AdS_6 \times S^5$
  \item $\AdS_5 \times S^6$
  \item $\AdS_5 \times \SLAG_3 \times S^1$
  \item $\AdS_5 \times S^5 \times S^1$
  \item $\AdS_5 \times S^3 \times S^3$
  \item $\AdS_3 \times S^8$
  \item $\AdS_3 \times \SU(3)$
  \item $\AdS_3 \times S^6 \times S^2$
  \item $\AdS_3 \times \SLAG_3 \times S^3$
  \item $\AdS_3 \times S^5 \times S^3$
  \item $\AdS_3 \times S^3 \times S^3 \times S^2$
  \item $\AdS_2 \times S^6 \times S^3$
  \item $\AdS_2 \times S^3 \times S^3 \times S^3$
  \item $\AdS_2 \times S^9$
\end{itemize}
\end{multicols}

\subsection{Geometries where $F \wedge F = 0$ implies $F= 0$}
\label{sec:geom-with-invar}

We now list $AdS_d \times M^{11-d}$ backgrounds which cannot appear because there are no nonzero invariant $4$-forms obeying $F \wedge F = 0$.  As usual, we only list the compact forms of the riemannian symmetric spaces, but the same holds for their noncompact duals.

\begin{itemize}
\item $\AdS_3 \times \CP^4$, $\AdS_3 \times \HP^2$ and $\AdS_3 \times \ASSOC$: Here $F = f \Omega$, where $\Omega$ is the unique invariant $4$-form on $\CP^4$, $\HP^2$ or $\ASSOC$.  Uniqueness means that $\Omega$ is either selfdual or antiselfdual, depending on orientation.  In either case $\Omega \wedge \Omega = \pm |\Omega|^2 \dvol$, whence $F \wedge F = 0$ implies that $F=0$.
\item $\AdS_3 \times G_\CC(2,4)$: The equation $F\wedge F = 0$ says that the class in $H^2(G_\CC(2,4),\RR)$ defined by $F$ has zero norm relative to the intersection form, and since the intersection form on $G_\CC(2,4)$ is positive-definite, the only solution is $F=0$.  To see that the intersection form is positive definite, we can argue as follows.  (I learnt the following from Elmer Rees.)  Recall that the cohomology of $G_\CC(2,4)$ is generated by the Chern classes $c_1,c_2$ of the tautological bundle $E$ subject to the relation that $E \oplus E^\perp$ is trivial, where $E^\perp$ is the bundle whose fibre is the perpendicular plane to the fibre of $E$.  Let $c_1^\perp,c_2^\perp$ denote the Chern classes of $E^\perp$.  Then letting $c(E)$ denote the total Chern class, the relations are encoded in the following
\begin{equation}
  c(E \oplus E^\perp) = c(E)c(E^\perp) = 1~,
\end{equation}
where $c(E)= 1 + c_1 + c_2$ and $c(E^\perp) = 1 + c_1^\perp + c_2^\perp$.  The first two relations (in degrees 2 and 4) allow us to solve for $c_1^\perp$ and $c_2^\perp$:
\begin{equation}
  c_1^\perp = - c_1 \qquad\text{and}\qquad c_2^\perp = c_1^2 - c_2~.
\end{equation}
The other two relations now become
\begin{equation}
  c_1^2 c_2 = c_2^2 \qquad\text{and}\qquad c_1^3 = 2 c_1 c_2~.
\end{equation}
Let $c_1^2$ and $c_2$ be a basis for $H^2$.  Then the intersection form is given by
\begin{equation}
  c_1^2 \cdot c_1^2 = 2 c_1^2 c_2\qquad c_2 \cdot c_2 = c_1^2 c_2 \qquad c_1^2 \cdot c_2 = c_1^2 c_2
\end{equation}
which, relative to the basis for $H^8$ given by $c_1^2 c_2$, has matrix
\begin{equation}
  \begin{pmatrix}
    2 & 1 \\ 1 & 1
  \end{pmatrix}~,
\end{equation}
which is clearly positive-definite.  In fact, $c_1^2 - c_2$ and $c_2$ are an orthonormal basis.  Alternatively, using the results in Appendix \ref{sec:g_cc2-4}, we see that $F = \lambda_1 \Omega^{(1)} + \lambda_2 \Omega^{(2)}$, and $|F \wedge F|^2 = 3 (\lambda_1^2 + \lambda_2^2)$, whence if $F \wedge F =0$, then $F=0$.

\end{itemize}

\subsection{Geometries where the Einstein equation implies $F=0$}
\label{sec:geometries-with-no}

We now list those geometries which do not give rise to backgrounds because of the failure of the Einstein equation.  The heuristic principle in Section \ref{sec:useful-heuristic} allows us to rule out any geometry of the form $\AdS_d \times T^{11-d}$.


A second class of geometries which we can rule out are those which have a flat direction and yet $|F|\geq 0$, for then the flat components of the Einstein equation implies that $F=0$.  These geometries are the following:

\begin{multicols}{3}
  \begin{itemize}
  \item $\AdS_7 \times S^3 \times S^1$
  \item $\AdS_7 \times S^2 \times T^2$
  \item $\AdS_6 \times S^4 \times S^1$
  \item $\AdS_6 \times \CP^2 \times S^1$
  \item $\AdS_6 \times S^3 \times T^2$
  \item $\AdS_6 \times S^2 \times S^2 \times S^1$
  \item $\AdS_6 \times S^2 \times T^3$
  \item $\AdS_5 \times \CP^2 \times T^2$
  \item $\AdS_5 \times S^4 \times T^2$
  \item $\AdS_5 \times S^3 \times S^2 \times S^1$
  \item $\AdS_5 \times S^3 \times T^3$
  \item $\AdS_5 \times S^2 \times T^4$
 \end{itemize}
\end{multicols}

The only case which might require some explanation is $\AdS_5 \times S^2 \times T^4$.  For that background $F = \omega \wedge \beta + f \tau$, where $\omega$ and $\tau$ are the volume forms on $S^2$ and $T^4$ and $\beta \in \Omega^2(T^4)$.  We can always choose an orthonormal coframe $\theta^i$ for $T^4$ relative to which $\tau = \theta^1 \wedge \theta^2 \wedge \theta^3 \wedge \theta^4$ and $\beta = \lambda \theta^1 \wedge \theta^2 + \mu \theta^3 \wedge \theta^4$.  Then $|F|^2 = \lambda^2 + \mu^2 + f^2$, whereas $\left<F_1,F_1\right> = \left<F_2,F_2\right> = \lambda^2 + f^2$ and $\left<F_3,F_3\right> = \left<F_4,F_4\right> = \mu^2 + f^2$.  The torus components of the Einstein equations are then equivalent to
\begin{equation}
  \left<F_1,F_1\right> = \tfrac13 |F|^2 \implies \mu^2 = 2 (\lambda^2 + f^2)
\end{equation}
and
\begin{equation}
  \left<F_3,F_3\right> = \tfrac13 |F|^2 \implies \lambda^2 = 2 (\mu^2 + f^2)
\end{equation}
and the two equations together force $\lambda = \mu = f = 0$ and hence $F=0$.

The following geometries cannot appear either:
\begin{multicols}{3}
  \begin{itemize}
  \item $\AdS_4 \times S^3 \times T^4$
  \item $\AdS_4 \times S^3 \times S^2 \times T^2$
  \item $\AdS_4 \times S^3 \times S^3 \times S^1$
  \item $\AdS_4 \times S^2 \times S^2 \times T^3$
  \item $\AdS_4 \times S^2 \times T^5$
  \end{itemize}
\end{multicols}

The geometry $\AdS_4 \times S^3 \times T^4$ is forced to have $F=0$, which contradicts the presence of the $\AdS_4$ factor.  Notice that $F = \lambda \nu_4 + \mu \sigma_3 \wedge d\vartheta$, where $\sigma_3$, $\nu_4$ are the volume forms on $S^3$ and $T^4$, respectively, and $d\vartheta$ is a unit-norm 1-form on $T^4$.  Then the Einstein equations in the $\theta$ direction say that $R_{\theta\theta} = \tfrac13 (\lambda^2 + \mu^2)$, whence $\lambda = \mu = 0$.  Similar arguments allows us to discard symmetric backgrounds with underlying geometries $\AdS_4 \times S^3 \times S^2 \times T^2$ and $\AdS_4 \times S^3 \times S^3 \times S^1$.

The geometry $\AdS_4 \times S^2 \times S^2 \times T^3$ too can be discarded since it forces $F=0$.  Notice that as discussed in Section \ref{sec:ads_4-backgrounds}, the equation $F \wedge F = 0$ forces $F$ to have either no legs along $\AdS_4$ or else be proportional to the volume form $\nu$ of $\AdS_4$.  In the latter case, the existence of flat directions in the geometry contradicts that $|F|<0$.  In the former case, choose flat coordinates $\theta^{1,2,3}$ for $T^3$ in such a way that the corresponding one-forms $d\vartheta^i$ have unit norm.  Let $\sigma$ and $\sigma'$ denote the area forms on the two spheres.  The most general form for $F$ is easily seen to be
\begin{equation}
  F = \alpha \sigma \wedge \sigma' + (\beta \sigma + \gamma \sigma') \wedge d\vartheta^2 \wedge d\vartheta^3 + \delta \sigma' \wedge d\vartheta^1 \wedge d\vartheta^3~,
\end{equation}
for some real numbers $\alpha,\beta,\gamma,\delta$.  One sees that $|F|^2 = \alpha^2 + \beta^2 + \gamma^2 + \delta^2$, whereas $|F_1|^2 = \delta^2$, $|F_2|^2 = \beta^2 + \gamma^2$ and $|F_3|^2 = \beta^2 + \gamma^2 + \delta^2$, where $1,2,3$ refer to the flat directions.  The Einstein equations along the flat directions imply the three equations $|F_i|^2 = \tfrac13|F|^2$~.  It is not hard to see that the only solution of the three equations is $\alpha=\beta=\gamma=\delta=0$.

Finally, the remaining geometry $\AdS_4 \times S^2 \times T^5$ can be ruled out as well.  Again, the existence of flat directions allows us to discard the case where $F$ is proportional to the $\AdS_4$ volume form.  To tackle the case where $F$ has no legs along $\AdS_4$, let $\theta^i$ denote flat coordinates on $T^5$ with $d\vartheta^i$ of unit norm and let $\sigma$ denote the area form on $S^2$.  Then we have that the most general $F$ is given by
\begin{equation}
  F = f_1 \sigma \wedge d\vartheta^{12} + f_2 \sigma \wedge d\vartheta^{34} + f_3 d\vartheta^{2345} + f_4 d\vartheta^{1245} + f_5 d\vartheta^{1234}~.
\end{equation}
To see that this is not obviously wrong, notice that $F = \sigma \wedge \alpha + \star \beta$ for $\alpha$ and $\beta$ a constant coefficient $2$-form and $1$-form, respectively, on $T^5$.  We are free to rotate the $\theta^i$ using $\SO(5)$.  Now $\SO(5)$ has dimension $10$ and the space $\Lambda^1 \RR^5 \oplus \Lambda^2 \RR^5$ has dimension $15$, whence we expect that the orbits should be labelled by at least 5 parameters.  It is not hard to argue that the above form of $F$ is generic by successively fixing its form.  Indeed, we can bring $\alpha$ to the form $f_1 d\vartheta^{12} + f_2 d\vartheta^{34}$, which still leaves the possibility of rotating in the $(12)$ and $(34)$ planes separately, to get rid of the $\theta^2$ and $\theta^4$ components in $\beta$.  Now $|F|^2 = \sum_{i=1}^5 f_i^2$, whereas $|F_1|^2 = \sum_{i=1,4,5} f_i^2$, $|F_2|^2 =\sum_{i=1,3,5} f_i^2$, $|F_3|^2 = \sum_{i=2,3,5} f_i^2$, $|F_4|^2 = \sum_{i=2,3,4,5} f_i^2$, and $|F_5|^2 = f_3^2 + f_4^2$.  The Einstein equations in the $T^5$ directions become $|F_i|^2 = \tfrac13 |F|^2$.  In particular, all the $|F_i|^2$ should be equal.  It is easy to see from their explicit expressions that the only way this can happen is if $f_i=0$ for all $i$, whence $F=0$.

The following geometries cannot appear either:

\begin{multicols}{3}
  \begin{itemize}
  \item $\AdS_3 \times S^7 \times S^1$
  \item $\AdS_3 \times S^6 \times T^2$
  \item $\AdS_3 \times \CP^3 \times T^2$
  \item $\AdS_3 \times G_\RR^+(2,5) \times T^2$
  \item $\AdS_3 \times \SLAG_3 \times S^2 \times S^1$
  \item $\AdS_3 \times S^5 \times S^2 \times S^1$
  \item $\AdS_3 \times \SLAG_3 \times T^3$
  \item $\AdS_3 \times S^5 \times T^3$
  \item $\AdS_3 \times S^4 \times S^2 \times T^2$
  \item $\AdS_3 \times S^4 \times T^4$
  \item $\AdS_3 \times \CP^2 \times T^4$
  \item $\AdS_3 \times S^3 \times S^2 \times T^3$
  \item $\AdS_3 \times S^4 \times S^3 \times S^1$
  \item $\AdS_3 \times \CP^2 \times S^3 \times S^1$
  \item $\AdS_3 \times S^2 \times S^2 \times S^3 \times S^1$
  \end{itemize}
\end{multicols}

Indeed, the background $\AdS_3 \times S^7 \times S^1$ cannot appear for the following reason.  Here $F = f \nu_3 \wedge \theta$, where $\theta$ is the coframe on $S^1$.  Then $|F|^2 = -f^2$ and the vanishing of the $S^1$ component of the Ricci tensor is inconsistent with the Einstein equations unless $f=0$: $R_{\theta\theta} = -\tfrac13 f^2$.  A very similar argument shows that $\AdS_3 \times S^6 \times T^2$ cannot appear.  Here $F = f\nu_3 \wedge \theta$ as well and the toroidal components of the Einstein equation are inconsistent unless $f=0$, which is ruled out on geometrical grounds.  Similar arguments also apply to eliminate $\AdS_3 \times \SLAG_3 \times S^2 \times S^1$, $\AdS_3 \times S^5 \times S^2 \times S^1$, $\AdS_3 \times \SLAG_3 \times T^3$ and  $\AdS_3 \times S^5 \times T^3$.

The related geometries $\AdS_3 \times \CP^3 \times T^2$ and $\AdS_3 \times G_\RR^+(2,5) \times T^2$ can be ruled out as follows.  In either case $F= f_0 \nu_3 \wedge d\vartheta^1 + f_1 \half \omega^2 + f_2 \omega\wedge d\vartheta^{12}$, with $\omega$ as before.  Then $F\wedge F = f_0f_1 \nu_3 \wedge d\vartheta^1 \wedge \omega^2 + f_1f_2 \omega^3\wedge d\vartheta^{12}$, whence $f_0f_1=0$ and $f_1f_2=0$.  Along the $T^2$ directions $F_1 = f_0 \nu_3 + f_2\omega\wedge d\vartheta^2$ and $F_2 = -f_2 \omega \wedge d\vartheta^1$, whence $|F_1|^2 = -f_0^2 + 3 f_2^2$ and $|F_2|^2 = 3 f_2^2$.  Now the $T^2$ Einstein equations say that $|F_1|^2 = |F_2|^2 = \tfrac13 |F|^2$, whence we see that $f_0=0$ and that $2f_2^2 = f_1^2$.  But since $f_1f_2=0$, we find that $f_1=f_2=0$ and hence $F=0$ contradicting the anti de Sitter geometry.

The geometry $\AdS_3 \times S^4 \times S^2 \times T^2$ can be ruled out as follows.  Here we can take $F = f_0 \nu_3 \wedge d\vartheta^1 + f_1 \sigma_4 + f_2 \sigma_2 \wedge d\vartheta^{12}$ without loss of generality.  The equation $F \wedge F = 0$ forces $f_0 f_1 =0$ and $f_1 f_2 = 0$.  The Einstein equation along the flat directions impose the relations $-f_0^2 + f_2^2 = f_2^2 = \tfrac13 (-f_0^2 + f_1^2 + f_2^2)$, whence we see that $f_0=0$ and $2f_2^2 = f_1^2$.  However since one of $f_1,f_2$ must vanish, both do.

The geometry $\AdS_3 \times S^4 \times T^4$ can be ruled out as follows.  Here $F=f_0 \nu_3 \wedge d\vartheta^1 + f_1 \nu_4 + f_2 d\vartheta^{1234}$ and $F\wedge F = 0$ imposes the conditions $f_0f_1=0$ and $f_1f_2=0$.  The norm $|F|^2 = -f_0^2 + f_1^2 + f_2^2$ whereas in the $T^4$ directions, $|F_1|^2 = -f_0^2 + f_2^2$ and $|F_j|^2 = f_2^2$, for $j=2,3,4$.  The $T^4$ components of the Einstein equations then force $f_0=0$ and $2f_2^2 = f_1^2$.  But since $f_1f_2=0$, this forces $f_1=f_2=0$ as well.

The geometry $\AdS_3 \times S^3 \times S^2 \times T^3$ can be ruled out as follows.  The most general $F$ is given by $F = \nu_3 \wedge \alpha + \sigma_3 \wedge \beta + \sigma_2 \wedge \gamma$, where $\alpha,\beta \in \Omega^1(T^3)$ and $\gamma \in \Omega^2(T^3)$.  The equation $F\wedge F =0$ says that $\alpha\wedge \beta = 0$, $\alpha \wedge \gamma = 0$ and $\beta \wedge \gamma =0$.  The former two equations says that $\beta$ and $\alpha$ are proportional (this holds trivially if they are both zero) and that $\gamma = \alpha \wedge \delta$, for some $\delta \in \Omega^1(T^3)$.  In other words, we can bring $F$ to the following form: $F=d\vartheta^1 \wedge (f_0 \nu_3 + f_1 \sigma_3 + f_2 \sigma_2 \wedge d\vartheta^2)$.  The absence of $d\vartheta^3$ in $F$ together with the Einstein equation for the $\theta^3$ direction says that $|F|^2=0$, whence $f_0^2 = f_1^2 + f_2^2$.  The $\theta^2$ component of the Einstein equation now says that $f_2=0$, whence $F$ has no legs along the $S^2$ direction.  In turn, that implies, via the Einstein equation along $S^2$, that $S^2$ is Ricci-flat, which is absurd.

The geometry $\AdS_3 \times S^4 \times S^3 \times S^1$ can be ruled out as follows.  Here $F=f_0 \nu_3 \wedge d\vartheta + f_1 \sigma_3 \wedge d\vartheta + f_2 \sigma_4$, whence $F\wedge F = 0$ implies that $f_0f_2=0$ and $f_1f_2=0$, while $|F|^2=-f_0^2 + f_1^2 +f_2^2$.  The Einstein equation in the $S^1$ direction, namely $|F_\theta|^2 = \tfrac13 |F|^2$, imposes $2f_1^2 = f_2^2 + 2 f_0^2$.  This means that $f_1$ cannot vanish, or else $F=0$.  But then the $F \wedge F = 0$ equations require $f_2=0$.  In turn this means $|F|^2=0$ and since $F$ has no legs along the $S^4$ direction, the condition $|F|^2=0$ says that this factor is Ricci-flat, which is absurd.  The same argument works for $\AdS_3 \times \CP^2 \times S^3 \times S^1$ and $\AdS_3 \times S^3 \times S^2 \times S^2 \times S^1$ because, although $\CP^2$ and $S^2 \times S^2$ have additional invariant forms, these cannot enter into the definition of $F$.

The following geometries are also ruled out:

\begin{multicols}{3}
  \begin{itemize}
  \item $\AdS_2 \times \SU(3) \times S^1$
  \item $\AdS_2 \times S^7 \times T^2$
  \item $\AdS_2 \times S^6 \times S^2 \times S^1$
  \item $\AdS_2 \times \CP^3 \times S^2 \times S^1$
  \item $\AdS_2 \times G_\RR^+(2,5) \times S^2 \times S^1$
  \item $\AdS_2 \times S^6 \times T^3$
  \item $\AdS_2 \times \SLAG_3 \times S^3 \times S^1$
  \item $\AdS_2 \times S^5 \times S^3 \times S^1$
  \item $\AdS_2 \times S^4 \times S^4 \times S^1$
  \item $\AdS_2 \times S^4 \times \CP^2 \times S^1$
  \item $\AdS_2 \times \CP^2 \times \CP^2 \times S^1$
  \item $\AdS_2 \times S^4 \times S^3 \times T^2$
  \item $\AdS_2 \times S^4 \times S^2 \times S^2 \times S^1$
  \item $\AdS_2 \times \CP^2 \times S^2 \times S^2 \times S^1$
  \item $\AdS_2 \times S^4 \times S^2 \times T^3$
  \item $\AdS_2 \times S^4 \times T^5$
  \item $\AdS_2 \times S^3 \times S^3 \times S^2 \times S^1$
  \end{itemize}
\end{multicols}

The geometry $\AdS_2 \times \SU(3) \times S^1$ can be ruled out as follows.  The $4$-form is $F = f H \wedge d\vartheta$, where $H$ is the invariant $3$-form on $\SU(3)$ given by the structure constants with indices lowered with the Killing form.  This means that $|F|^2=f^2 |H|^2$ but also $|F_\theta|^2 = f^2 |H|^2$.  The $S^1$-component of the Einstein equation force $|F_\theta|^2 = \tfrac13 |F|^2$, which means $H=0$ and hence forces $F=0$.

The same calculation which showed the existence of the $\AdS_2 \times H^2 \times S^7$ background shows that we can rule out the geometry $\AdS_2 \times S^7 \times T^2$, since the Einstein equation for $T^2$ is the same as for $\AdS_2$, which is absurd.  In a similar way we can rule out $\AdS_2 \times S^6 \times S^2 \times S^1$, since $F$ takes the same form, yet the $S^1$ component of the Einstein equation force it to have zero norm, whence $F=0$.

The geometries $\AdS_2 \times \SLAG_3 \times S^3 \times S^1$ and $\AdS_2 \times S^5 \times S^3 \times S^1$ can be discarded as follows.  In both geometries, the $4$-form takes the form $F = f \sigma_3 \wedge d\vartheta$, with $\sigma_3$ the volume form on $S^3$.  Although $F \wedge F = 0$ is satisfied, the $S^1$ component of the Einstein equation sets $f^2=0$ and hence $F=0$.

The geometry $\AdS_2 \times S^4 \times S^4 \times S^1$ is discarded because the only possible $F$ has $|F|^2 \geq 0$, yet has no legs along the $S^1$.  This means that the $S^1$ component of the Einstein equation sets $|F|^2=0$, whence $F=0$.

The geometries $\AdS_2 \times \CP^3 \times S^2 \times S^1$ and $\AdS_2 \times G_\RR^+(2,5) \times S^2 \times S^1$ can be ruled out by the following argument.  The most general invariant $4$-form is given by $F = f_0 \nu\wedge\omega + f_1 \nu \wedge \sigma_2 + f_2 \omega \wedge \sigma_2 + f_3 \half \omega^2$, where $\omega$ is the Kähler form and $\sigma_2$ the area form on $S^2$.  The equation $F\wedge F = 0$ translates into the relations
  \begin{equation}\label{eq:k6s2s1rels}
    f_0f_3 = 0 \qquad f_2f_3 = 0 \qquad 2 f_0f_2 + f_1f_3 = 0~,
  \end{equation}
whereas the norm $|F|^2 = -3f_0^2 - f_1^2 + 3 f_2^2 + 3 f_3^2$.  Since $F$ has no legs along $S^1$, this component of the Einstein equation says $|F|^2=0$.  The Einstein equations for $\AdS_2$, $S^2$ and $\CP^3$ (or $G_\RR^+(2,5)$) are, respectively,
\begin{equation}
  R_{\mu\nu} = -\tfrac32 (f_2^2 + f_3^2) g_{\mu\nu} \qquad R_{ij} = \tfrac32 (f_0^2 - f_3^2) g_{ij} \qquad\text{and}\qquad R_{ab} = \half (-f_0^2 + f_2^2 + 2f_3^2) g_{ab}~.
\end{equation}
In satisfying the relations \eqref{eq:k6s2s1rels}, if $f_3=0$ then the first two equations are satisfied and the third says that either $f_0$ or $f_2$ must vanish, but in either case then either $\AdS_2$ or $S^2$ would be Ricci-flat, which is absurd.  Therefore $f_3\neq 0$.  In that case, equations \eqref{eq:k6s2s1rels} are only satisfied provided that $f_0=f_1=f_2=0$, but then the condition $|F|^2=0$ forces $f_3=0$ as well.

The geometry $\AdS_2 \times S^4 \times \CP^2 \times S^1$ can be ruled out as follows.  The most general invariant $4$-form is given by $F = f_0 \nu \wedge \omega + f_1 \sigma_4 + f_2 \half \omega^2$.  The equation $F \wedge F = 0$ imposes the relations $f_0 f_1 = 0$ and $f_1 f_2 = 0$.  Since $F$ has no legs along the $S^1$, the $S^1$ Einstein equation implies the vanishing of the norm $|F|^2 = -2 f_0^2 + f_1^2 + f_2^2$, whence $2 f_0^2 = f_1^2 + f_2^2$.  Now, if $f_0=0$, then $f_1 = f_2 = 0$, whence $F=0$, so $f_0\neq 0$ and thus $f_1=0$.  But now the $S^4$ Einstein equation says that it is Ricci-flat, which is absurd.

The geometry $\AdS_2 \times \CP^2 \times \CP^2 \times S^1$ can be ruled out as follows.  The most general invariant $4$-form is given by
  \begin{equation}
    F = f_0 \nu \wedge \sigma_1 + f_1 \nu \wedge \sigma_2 + f_2 \sigma_1 \wedge \sigma_2 + f_3 \half \sigma_1^2 + f_4 \half \sigma_2^2~,
  \end{equation}
whence the equation $F\wedge F = 0$ imposes the following relations
\begin{equation}
  \label{eq:AdS2CP2CP2S1rels}
  2 f_0 f_2 + f_1 f_3 = 0 \qquad 2 f_1 f_2 + f_0 f_4 = 0 \qquad 2 f_2^2 + f_3 f_4 = 0~.
\end{equation}
Since $F$ has no legs along the $S^1$, its Einstein equation sets $|F|^2 = 0$, which turns into the relation
\begin{equation}
  \label{eq:S1EECP2CP2S1}
  2 f_0^2 + 2 f_1^2 = 4 f_2^2 + f_3^2 + f_4^2~.
\end{equation}
The Einstein equations for the two $\CP^2$s are
\begin{equation}
  R_{ab} = \half (-f_0^2 + 2 f_2^2 + f_3^2) g_{ab} \qquad\text{and}\qquad R_{a'b'} = \half (-f_1^2 + 2 f_2^2 + f_4^2) g_{a'b'}~.
\end{equation}
It is now straight-forward to show that the only nonzero $F$s satisfying the relations \eqref{eq:AdS2CP2CP2S1rels} and 
\eqref{eq:S1EECP2CP2S1} imply that one or the other of the $\CP^2$s has zero Ricci curvature, which is absurd.
The geometry $\AdS_2 \times S^4 \times S^3 \times T^2$ can be ruled out as follows.  Here $F = f_0 \nu \wedge d\vartheta^{12} + f_1 \sigma_4 + f_2 \sigma_3 \wedge d\vartheta^1$ without loss of generality.  The equation $F \wedge F = 0$ implies the relations $f_0f_1 = 0$ and $f_1f_2 = 0$.  The Einstein equation for the $T^2$ directions say that $-f_0^2 = -f_0^2 + f_2^2 = \tfrac13 |F|^2$, where $|F|^2 = -f_0^2 + f_1^2 + f_2^2$.  The first equality sets $f_2=0$, whereas the second one says that $2 f_0^2 + f_1^2 = 0$, which forces $F=0$.

The geometry $\AdS_2 \times S^4 \times S^2 \times S^2 \times S^1$ can be ruled out as follows.  Here $F = f_0 \nu \wedge \sigma_1 + f_1 \nu \wedge \sigma_2 + f_2 \sigma_1 \wedge \sigma_2 + f_3 \sigma_4$, with $\sigma_1,\sigma_2$ the area forms on the two $S^2$s and $\sigma_4$ the volume form on $S^4$.  The equation $F \wedge F =0$ imposes the relations $f_0 f_3 = 0$, $f_1 f_3 = 0$ and $f_2 f_3 = 0$.  Since $F$ has no legs along the $S^1$, the Einstein equation along that direction becomes the vanishing of $|F|^2 = -f_0^2 - f_1^2 + f_2^2 + f_3^2$.  The Einstein equation along $S^4$ now becomes $\Ric = \half f_3^2 g$, which means that $f_3 \neq 0$.  However, this implies that $f_0 = f_1 = f_2 = 0$, which is inconsistent with the vanishing of $|F|^2$.

The geometry $\AdS_2 \times S^4 \times S^2 \times T^3$ can be ruled out as follows.  The most general $F = f_0 \nu \wedge \sigma_2 + \nu \wedge \alpha + f_1 \sigma_4 + \sigma_2 \wedge \beta$, for some $\alpha,\beta \in \Omega^2(T^3)$.  The equation $F \wedge F = 0$ imposes the relations $f_0f_1=0$, $f_1 \alpha =0$ and $f_1\beta =0$.  If $f_1\neq 0$, then $f_0 = \alpha =\beta = 0$ and the $T^3$ Einstein equations set $F=0$.  Hence $f_1 =0$.  Without loss of generality we can bring $F$ to the form
  \begin{equation}
    F = f_0 \nu \wedge \sigma_2 + f_2 \nu \wedge d\vartheta^{12} + f_3 \sigma_2 \wedge d\vartheta^{12} + f_4 \sigma_2 \wedge d\vartheta^{13}~,
  \end{equation}
with norm $|F|^2 = -f_0^2 - f_2^2 + f_3^2 + f_4^2$.  The $T^3$ Einstein equations impose the relations
\begin{equation}
  -f_2^2 + f_3^2 + f_4^2 = -f_2^2 + f_3^2 = f_4^2 = \tfrac13 |F|^2~.
\end{equation}
From the first equality we see that $f_4=0$ and from the second that $f_2^2 = f_3^2$.  The last equality says that $f_0=0$, whence $F = (f_0 \nu + f_3 \sigma_2) \wedge d\vartheta^{12}$ with $|F|^2=0$.  But now, since $F$ has no legs along the $S^4$, the Einstein equations say that $S^4$ is Ricci-flat, which is absurd.

The geometry $\AdS_2 \times S^4  \times T^5$ can also be ruled out.  Here we may bring $F$ to the following form:
  \begin{equation}
    F = f_0 \nu \wedge d\vartheta^{12} + f_1 \nu \wedge d\vartheta^{34} + f_2 \sigma_4 + f_3 d\vartheta^{1234} + f_4 d\vartheta^{1245} + f_5 d\vartheta^{2345}~.
  \end{equation}
The equation $F\wedge F = 0$ imposes the relations $f_if_2 =0$ for all $i\neq 2$.  The Einstein equations along the flat directions become
\begin{multline}
  -f_0^2 + f_3^2 + f_4^2 = -f_0^2 + f_3^2 + f_4^2 + f_5^2 = -f_1^2 + f_3^2 + f_5^2 = -f_1^2 + f_3^2 + f_4^2 + f_5^2 = f_4^2 + f_5^2\\
  = \tfrac13 (-f_0^2 - f_1^2 + f_2^2 + f_3^2 + f_4^2 + f_5^2)~.
\end{multline}
These equations imply that $f_5=f_4=0$ and that $f_0^2 = f_1^2 = f_2^2 = f_3^2$, but then $f_2f_3=0$ would say that $F=0$.

The geometry $\AdS_2 \times S^3 \times S^3 \times S^2 \times S^1$ can also be ruled out.  Here $F = f_0 \nu \sigma_2 + f_1 \sigma_3 \wedge d\vartheta + f_2 \omega'_3 \wedge d\vartheta$, in the obvious notation.  The equation $F\wedge F = 0$ implies the following relations: $f_0 f_1 = 0$  and $f_0 f_2 = 0$.  The $S^1$ component of the Einstein equation says that $3(f_1^2 + f_2^2) = -f_0^2 + f_1^2 + f_2^2$, or equivalently, $f_0^2 + 2 f_1^2 + 2 f_2^2 = 0$, which implies $F=0$.

The geometry $\AdS_2 \times \CP^2 \times S^2 \times S^2 \times S^1$ can be ruled out.  The most general $F$ is given by
  \begin{equation}
    F = f_0 \nu \wedge \omega + f_1 \nu \wedge \sigma_1 + f_2 \nu \wedge \sigma_2 + f_3 \half \omega^2 + f_4 \omega \wedge \sigma_1 + f_5 \omega \wedge \sigma_2 + f_6 \sigma_1 \wedge \sigma_2
  \end{equation}
in the obvious notation.  The equation $F\wedge F =0$ imposes the following relations
\begin{equation}\label{eq:AdS2CP2S2S2S1rels}
  2 f_4 f_0 + f_1 f_3 = 0, \quad 2 f_5 f_0 + f_2 f_3 = 0, \quad 2 f_4 f_5 + f_3 f_6 = 0, \quad f_0 f_6 + f_2 f_4 + f_1 f_5 = 0~,
\end{equation}
whereas the $S^1$ Einstein equation says that $|F|^2=0$, since $F$ has no legs along the $S^1$.  This means that
\begin{equation}\label{eq:AdS2CP2S2S2S1ee}
  2f_0^2 + f_1^2 + f_2^2 = f_3^2 + 2f_4^2 + 2 f_5^2 + f_6^2~.
\end{equation}
The Einstein equations for the $\AdS_2$, $\CP^2$ and the two $S^2$s are given by $\Ric = \half \lambda g$, where
\begin{equation}
  \begin{aligned}[m]
    \lambda_{\CP^2} &= -f_0^2 + f_3^2 + f_4^2 + f_5^2\\
    -\lambda_{\AdS_2} &= 2 f_0^2 + f_1^2 + f_2^2
  \end{aligned}
 \qquad\qquad
  \begin{aligned}[m]
    \lambda_{S^2_{(1)}} &= - f_1^2 + 2 f_4^2 + f_6^2\\
    \lambda_{S^2_{(2)}} &= - f_2^2 + 2 f_5^2 + f_6^2
  \end{aligned}
\end{equation}
It can be shown that there are no values of $f_i$ satisfying the equations \eqref{eq:AdS2CP2S2S2S1rels} and \eqref{eq:AdS2CP2S2S2S1ee} for which none of the $\lambda$s vanish.  Indeed, let us first assume that $f_0\neq 0$.  Then using all but the third relation in \eqref{eq:AdS2CP2S2S2S1rels}, we can solve for $f_4,f_5,f_6$ in terms of $f_0,f_1,f_2,f_3$; namely,
\begin{equation}
  f_4 = - \frac{f_1 f_3}{2f_0} \qquad f_5 = - \frac{f_2 f_3}{2f_0} \qquad\text{and}\qquad  f_6 =  \frac{f_1 f_2 f_3}{f_0^2}~.
\end{equation}
The third relation then says $f_1f_2f_3 =0$, which implies $f_6=0$.  Inserting these expressions into equation \eqref{eq:AdS2CP2S2S2S1ee} we find $f_3^2 = 2f_0^2$, whence in particular $f_3 \neq 0$.  This implies that $f_1 f_2 =0$, so that at least one of $f_1,f_2$ is zero.  If $f_1=0$, then $f_4 =0$ as well and $\lambda_{S^2_{(1)}} = 0$.  If $f_2=0$, then $f_5=0$ as well, and now it is $\lambda_{S^2_{(2)}} = 0$.  Let us now consider the case $f_0=0$.  Then the relations \eqref{eq:AdS2CP2S2S2S1rels} become
\begin{equation}
 f_1 f_3 = 0, \quad f_2 f_3 = 0, \quad 2 f_4 f_5 + f_3 f_6 = 0, \quad f_2 f_4 + f_1 f_5 = 0~.
\end{equation}
If $f_3 \neq 0$, then $f_1 = f_2 = 0$, which by equation \eqref{eq:AdS2CP2S2S2S1ee} implies that $f_3 =0$, contradicting the hypothesis.  So let us take $f_3=0$.  Then we have are left with $f_4 f_5 = 0$, $f_2 f_4 + f_1 f_5 = 0$ and 
$f_1^2 + f_2^2 = 2f_4^2 + 2 f_5^2 + f_6^2$.  If $f_4 \neq 0$, then $f_5 =0$ and $f_2=0$, but then the Einstein equation $f_1^2 = 2f_4^2 + f_6^2$ implies the vanishing of $\lambda_{S^2_{(1)}}$.  If $f_4=0$, then we are left with the equations $f_1 f_5 =0$ and $f_1^2 +f_2^2 = 2 f_5^2 + f_6^2$.  If $f_5 \neq 0$, then $f_1=0$ and $f_2^2 = 2 f_5^2 + f_6^2$ is precisely the vanishing of $\lambda_{S^2_{(2)}}$.  Finally, if $f_5=0$, then together with the vanishing of $f_0,f_3$ and $f_4$ we see that $\lambda_{\CP^2}=0$: a contradiction.

\bibliographystyle{utphys}
\bibliography{AdS,AdS3,ESYM,Sugra,Geometry,CaliGeo}

\end{document}